\documentclass[twocolumn]{aastex631}
\usepackage{tabularx}
\usepackage{longtable}
\LTcapwidth=\textwidth
\usepackage{amsmath}
\usepackage{rotating}
\begin{document}

%\title{Exploring the diversity of type IIn supernovae}
\title{Unveiling the Diversity of Type IIn Supernovae via Systematic Light Curve Modeling}

\author[0000-0003-4175-4960]{C.~L.~Ransome}
\affiliation{Center for Astrophysics \textbar{} Harvard \& Smithsonian, 60 Garden Street, Cambridge, MA 02138-1516, USA}

\author[0000-0002-5814-4061]{V.~A.~Villar}
\affiliation{Center for Astrophysics \textbar{} Harvard \& Smithsonian, 60 Garden Street, Cambridge, MA 02138-1516, USA}
\affiliation{The NSF AI Institute for Artificial Intelligence and Fundamental Interactions}

\begin{abstract}

Type IIn supernovae (SNe\,IIn) are a highly heterogeneous subclass of core-collapse supernovae, spectroscopically characterized by signatures of interaction with a dense circumstellar medium (CSM). Here we systematically model the light curves of 142 archival SNe\,IIn using \texttt{MOSFiT} (the Modular Open Source Fitter for Transients). We find that the observed and inferred properties of SN\,IIn are diverse, but there are some trends. The typical SN CSM is dense ($\sim$\,10$^{-12}$\,g\,cm$^{-3}$) with highly diverse CSM geometry, with a median CSM mass of $\sim$\,1\,M$_\odot$. The ejecta are typically massive ($\gtrsim10$\,M$_\odot$), suggesting massive progenitor systems. We find positive correlations between the CSM mass and the rise and fall times of SNe\,IIn. Furthermore there are positive correlations between the rise time and fall times and the $r$-band luminosity. We estimate the mass-loss rates of our sample (where spectroscopy is available) and find a high median mass-loss rate of $\sim$10$^{-2}$\,M$_\odot$\,yr$^{-1}$, with a range between 10$^{-4}$\,--\,1\,M$_\odot$\,yr$^{-1}$. These mass-loss rates are most similar to the mass loss from great eruptions of luminous blue variables, consistent with the direct progenitor detections in the literature. We also discuss the role that binary interactions may play, concluding that at least some of our SNe\,IIn may be from massive binary systems. Finally, we estimate a detection rate of 1.6\,$\times$\,10$^5$\,yr$^{-1}$ in the upcoming Legacy Survey of Space and Time at the Vera C. Rubin Observatory.

\end{abstract}


\section{Introduction} \label{sec:intro}

As a massive star exhausts its nuclear fuel, radiation pressure can no longer support the stellar core, leading to a core-collapse supernova (CCSN). The underlying engines, progenitor systems and immediate environments of stars shape the observed CCSN properties, leading to a zoo of spectral and photometric classes. Arguably, one of the most heterogeneous CCSN classes is the type IIn class (SN\,IIn), named such due to narrow features superimposed on emission lines. These narrrow features are notably (but not exclusively) seen on the hydrogen Balmer lines \citep[perhaps most obviously seen on the H$\alpha$ profile,][]{Filippenko_1989, Schlegel_1990, Filippenko_1997, Ransome_2021}. 

This unique spectroscopic signature points to a complex circumstellar environment. Narrow emission features arise when the SN ejecta collides with a pre-existing dense and slow circumstellar medium \citep[CSM;][]{Chugai_1991, Chugai_2004}. The resultant complex emission profiles in the Balmer series can be decomposed into a narrow component with a full width at half maximum (FWHM) of $\sim$\,10$^{2}$\,kms$^{-1}$, an intermediate width component (FWHM $\sim$\,10$^{3}$\,kms$^{-1}$) and a broad component (FWHM $\sim$\,10$^3$\,--\,10$^4$\,kms$^{-1}$), each arising from a unique component of the CSM or SN itself \citep{Chugai_2001, Chugai_2004, Dessart_2009, Humphreys_2012, Huang_2018}. Early CSM interaction features are common among CCSNe, arising from the photoionization of a confined CSM, or inflated H-rich envelope \citep[photoionized by high energy photons from the shock break out,][]{Khazov_2016, Jacobson-Galan_2024, Fuller_2024}. However, these features fade within roughly a week and are distinct from the CSM interaction characteristic of SNe\,IIn that are powered by shock interaction and endure much longer.

The dense CSM surrounding SN\,IIn must be created by the progenitor (or perhaps a companion) through mass-loss events towards the end of the star's life \citep{Smith_2014}. The progenitor mass loss may manifest from a diverse set of mechanisms. These mechanisms range from line-driven winds to dramatic outbursts, such as the great eruption of $\eta$\,Car in the 1880s in which the progenitor lost mass at a rate of $\dot{M}\,\sim$\,1\,M$_\odot$\,yr$^{-1}$. This event produced a massive CSM, in excess of 10\,M$\odot$ \citep{Smith_2003, Smith_2010}. Alternatively, the mass loss may be due to interactions in a close binary system \citep{Soker_2013, Kashi_2013, Smith_2014, Ercolino_2024}. A combination of these mechanisms may then produce a highly complex circumstellar environment. In some cases, mass-loss episodes are observed to precede the SN in the years or months prior to the terminal explosion \citep{Smith_2014_2009ip, Ofek_2014, Strotjohann_2020, Pessi_2021, Reguitti_2024}. Other mass-loss mechanisms may include unstable burning \citep{Smith_2014}, gravity wave driven pulsations \citep{Shiode_2014, Wu_2022} and outbursts due to pair instability \citep[for very massive progenitors, with an initial mass in excess of 100\,M$_\odot$][]{Woosley_2007, Blinnikov_2010, Woosley_2017}. However, for SNe\,IIn, high mass-loss rates are required to produce the solar masses of CSM in a relatively short period of time (the CSM interaction is typically prompt so the CSM is not distant). The required mass-loss rates are between 10$^{-3}$\,--\,10\,M$_\odot$\,yr$^{-1}$, which precludes the classic line driven winds of red supergiants \citep{Smith_2014, Smith_2017_review,Beasor_2020, Hiramatsu_2024}. 

A common suggested SN\,IIn progenitor system are luminous blue variables (LBVs), massive evolved stars that undergo dramatic mass loss episodes \citep{Kotak_2006, Smith_2014}. A small number of SNe\,IIn have direct progenitor detections from pre-explosion images. For example, SN\,2005gl had archival, pre-explosion \textit{Hubble Space Telescope} imaging, and it was found the progenitor was an LBV with mass in excess of 50\,M$_\odot$ \citep{Gal-Yam_2007, GalYam_2009}. It has recently been found, however, that the environments of SNe\,IIn are inconsistent with the environments of single massive stars \citep[i.e. they do not trace star formation, generally,][]{and12, hab14, Ransome_2022}, suggesting multiple progenitor types. Similarly, spectroscopic studies of the hosts of SNe\,IIn have unveiled that there may be a population of SNe\,IIn in a young environment, and a population in an older environment \citep[with peaks in age bins of 0\,--\,20\,Myr and 100\,--\,400\,Myr][]{Galbany_2018}. Proposed lower-mass progenitors include stars on the lowest end of the CCSN range (8\,--\,10\,M$_\odot$) that end their lives as an electron capture supernova \citep[ecSN, e.g. SN\,2011ht][]{Smith_2013}. It should be noted, however, that some ecSN candidates are more consistent with flash ionization than CSM-ejecta interaction due to the fleeting interaction features \citep[e.g. SN\,2018zd,][]{Zhang_2020, Hiramatsu_2021_18zd}. The more commonly observed CCSN progenitors, RSGs \citep[i.e. the progenitors of SNe\,IIP and IIL][]{Smartt_2009_IIP, Smartt_Review} may make up a proportion of the SN\,IIn progenitor channel. Based on environmental analyses, the distribution of the star-formation association of SNe\,IIn can be recreated with a mix of the environments of LBVs and RSGs \citep[][]{Kangas_2017}. However for enough material to be stripped from the star, a higher mass-loss rate than the canonical RSG wind, or a different mass loss mechanism is required \citep{Smith_2006}. \citet{Smith_2009_RSGprog} found that the mass loss of the galactic RSG, VY\,CMa was episodic. High resolution IR spectra of the circumstellar environment of this RSG showed that there are CO clouds from mass ejections occurring $\sim\,$hundreds of years ago. The mass-loss rate, luminosity, wind speed and mass were also found to be larger than that of Betelgeuse. Those authors conclude that the CSM produced by this star may result in a SN\,IIn.

While there are some commonalities between members of the SN\,IIn class, such as the narrow components on the Balmer profile, a blue continuum \citep[][]{Stathakis_1991, Turrato_1993} and being generally more luminous than other SN\,II with an average observed peak of $M_B$\,=\,--18.7\,mag \citep{Kiewe_2012}, the SN\,IIn class is highly heterogeneous. Some SNe\,IIn are intrinsically faint, peaking at $M_V$\,$\approx$\,--14\,mag \citep[e.g. SN\,2008S;][]{Botticella_2009, Adams_2016}. Other SNe\,IIn lay within the `standard' CCSN region in the luminosity-timescale space of exploding transients \citep{kas11, Villar_2017} with peak absolute magnitudes in the range --17\,--\,--19\,mag \citep{Li_2011, Kiewe_2012, tadd13}. Furthermore on the extreme of the luminosity axis of the luminosity-timescale, there are the superluminous SNe (SLSNe), where some type II SLSNe (SLSNe-II) may be powered by CSM interaction similarly to SNe\,IIn \citep[i.e. a SLSN-IIn][]{Smith_2007, Wang_2019}. Indeed, the SN\,IIn, SN\,2006gy was the most luminous SN observed at the time of discovery with peak absolute magnitude of $M_R$\,$\sim$\,--21\,mag \citep[][]{Smith_2007}.

Furthermore, the light curve morphologies of SNe\,IIn are highly diverse. On the extreme of the timescale axis of the luminosity-timescale phase space, some SNe\,IIn are long-lived in H$\alpha$ emission due to ongoing CSM interaction. Some SNe\,IIn remain observable for decades post explosion--for example, SN\,1978K which was only discovered in 1990, thought to be a classical nova initially but the SN was foudn in archival data \citep{1978k, Ryder_1993, Chugai_1995}. The earliest example of a long-lived SN\,IIn was one of the original three SNe\,IIn in the sample of \citet{Schlegel_1990}, SN\,1988Z \citep[with long-lived SNe\,IIn sometimes being termed SN\,1988Z-like or IIn-E,][]{hab14, Branch_2017}. Perhaps the most well known long-lived SN\,IIn is SN\,2005ip. \citet{Stritzinger_2012} found that the CSM interaction was still ongoing more than six years post-explosion. The light curve decline stalled in a plateau which remained at an almost constant brightness of $\sim$\,--15\,mag for $\sim$\,5000 days. Furthermore, three years post-explosion, SN\,2005ip remained the strongest H$\alpha$ source in its host, NGC\,2906 \citep{hab14}. SN\,2005ip only began to decline more rapidly 5000 days post-explosion \citep{Fox_2020}. These long-lived SNe\,IIn seem to form a small but interesting sub-group which may have a more extended CSM compared to other SNe\,IIn or perhaps a central engine driving ongoing CSM interaction \citep[for a recent example of a long-lived SN\,IIn, see SN\,2017hcc;][]{Smith_2020_17hcc, Moran_2022, Chandra_2022}. Further to this long-lived behavior seen in some SNe\,IIn \citep[see also; ][]{Fox_2013}, `bumpy' light curves be observed which may indicate interaction with denser regions of CSM, perhaps from more discrete mass loss events from the progenitor \citep[e.g. SN\,2006jd and iPTF13z][]{Stritzinger_2012, Nyholm_2017}. The light curve shapes that characterize the SN\,IIP (with recombination-wave driven plateaus) and SN\,IIL (linearly declining) classes are seen in some SNe\,IIn, designated SN\,IIn-P \citep[e.g. SN\,2011ht and PTF\,11iqb][]{Mauerhan13a, Smith_2015} and SN\,IIn-L \citep[e.g. SN\,1998S and SN\,1999el][]{DiCarlo_2002, tad15}. These may be considered members of a continuum of pre-supernova mass loss, between that of more standard SNe\,II and `classical' SNe\,IIn.

The SN\,IIn class is highly heterogeneous, with great photometric, spectral and environmental diversity. With the advent of large transient surveys in the past couple of decades, we now have a large population of SNe\,IIn including hundreds of events. In this paper, we perform the first systematic physical modeling of a large sample of multi-band SNe\,IIn light curves, with the aim to infer their physical properties and the CSM that surround them. Throughout this work, we assume a standard reddening law, with $R_V = A_V / E(B-V) = 3.1$ according to \citet{Cardelli_1989}. We assume a standard flat $\Lambda$CDM cosmology with H$_0$\,=\,67.7\,kms$^{-1}$Mpc$^{-1}$ and $\Omega_\mathrm{M}$\,=\,0.307 \citep{Planck_2015}. This paper is organized as follows: in Section\,\ref{sec:model}, we describe the models of SN\,IIn light curves and our approach to modeling our large sample, in Section\,\ref{sec:sample}, we outline our sample of 142 SNe\,IIn collected from various sources. In Section\,\ref{sec:curves} we present our results and analysis, including our light curves, inferred parameter distributions, tests for clustering of parameters and exploring correlations between parameter pairs and then discuss the diversity of SN\,IIn light curves. We outline the photometric groupings of our sample in Section\,\ref{sec:groups}. The mass-loss rates of the progenitors are presented in Section\,\ref{sec:mdot}. In Section\,\ref{sec:progenitors} we discuss the implications of our findings for the possible progenitor routes of our SNe\,IIn. In Section\,\ref{sec:future}, we describe future developments and present an estimate on the rates of SNe\,IIn in the upcoming wide field survey at the Vera C. Rubin Observatory. Finally, in Section\,\ref{conc}, we summarize our findings.

\section{Modeling the Light Curves of Type IIn Supernovae} \label{sec:model}

We adopt the model for ejecta-CSM interaction driven light curves that are used in the Modular Open Source Fitter for Transients \citep[\texttt{MOSFiT},][]{Guillochon_2017} developed by \citet{Villar_2017} and \citet{Jiang_2019} which builds on earlier work by \citet{Arnett_1980, Arnett_1982, Chevalier_1982, Chatzopoulos_2012}. \texttt{MOSFiT}\footnote{\url{https://mosfit.readthedocs.io}} is an open-source Python-based package that performs Bayesian inference on the multi-band light curves of SNe \citep[and other thermal transients;][]{Guillochon_2017}. 

The ejecta-CSM interaction model follows the reverse and forward shocks produced by the interaction between the SN ejecta and CSM, both of which are assumed to be spherically symmetric. This interaction converts kinetic energy into thermalized radiation, 

\begin{equation}
    \begin{split}
   L & = \epsilon \frac{dE_{\mathrm{kin}}}{dt} = \frac{\epsilon}{2} \frac{d}{dt}(M_{\mathrm{sw}} v_{\mathrm{sh}}^2) \\
    & = \epsilon \left(M_{\mathrm{sw}} v_{\mathrm{sh}} \frac{dv_{\mathrm{sh}} }{dt} + \frac{1}{2}\frac{dM_{\mathrm{sw}}}{dt} v_{\mathrm{sh}}^2\right)   
    \end{split}
\end{equation}
where $\epsilon$ is the efficiency coefficient of the conversion of kinetic energy to radiation in the CSM-ejecta interaction. Here, $\epsilon=0.5$, is assumed (consistent with typical literature assumptions, e.g., \citealt{vanMarle_2010, Moriya_2013_6gy, moriya13, Dessart_2015, Villar_2017}). $v_{\mathrm{sh}}$ is the shock velocity, and the swept up mass is M$_\mathrm{sw}$. 

Using the sample outlined in Table.\,\ref{sample} and Section\,\ref{sec:sample}, along with the parameters described in Table.\,\ref{tab:params}, we use \texttt{MOSFiT} to fit the light curves of our sample of SNe\,IIn. The parameters in the CSM-ejecta interaction model are as follows: $M_{\mathrm{CSM}}$ is the CSM mass, $M_{\mathrm{ej}}$ is the mass of the SN ejecta, $s$ is the CSM density profile exponent where $\rho_{\mathrm{CSM}}$\,$\propto$\,$r^{-s}$ where $s\,=\,0$ would represent a shell-like CSM and $s\,=\,2$ would represent a wind like CSM profile. $n$ is the inner ejecta density profile parameter and $\delta$ is the outer ejecta density profile parameter (which we fix at 0) and are both indices in a power law that describes the SN ejecta density profile. The value of $n$ is determined by the progenitor type, $n\,=\,7$ would suggest a compact progenitor, and $n\,=\,12$  would represent a more extended progenitor. \citet{Chevalier_1982} describe the ejecta density profile as a broken power law dependent on $n$ and $\delta$ as indices with $\rho_{\mathrm{SN}}\,=\,g^{n}t^{n - 3}r^{-n}$ where $g^n$ is given by,

\begin{equation} \label{eq:g}
    g^n = \frac{1}{4 \pi (\delta - n)} \frac{[2(5 - \delta)(n - 5)E_{\mathrm{KE}}]^{\frac{n-2}{2}}}{[(3 - \delta)(n - 3) M_{\mathrm{ej}}]^{\frac{n-5}{2}}}
\end{equation}
where E$_{\mathrm{KE}}$ is the SN kinetic energy. 

The inner ejecta density parameter, $n$, may determine whether the input luminosity increases or decreases for $t\,<\,t_{\mathrm{FS}}$, where $t_{\mathrm{FS}}$ is the shock crossing time. This is also dependent on if the CSM geometry is shell- or wind-like respectively (as $L_\mathrm{inp}\,\propto\,\frac{2n + 6s - ns - 15}{n - s}$). Brighter transients result if $n = 12$ for a shell-like CSM geometry ($s\,=\,0$) compared to the wind-like ($s\,=\,2$) case. $r_0$ is the inner CSM radius.  $v_{\mathrm{ej}}$ is the representative velocity of the SN ejecta and $\rho_0$ is the density of the CSM at the inner radius of the CSM.  $t_{\mathrm{exp}}$ is the time of explosion, relative to the first observation. $T_{\mathrm{min}}$ is the minimum photospheric temperature--i.e., the temperature at which the photosphere begins to recede in our models. Finally, $N_\mathrm{H}$ is the host hydrogen column density.

We used nested sampling via \texttt{dynesty} \citep{dynesty} with 1000 live points to fit each light curve. Convergence is reached when when the marginalized likelihood between samples stabilizes for a given stopping value. This is when the change in the log-evidence (dlogz) is below a given threshold (in this case, dlogz$\,=\,0.1$). The redshift and Milky Way extinction are fixed using the values recorded in the literature or YSE-PZ \citep{YSE-PZ} which uses \texttt{dustmaps} \citep{Green_2018} and \texttt{extinction}\footnote{\url{https://github.com/kbarbary/extinction}} to calculate these values.

Additional quantities may be derived from the parameters inferred by \texttt{MOSFiT}, we calculate the mass-loss rate and the outer CSM radius. The CSM extent is calculated using,

\begin{equation} \label{eq:rcsm}
    \begin{split}
     R_{\mathrm{CSM}} = \left(\frac{3M_{\mathrm{CSM}}}{4 \pi \rho_0  r_0^s} + r_0^{3}\right)^\frac{1}{3} \\
    \end{split}
\end{equation}

This CSM extent can be thought of as a lower limit, as we are only able to probe CSM that interacts with the SN ejecta within the observational window. Indeed, radio observations may also reveal interaction with distant, more diffuse CSM (e.g., \citealt{Chandra_2012}). Furthermore, \citet{Fox_2011} found that around 15\% of SNe\,IIn exhibit late-time emission from pre-existing dust. This dust was formed from material ejected from the progenitors in LBV-like eruptions.

Finally, we estimate the CSM velocity using the full width at half maximum intensity (FWHM) of the narrow component of the H$\alpha$ emission line in velocity space \citep[as per][]{Ransome_2021} so that we can estimate mass-loss rates. The CSM velocity may vary considerably, from $\sim$10\,--\,10$^3$\,km s$^{-1}$, depending on the progenitor system. As we do not assume a constant mass-loss rate (i.e., $s\,=\,2$), the average mass-loss rate may be expressed as \citep[e.g.][]{Takei_2020},

\begin{equation} \label{eq:avmdot}
    \begin{split}
     \langle\dot{M}\rangle=\frac{4\pi v_{\mathrm{CSM}}}{R_{\mathrm{CSM}}}\int^{R_{\mathrm{CSM}}}_{r_0}r^2\rho_{\mathrm{CSM}}dr\\
     \approx\,\frac{4\pi Dv_{\mathrm{CSM}}}{3 - s}R_{\mathrm{CSM}}^{2 - s}
    \end{split}
\end{equation}
where we have assumed that the inner radius of the CSM is much smaller than the CSM extent (i.e. r$_0$\,$\ll$\,R$_{\mathrm{CSM}}$), and where $D\,=\,\rho_0r_0^s$. 

Furthermore, we can estimate mass-loss rates with other observational quantities, such as the luminosity. Utilizing the luminosity arising from the CSM-ejecta interaction, the mass-loss rate can be represented as, 

\begin{equation} \label{eq:smithmdot}
    \dot{M} = 2 L \frac{v_{\mathrm{CSM}}}{v_{\mathrm{sh}}^3}
\end{equation}
where $L$ is the luminosity, $v_{\mathrm{CSM}}$ is the CSM velocity, and $v_{\mathrm{sh}}$ is the velocity of the shock \citep[e.g.][]{Smith_2017_review, Dickinson_2024, Dukiya_2024}. Those authors assume the shock velocity to be 2500\,km\,s$^{-1}$ \citep[see also,][]{Smith_2014}. In this work, however, we estimate the shock velocity from our model in \texttt{MOSFiT}, which traces the position of the shock front (both forward and reverse shock fronts) over time.

 The shock velocity can be directly calculated from spectral modeling \citep[e.g.][]{Brennan_2023}. However, a relatively high-resolution spectroscopic measurement is required for these calculations. In lieu of these measurements, we can take the time derivative of the position of the reverse shock, as calculated in \texttt{MOSFiT}, which uses the solutions of \citet{Chatzopoulos_2012}, 

\begin{equation}
    \frac{dr_{\mathrm{RS}}}{dt} = \frac{d}{dt}\Big(r_0 + \beta_r\Big[\frac{(Ag^n)}{D}\Big]^{\frac{1}{n - s}} t^\frac{n - 3}{n - s}\Big)
\end{equation}
where $r_{\mathrm{RS}}$ is the position of the reverse shock front at time $t$. $\beta_r$ and $A$ are constants that are dependent on $s$ and $n$. We assume a constant v$_{\mathrm{sh}}$ for these mass-loss rate calculations, taking the average value over the first 100 days after $t_{\mathrm{exp}}$, the time of explosion.

These inferred parameters, as well as being informative on the SN characteristics, may also inform on progenitor properties such as the mass-loss rates. These mass-loss rates may be linked to the progenitors themselves \citep[e.g.][]{Smith_2017_review}. Observed features can also be used to infer the physical parameters of SNe\,IIn. For example, the rise times, and peak luminosities can be used to constrain the CSM density, SN ejecta energy and ejecta mass \citep{moriya13, Moriya_2014}. The model used by those authors is similar to those of \citet{Chatzopoulos_2012} but differs in that that their models of the bolometric light curve assume that the unshocked CSM is optically thin (i.e., the diffusion time is much shorter than the shock-crossing time). Therefore this model traces the evolution of the shock more directly.

Recent efforts have also explored more complex light curve morphologies observed in SNe\,IIn such as bumps and rebrightenings which can not be reproduced in our model. \citet{Khatami_2023} approach this problem by considering different ejecta mass to CSM mass ratios and a `breakout' parameter. The breakout parameter is related to where shock breakout occurs, where the shock emerges from the outer CSM, and the degree to the deceleration of the shock. Those authors demonstrate that different sets of parameters (e.g. more CSM than ejecta or vice versa) can reproduce features such as the secondary `humps' seen in some SNe\,IIn such as SN\,2019zrk \citep[][]{Fransson_2022}. \citet{Chiba_2024} found that for $s\,\neq\,2$, models that use a dense CSM and follow the evolution of the shocks and where the shocks emerge from the CSM can reproduce various light curve shapes. These models can also reproduce the long plateaus characteristic of the long-lived SNe\,IIn. A notable example of a SN\,IIn with a clear rebrightening is SN\,2021qqp \citep{Hiramatsu_2024}. The secondary bump in this object was likely due to the SN ejecta interacting with material ejected from a violent precursor mass loss event, and was reproduced by pre-SN eruption models adapted from \citet{Matsumoto_2022}.

Another photometric category of SNe\,IIn that has been explored are superluminous SNe\,IIn. In particular, \citet{Dessart_2015} present radiation-hydrodynamic simulations of interaction powered superluminous SNe (i.e. SLSNe-IIn) in 1- and 2-D. Those authors use a multi-group radiation transport approach implemented in the \texttt{HERACLES} code in order to model the photometric and spectroscopic properties with a more realistic treatment of opacities of SLSNe\,IIn. The models of \citet{Dessart_2015} reproduce the evolution of photometry, spectroscopy and polarimetry of SLSNe\,IIn based on input explosion energetics, ejecta characteristics, mass-loss rates and CSM configuration. These simulations were able to reproduce the observed photometric, spectroscopic and polarization characteristics of the well known SNe\,IIn SN\,2006gy and SN\,2010jl. Given the simplifying assumptions of our model, we directly compare our models to those of \citet{Dessart_2015}.  We find a general agreement in the inferred CSM masses, presented in Appendix\,\ref{sec:dessart}, suggesting that \texttt{MOSFiT} provides reliable CSM estimates for the range explored here.

\section{The Sample} \label{sec:sample}

\begin{table*}
	\centering
	\caption{Sources of our sample of spectroscopically confirmed and photometrically monitored SNe. The number of SNe\,IIn are recorded along with the reference for the source. Some of the YSE sample overlaps with the ZTF sample. In these cases, we consider then as YSE SNe.}\label{sample}
	
	\begin{tabular}{llllr} 
		\hline
		Source &Initial sample& Number & Cut& Reference\\
		\hline
        ZTF Bright Transient Survey& 118 & 85 & 33 & \citet{Fremling_2020, Perley_2020_sndemo}\\
        Palomar Transient Factory &42& 25 & 17 & \citet{Nyholm_2020}\\
        Gold SNe\,IIn &37& 8 & 29 & \citet{Ransome_2021} \\ 
        Pan-STARRS1 Medium Deep &24& 16 & 8& \citet{villar2019supernova} \\
        Young Supernova Experiment (DR1) &13& 9 &4 & \citet{ysedr1} \\	
		\hline
		\textbf{Total} &234& 142& 92& \\
		\hline
		
	\end{tabular}
\end{table*}

\begin{figure}[!t]
	\includegraphics[width=0.99\columnwidth]{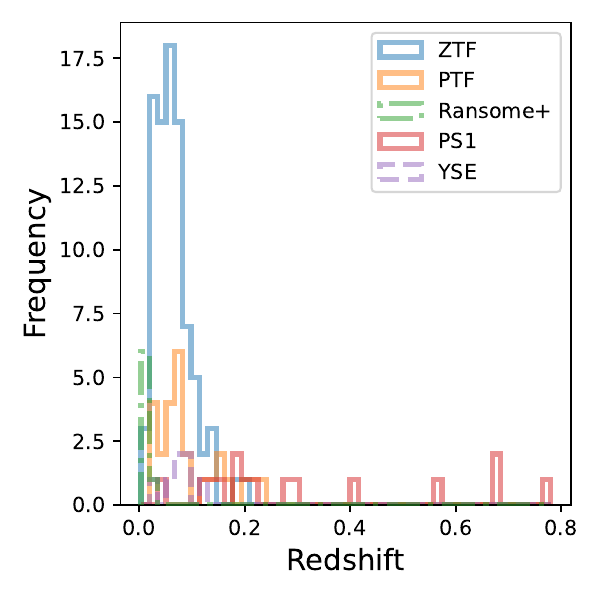}
    \caption{The redshift distribution of our SN\,IIn sample. The majority of our sample is relatively nearby with z\,$\textless$\,0.1 but this range extends to more distant redshift (z$\,\simeq$\,0.8) with the inclusion of SNe\,IIn discovered by the Pan-STARRS1 Medium Deep survey.}
    \label{fig:redshift}
\end{figure}

\begin{figure*}[!t]
	\includegraphics[width=0.99\textwidth]{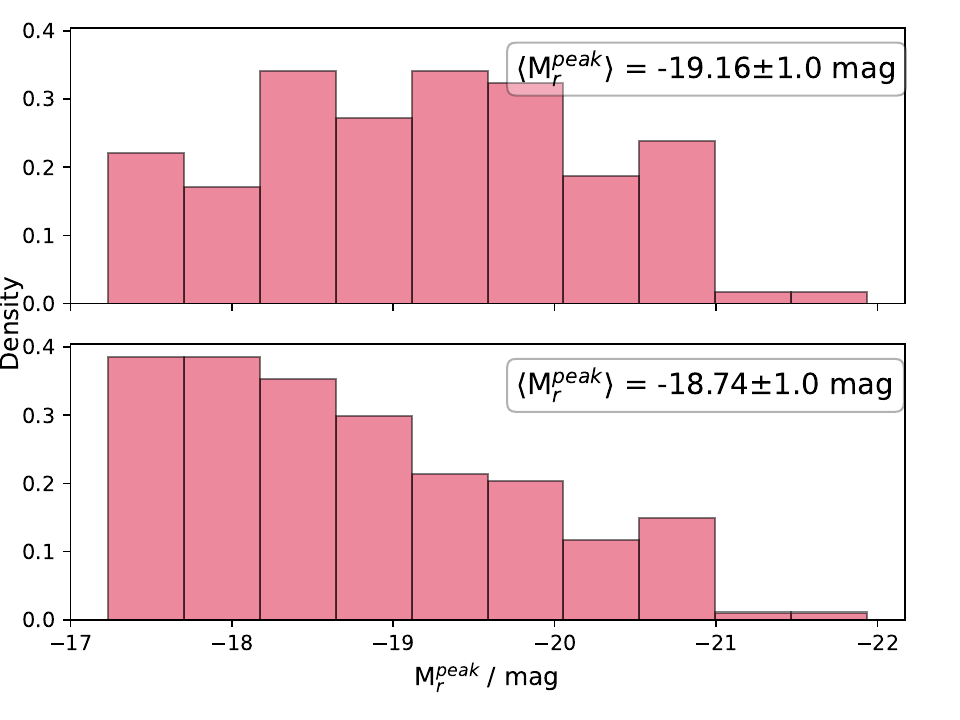}
    \caption{The peak $r$-band absolute magnitude distribution for our sample of SNe\,IIn. These distributions are corrected for host and Milky Way extinction. The top panel is the observed distribution and the bottom panel is Malmquist-corrected. }
    \label{fig:absr}
\end{figure*}

We select our sample following the schema described by \citet{Ransome_2021}. Our sample of SNe\,IIn span in explosion date from 1989\,--\,2023, covering a redshift range 0.003\,--\,0.780 (shown in Figure.\,\ref{fig:redshift}). The observed (uncorrected for Malmquist bias) mean absolute peak $r$-band magnitude of our sample is --19.2 with a standard deviation of 1.0\,mag; the peak $r$-band absolute magnitude distribution is shown in Figure \ref{fig:absr}. We also show the Malmquist corrected distribution in Figure\,\ref{fig:absr}; we describe this correction in Appendix\,\ref{sec:malm}. These SNe\,IIn have reliable spectroscopic classifications such that we can rule out misclassified SNe which exhibit early time flash ionization features. Our selected sample also requires sufficient photometric coverage for our inference (i.e., we exclude events with small observing baselines or with significant temporal gaps). Finally, we do not include objects where the reduced photometric data are not publicly accessible. Our sample and the sources are summarized in Table\,\ref{sample}. 

After thorough inspection of the reported SNe\,IIn, we note that some objects are likely nuclear transients (such as AGN), based on the object being highly central to their hosts and red color-evolution inconsistent with SNe\,IIn. If there is insufficient spectral information to confirm a SN\,IIn classification, we exclude the transient from our sample. Moreover, we exclude bona-fide SNe\,IIn which exhibit multiple peaks in their light curve \citep[e.g. SN\,2019zrk,][]{Fransson_2022, Soker_2022} or are extremely long-lasting (such as SN\,2005ip) as these cases are not an appropriate application of the \citet{Chatzopoulos_2012} models.

Our general selection criteria can be summarized as follows:
\begin{enumerate}
    \item There is at least one spectrum taken over a week post-discovery to rule out flash-ionization SNe \citep[see][]{Khazov_2016, Jacobson-Galan_2024}.
    \item The H$\alpha$ profiles should be complex (with at least 2 Gaussian components), Moreover the spectra should be SN spectra, rather than narrow host lines misinterpreted as interaction signatures \citep[see][]{Ransome_2021}.
    \item There is data covering a baseline of at least 50 days for our inference. 
    \item The SNe should not have clear `bumps' in the light curves \citep[e.g., dip and rise again as in the case of SN2021qqp;][]{Hiramatsu_2024}.
\end{enumerate}

Our SNe\,IIn were selected from both the literature and large optical surveys. Our largest source for SNe\,IIn (with 85 members) is the Zwicky Transient Facility Bright Transient Survey (ZTF\,BTS). The SNe\,IIn from the ZTF\,BTS \footnote{https://sites.astro.caltech.edu/ztf/bts/bts.php} are spectroscopically confirmed \citep[e.g. using the SEDMachine][]{Blagorodnova_2018} and are typically observed in two filters ($g$- and $r$-bands) \citep{Fremling_2020, Perley_2020_sndemo}. The ZTF\,BTS spectroscopic sample includes only transients with apparent magnitudes $<$\,19\,mag. The transients in this sample extend to a redshift of $\sim$\,0.2. Of the excluded transients from this subsample (of a total of 119), 15 are excluded for having a short baseline, 13 have either no publicly available spectra or the available spectra were within 7 days post-discovery and 5 had bumps or double peaks. One additional object is central to its host and has a very red peak (with $g-r$\,$\approx$\,1\,mag), indicating this transient may be an AGN rather than a SN\,IIn.

Our next largest source of events is the Palomar Transient Factory \citep[PTF][]{Cenko_2006} and this sample is presented in the light curve analysis of \citet{Nyholm_2020}. The majority of the full sample of 42 from \citet{Nyholm_2020} have a single classification spectrum but the majority of these do have a spectrum after our 7 day cut. From these 42 SNe\,IIn, we use 25 in our analysis. The photometry for this subsample were obtained from WISErEP\footnote{\url{https://www.wiserep.org}} \citep{wisrep}. These SNe extend to a redshift of $\sim$\,0.2. Of this sample, 12 of the SNe had a short baseline or data quality issues (one transient had large uncertainties on all points); 4 had an ambiguous classification spectrum (either within our cut off range or exhibiting strong host lines) and one had a bumpy light curve.

We incorporate 8 SNe\,IIn which were presented by \citet{Ransome_2021} from their `gold' sample. These objects are from the literature and are well-studied transients with well-sampled spectroscopy, exhibiting CSM interaction signatures for multiple epochs. All events in this sample are within $z\,<=\,0.02$. These observations include data from numerous filters from multiple instruments and telescopes. In their full gold sample, \citet{Ransome_2021} identify 37 transients. We cut 29 of these objects from our sample. Of these cut SNe, 22 had no easily obtainable public photometric data, 4 were long-lived SNe\,IIn, 2 were thermonuclear and one is possible a SN impostor. 

The 16 SNe\,IIn from the Pan-STARRS1 Medium Deep Survey \citep{Chambers_2016} are the spectroscopically confirmed sample of SNe\,IIn \citep{villar2019supernova}. These SNe\,IIn tend to be more distant (out to redshift  $\sim$\,0.78) than the relatively nearby transients in our other sources and are observed in 5 bands, Pan-STARRS (\textit{grizy}). The full Pan-STARRS1 sample consisted of 24 SNe. Out of the 8 excluded transients from this sample, 7 had short baselines and another had an incorrect redshift measurement.

Finally, we incorporate an additional 8 SNe\,IIn from the first data release (DR1) of the Young Supernova Experiment \citep[YSE;][]{YSE, ysedr1}. Of these, 3 overlap with ZTF\,BTS in this sample but these 8 SNe were discovered by YSE in the Pan-STARRS $griz$ filters. In total there are 13 SNe\,IIn from the YSE\,DR1 subsample. Out of these 5 excluded SNe, 3 have a spectrum within one week of discovery and one has a short baseline.

As our data originate from various sources, a small amount of cleaning was required. This mostly takes the form of treating data points with a signal-to-noise ratio of less than 3 as an upper limit. We also truncate earlier times where precursor emission was identified, but including the `main' event (which would be much brighter, this was only seen in two objects). After data filtering, our sample consists of 142 SNe\,IIn (with 92 transients removed from an initial sample of 234, outlined in table\,\ref{sample}).

\begin{table*}
	\centering
	\caption{The \texttt{MOSFiT} CSM model prior distributions used in the fitting of our SN\,IIn light curves. The median values from the joint posterior distributions are also shown with spread from the 16 and 84 percent confidence intervals.}\label{tab:params}
	
	\begin{tabular}{lllr} % four columns, alignment for each
		\hline
		Parameter & Value or range of values & Log-uniform sampling & Median and confidence intervals\\
		\hline
		$N_\mathrm{H}$ & 10$^{16}-$-10$^{23}$\,cm$^{-3}$ & True& 18.9$^{+1.9}_{-1.9}$\,cm$^{-3}$\\ 
		$t_{\mathrm{exp}}$ & -20\,--\,0\,days &False&-10.4$^{+5.7}_{-7.7}$\,days\\
		$T_{\mathrm{min}}$ & 1\,--\,10$^4$\,K & True& 3760$^{+3990}_{-3740}$\,K\\
		$s$ & 0\,--\,2 & False& 1.27$^{+0.55}_{-0.88}$\\
		$n$ & 7\,--\,12 & False& 9.27$^{+1.82}_{-1.78}$\\
		$\delta$ & 0 & Fixed& -- \\
		$r_\mathrm{0}$ & 1\,--\,100\,AU & True&12.3$^{+33.7}_{-9.0}$\,AU\\
		$M_{\mathrm{CSM}}$ & 0.1\,--\,50\,M$_\odot$ & True& 0.08$^{+0.69}_{-0.57}$\,M$_\odot$\\
		$M_{\mathrm{ej}}$ & 1.0\,--\,50\,M$_\odot$ & True& 20.0$_{-15.0}^{+19.5}$\,M$_\odot$\\
		$v_{\mathrm{ej}}$ & 10$^3$\,--\,10$^5$\,km\,s$^{-1}$ & True& 4810$^{+3454}_{-2082}$\,km\,s$^{-1}$\\
		$\rho_\mathrm{0}$ & 10$^{-15}$\,--\,10$^{-11}$\,g\,cm$^{-3}$ & True& --11.3$^{+0.83}_{-1.05}$\,g\,cm$^{-3}$\\
        $\epsilon$ & 0.5 & Fixed& --\\
        $\sigma$ & 10$^{-5}$\,--\,1.0\,mag & True& -1.2$^{+0.4}_{-1.3}$\\
		\hline

	\end{tabular}
\end{table*}

\section{Exploring the Sample} \label{sec:curves} 

\begin{figure*}
	\includegraphics[width = \textwidth]{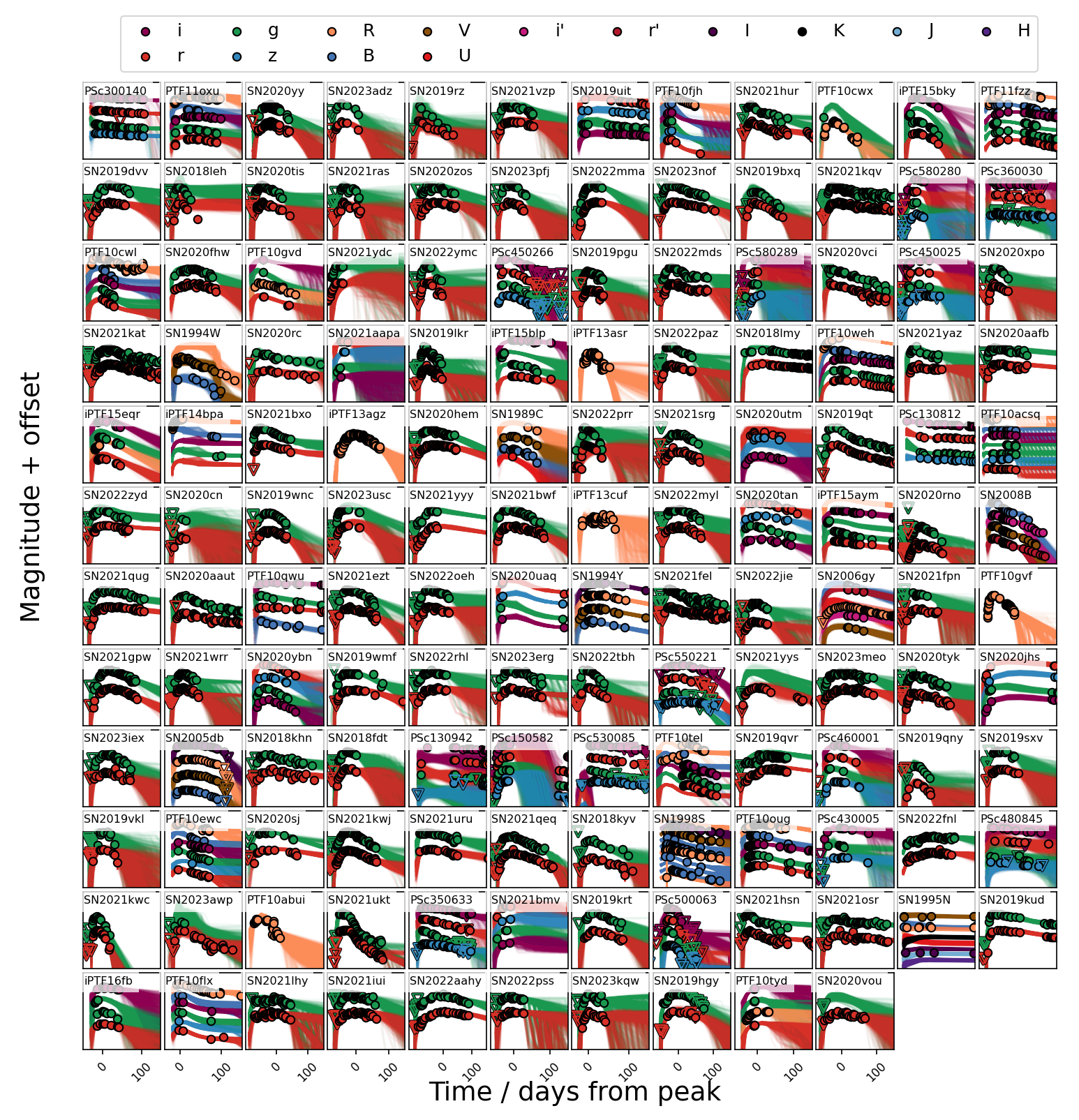}
    \caption{The photometric data and the \texttt{MOSFiT} fits for each of our 142 SNe\,IIn. Each band that is fit is offset for clarity. The time range covers the full range of the observed data. Detections are shown as solid circles and upper limits are shown as upended unfilled triangles. Note that the x-axes are not aligned for each subplot. The x-axis labels show time relative to peak.}
    \label{fig:lc}
\end{figure*}

\begin{figure*}
	\includegraphics[width=0.99\textwidth]{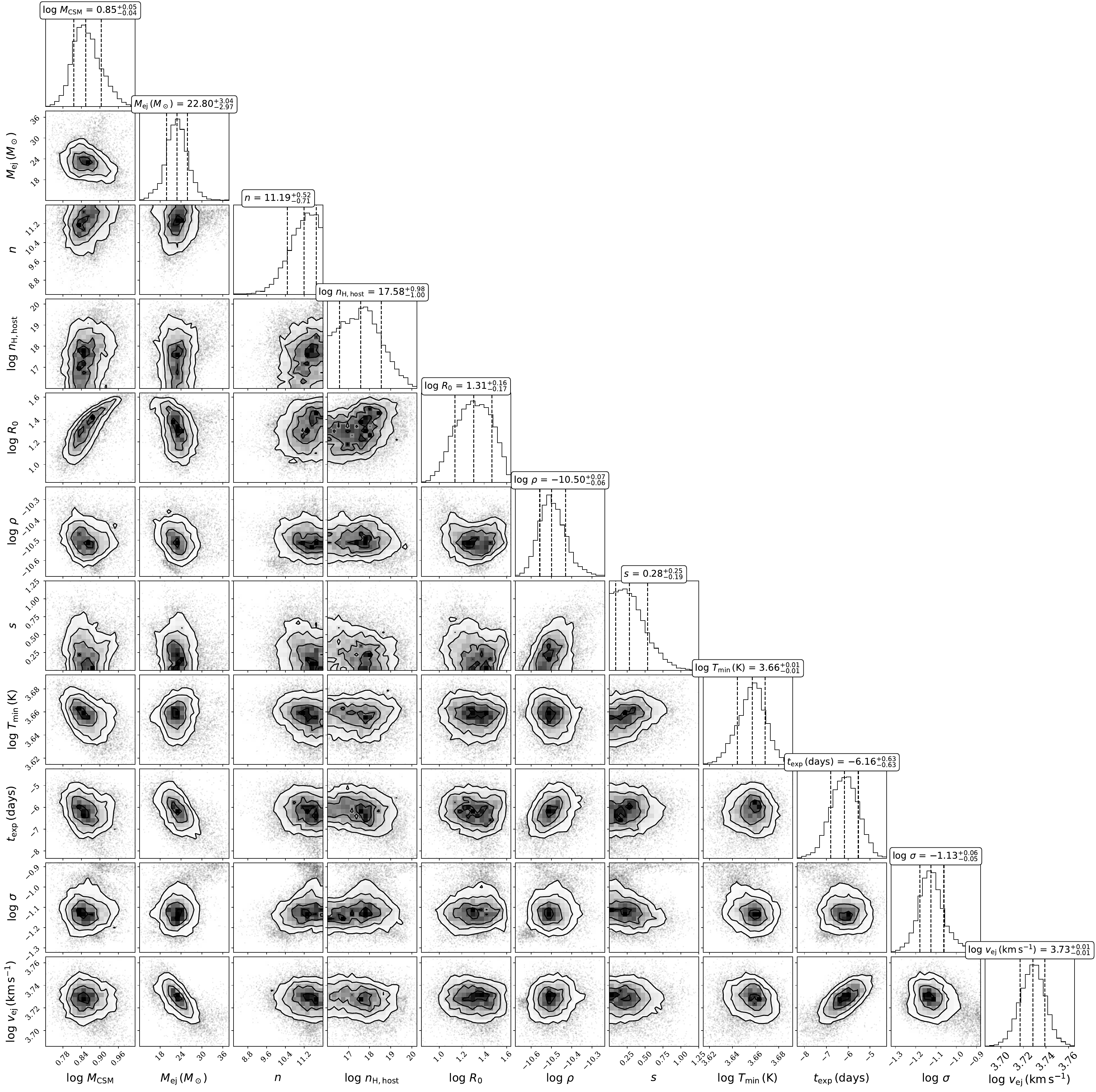}
    \caption{Corner plot for SN\,2021qug, one of the SNe\,IIn from the ZTF\,BTS subsample with data in the \textit{gr}-bands. Shown are the posterior distributions of each parameter used in the model implemented by \texttt{MOSFiT}, with the median, and 16 and 84 percent confidence intervals on the distribution.}
    \label{fig:corner}
\end{figure*}

Our full set of light curve fits and inferred parameters for our sample of 142 SNe\,IIn are presented in online tables and figure sets. We also present the light curves of our sample in Figure\,\ref{fig:lc}. These light curves show the data and models for each filter per SN with an offset for legibility. Also plotted are the realizations (independent samples from the posterior distribution) inferred by \texttt{MOSFiT}. Complementary to these light curves is an example parameter corner plot (Figure\,\ref{fig:corner}), showing the marginalized posterior distributions inferred by our \texttt{MOSFiT} light curve modeling. 

\subsection{Inferred Supernova Properties} \label{sec:params}

We first examine the joint posterior distributions for the inferred SN and CSM properties from \texttt{MOSFiT}. The median parameter values and the 16th/84th percentile spread in the distribution are shown in Table\,\ref{tab:avparams}. Note that these values are calculated using \textit{all} posterior realizations. Furthermore, to assess the quality of our fits, we consider a white noise parameter, $\sigma$ (in units of magnitude), which is inferred by \texttt{MOSFiT}. This gauges the underestimation of uncertainty in the data. The median white noise value is log($\sigma$)\,=\,$-1.2$\,mag. This metric suggests that there may be some unaccounted for uncertainties or short term variability in the data, which is then compensated for with the $\sigma$ parameter in our fits. We next explore the most interesting physical parameters in more detail.

\begin{figure*}
    \centering
	\includegraphics[width=0.95\columnwidth]{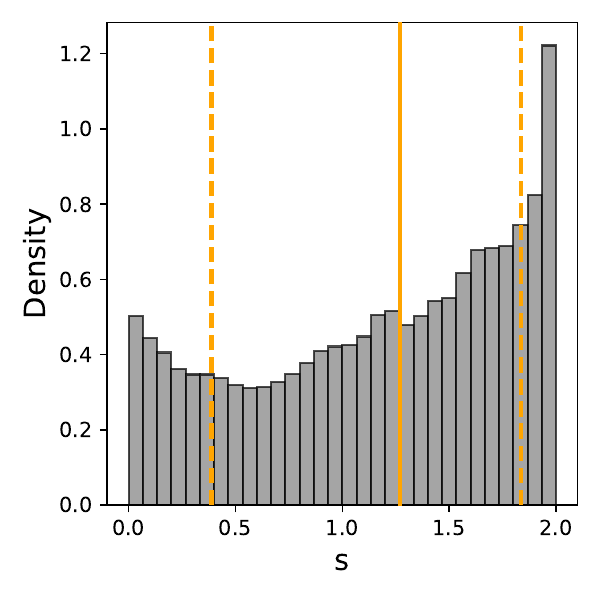}
    \includegraphics[width=0.95\columnwidth]{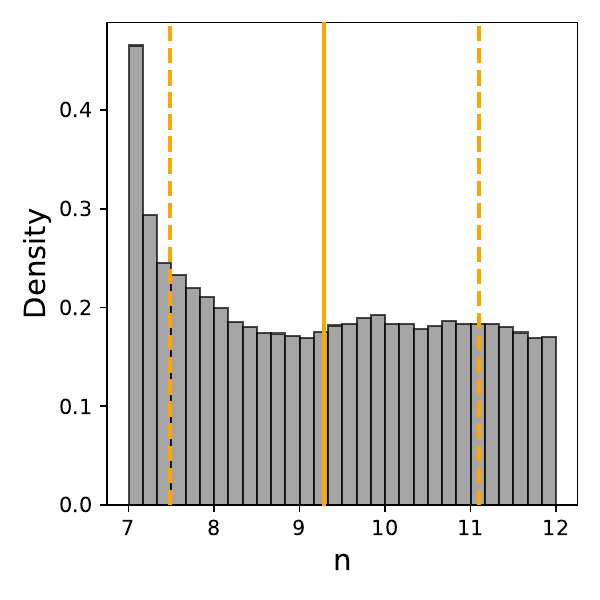}
    \includegraphics[width=0.95\columnwidth]{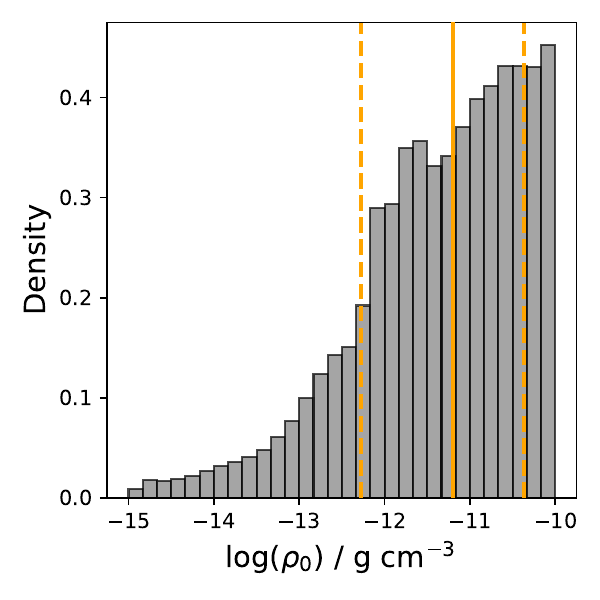}
    \includegraphics[width=0.95\columnwidth]{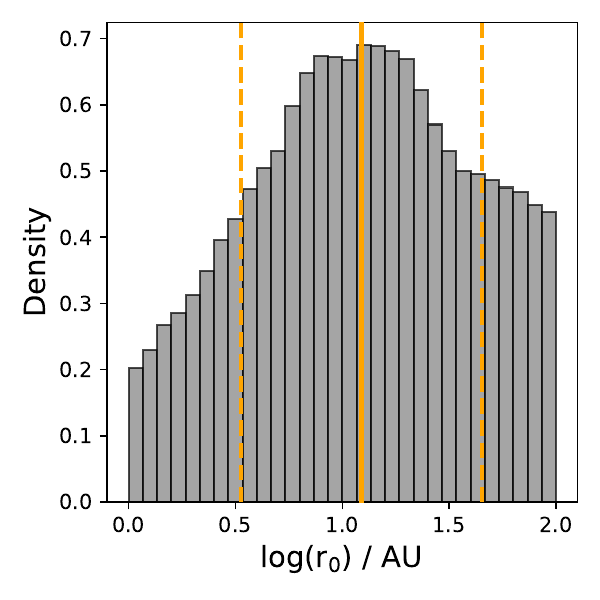}
    \caption{Histograms showing parameter distributions for our SNe\,IIn, calculated by \texttt{MOSFiT} light curve modeling. \textit{Top left:} the distribution of the geometric parameter that described the CSM density profile, $s$. \textit{Top right:} the distribution of $n$, which governs the inner SN ejecta density profile. \textit{Bottom left:} the distribution of the inner density of the CSM in log-space. \textit{Bottom left:} the distribution of the inner CSM radius.}
    \label{fig:s_n_rho_r0_hist}
\end{figure*}

The parameter that governs the CSM density profile, $s$, is left as a free parameter and varied between 0\,--\,2. This parameter is informative on the progenitor mass loss mechanisms, although it is often fixed to either $s\,=\,0$ and $2$ or assumed to be the latter \citep{moriya13, Ofek_2014_lcs}. Note that $s\,=\,0$ corresponds to an eruptive mass-loss episode, while $s\,=\,2$ corresponds to a constant (`wind-like') mass loss. The full distribution of inferred $s$ values is shown in the top left panel of Figure \,\ref{fig:s_n_rho_r0_hist}. It can be seen from this distribution that there is a peak towards higher values of $s$, i.e. more wind-like; however, the joint distribution spans the prior range, showing great diversity. Generally, $s$ is well constrained when the individual transient posterior distributions are considered\footnote{Presented in Appendix\,\ref{sec:indposts} in Figure\,\ref{fig:ind_post}.}. This joint posterior distribution has a median and spread of 1.27$^{+0.55}_{-0.88}$, with a typical uncertainty of $\sim$\,0.2 for individual events. While there is a slight preference towards `wind-like' constant mass loss, generally, the SN IIn sample does not undergo mass loss at a steady rate. 

The distribution of the SN ejecta geometric parameter, $n$, which was varied between 7 and 12, is shown in the top right panel of Figure \ref{fig:s_n_rho_r0_hist}. This parameter is related to the polytropic index of the core of the progenitor. For example, a red supergiant-like progenitor is consistent with $n\,=\,12$, and an LBV or Wolf-Rayet-like progenitors may have $n$ values between 7 and 10 \citep[e.g.][]{Colgate_1969, Matzner_1999}. We find a fairly uniform spread in the distribution of $n$; however there is a preference to lower values.  Similarly to $s$, in the literature, $n$ is often fixed. Typically, values of 7, 10 or 12 are assumed \citep[e.g.][]{Chevalier_2011,Moriya_2013, Moriya_2014}. The median value and spread of the joint posterior distribution of $n$ are 9.27$^{+1.82}_{-1.78}$. The typical uncertainty for individual SNe is $\sim$\,1, i.e. our posteriors for $n$ are broad and therefore only somewhat informative. Therefore, the wide spread in $n$ seen in Figure \ref{fig:s_n_rho_r0_hist} may be due to model uncertainty.

The bottom left panel of Figure \ref{fig:s_n_rho_r0_hist} shows the full posterior distribution of the inner CSM density, $\rho_0$. We vary $\rho_0$ between 10$^{-15}$\,--\,10$^{-10}$\,g\,cm$^{-3}$, consistent with theoretical studies of CSM interaction powered SNe \citep[e.g.,][]{Dessart_2015, Yaron_2017, Tsuna_2023}. This range also covers the higher densities found for interaction powered SLSNe-II(n) \citep[for example, the SLSN-IIn, ASSASN-15ua][was found to have a CSM density of $\approx$\,10$^{-10}$\,g\,cm$^{-3}$]{Dickinson_2024}. The majority of our SN\,IIn sample require high CSM densities, with the distribution peaking at $\sim$\,10$^{-12}$\,g\,cm$^{-3}$. The median inner CSM density and spread are log($\rho_0$)\,=\,--11.3$^{+0.8}_{-1.1}$\,g cm$^{-3}$ with a typical individual uncertainty of $\sim\,0.3$\,(g cm$^{-3}$). This distribution is skewed to the denser end of our prior distribution. While this distribution does tend towards a high density CSM, the individual posterior distributions (shown in Appendix\,\ref{sec:indposts}) generally do not cut off at the prior edge and are typically well constrained.  

The distribution of the inner CSM radius, r$_0$ is shown in the bottom right panel of Figure \ref{fig:s_n_rho_r0_hist}. This parameter is varied between 1\,--\,100\,AU, representative of the radii of evolved stars \citep[assuming the inner CSM radius is similar to the progenitor radius, e.g.][]{Fassia_2001, Morozova_2017, Moriya_2017, Dessart_2022}. This range is also consistent with the prior distribution used by \citet{Villar_2017}. It is apparent that the inner radius of the CSM is typically a few 10s\,AU, or of the order 10$^{14}$\,cm with a median $r_0$ and spread of 12.3$^{+33.7}_{-9.0}$\,AU. The typical uncertainty is $\sim$\,0.3\,AU. 

\begin{figure*}
    \centering
	\includegraphics[width=0.95\columnwidth]{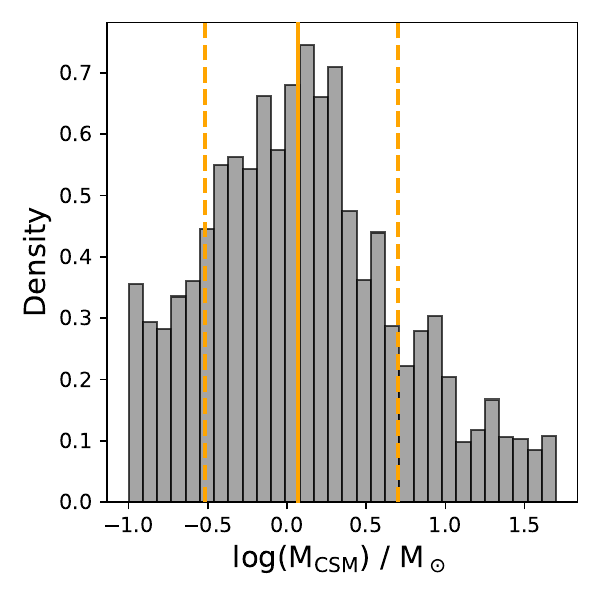}
    \includegraphics[width=0.95\columnwidth]{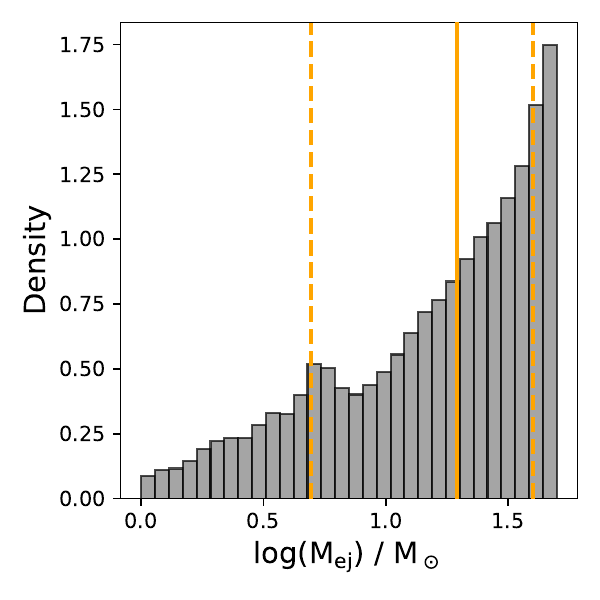}
    \includegraphics[width=0.95\columnwidth]{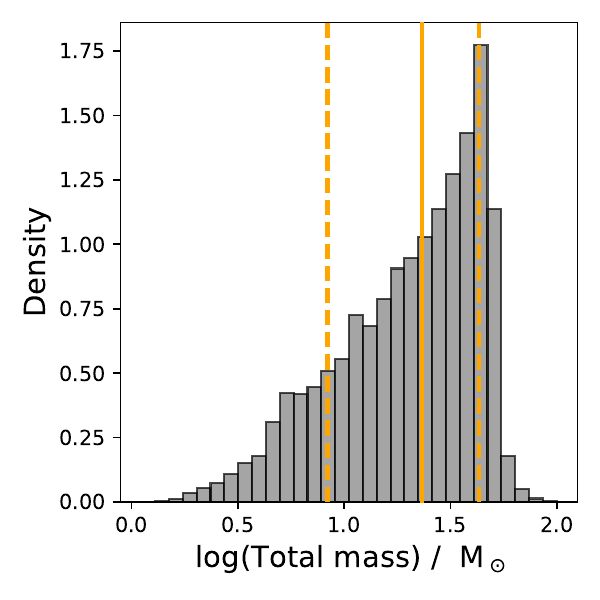}
    \includegraphics[width=0.95\columnwidth]{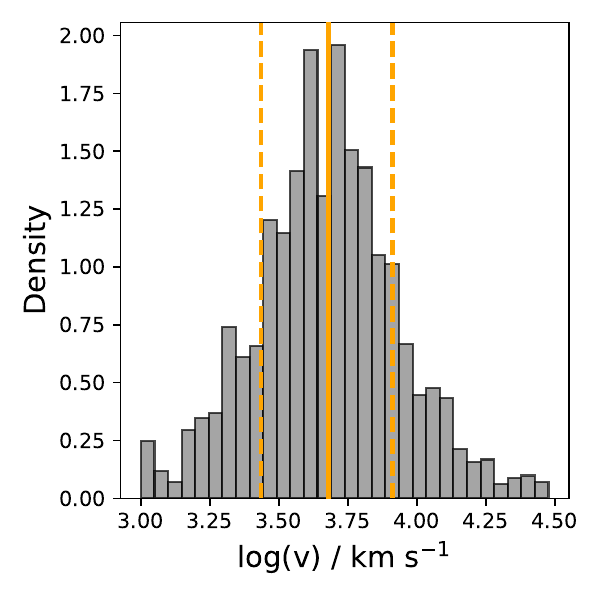}
    \caption{Histograms showing parameter distributions for our SNe\,IIn, calculated by \texttt{MOSFiT} light curve modeling. (\textit{Top left:}) the distribution of the CSM mass in log-space. (\textit{Top right:}) the distribution of the ejecta mass. (\textit{Bottom left:}) the distribution of the total mass (CSM mass + SN ejecta mass), roughly the preexplosion progenitor mass minus the mass of a compact remnant such as a neutron star or black hole. (\textit{Bottom left:}) the distribution of the outer CSM radius.}
    \label{fig:mcsm_mej_totmass_rcsm_hist}
\end{figure*}

The CSM mass distribution is shown in the top left panel of Figure \,\ref{fig:mcsm_mej_totmass_rcsm_hist}. We vary the CSM mass distribution between 0.1\,--\,50\,M$_\odot$. The lower end of this prior range is consistent with that used by \citet{Villar_2017} and probes examples of SNe\,IIn in the literature with lower CSM masses such as SN\,1994W \citep{Chugai_1994, Dessart_2009}. We extend the upper range of our prior distribution compared to \citet{Villar_2017}, however, to account for the high mass CSM seen around LBVs. The famous $\eta$\,Car, for example, has 10s of solar masses of CSM, \citep[e.g.][]{Smith_2003}. Other SNe\,IIn with estimated CSM masses in the literature include, SN\,2005gl \citep[0.03\,M$_\odot$][]{GalYam_2009, Smartt_Review}, SN\,2005ip \citep[a few M$_\odot$]{Smith_2009_05ip, Smith_2017_05ip, Fox_2020}, SN\,2010jl \citep[3\,--\,10\,M$_\odot$][]{Ofek_2014_10jl, Fransson_2014}, and SN\,2017hcc \citep[$\sim$\,10\,$_\odot$][]{Smith_2020_17hcc}. In our distribution, there is a clear peak around 1\,M$_{\odot}$ (with a median and spread of log(M$_{\mathrm{CSM}}$)\,=\,0.08$^{+0.69}_{-0.57}$\,M$_\odot$). For an individual SN, the typical uncertainty is around 0.3\,M$_\odot$, implying most are well-constrained. This CSM mass distribution is largely consistent with literature values. 

In the top right panel of Figure \,\ref{fig:mcsm_mej_totmass_rcsm_hist}, we present the joint posterior distribution for the ejecta mass. Our prior ranges between 1\,--\,50\,M$_\odot$. This range was chosen to account for progenitors which are stripped, and therefore have lower ejecta masses \citep[e.g. the low ejecta masses inferred for SN\,1988Z, SN\,2011ht and SN\,2020pvb][]{Chugai_1994_88z, Chugai_2016, Elias-Rosa_2024}, and also massive progenitors which may have large ejecta masses \citep[e.g. ASASSN-14il][]{Dukiya_2024}. The ejecta mass distribution peaks at the highest values of our prior. The median value of the ejecta mass and spread are 20.0$^{+19.5}_{-15.0}$\,M$_\odot$, consistent with more massive progenitor systems. The typical spread for any individual event is $\sim$\,7\,M$_\odot$, suggesting that the marginal posterior distributions are broad. 

The bottom left panel of Figure \,\ref{fig:mcsm_mej_totmass_rcsm_hist} shows the distribution of the total mass, which we define as,

\begin{equation}
    M_{\mathrm{tot}} = M_{\mathrm{CSM}} + M_{\mathrm{ej}}
\end{equation}

The total mass describes the lower limit of the mass of the progenitor prior to its death. It excludes the mass of a remnant and diffuse CSM beyond the limits our light curve probes. In Figure \,\ref{fig:mcsm_mej_totmass_rcsm_hist}, one can see that the population distribution roughly follows that of the ejecta mass, with a median total mass and spread of $\sim$24$^{+19}_{-16}$\,M$_\odot$. This is consistent with typical literature estimates for SN\,IIn progenitors \citep{Smith_2011,  Boian_2018, Brennan_2022_progenitor}.

The final parameter of note we discuss is the representative ejecta velocity. The velocity distribution is shown on the bottom right of Figure \,\ref{fig:mcsm_mej_totmass_rcsm_hist}. This distribution is apparently normally distributed around a median and spread of 4810$^{+3454}_{-2082}$\,kms$^{-1}$. The typical uncertainty in velocity is $\sim$\,400\,kms$^{-1}$ and the individual posteriors shown in Figure\,\ref{fig:ind_post} are well constrained. These velocities are typical of the ejecta velocities of SNe\,IIn measured from broad line profiles in spectra \cite[e.g.][]{Ransome_2021}.

Finally, we search for multi-modality within the parameter distributions described in this section. As SNe\,IIn are highly heterogeneous, probing for distinct groupings may prove informative on possible progenitor paths. In order to test for clustering in the SN\,IIn parameter distributions, we utilize a Gaussian mixture model approach \citep[GMM;][]{reynolds2009gaussian} with \texttt{scikit-learn} \citep[v1.4.2][]{scikit-learn}. The GMM attempts to describe our distributions with $k$ independent Gaussian distributions, each with differing means and variances. Using the Bayesian Information Criterion (BIC), we can determine the number of Gaussian distributions that best fit our data without overfitting. Here, we restrict our search to a combination of (at most) $k\,=\,3$ Gaussians, similarly to the analysis of \citet{Nyholm_2020}. We consider distributions to be multimodal if the BIC score for $k\,=\,2$ or $k\,=\,3$ is significantly lower ($\Delta$\,BIC\,$>\,10$) than the $k=1$ case. We note that this is a more stringent limit than used in previous work \citep[e.g.,][]{Sollerman_2009, Nyholm_2020}. 

As the individual posterior distributions themselves are not necessarily normally distributed, we test the validity of using the median posterior values for our GMM analysis. We pseudo-bootstrap our sample by taking each SN\,IIn light curve fit and randomly sampling one realization (i.e. the individual light curve fits) for each event. This is then repeated $10^3$ times. For each resampling, the GMM analysis is repeated, with the mean of each Gaussian component from the GMM being recorded along with the variance. We find that the number of Gaussian components for each physical parameter is consistent with the number of components when we use the median value of each posterior distribution. 

\begin{figure}[!h]
    \includegraphics[width=0.99\columnwidth]{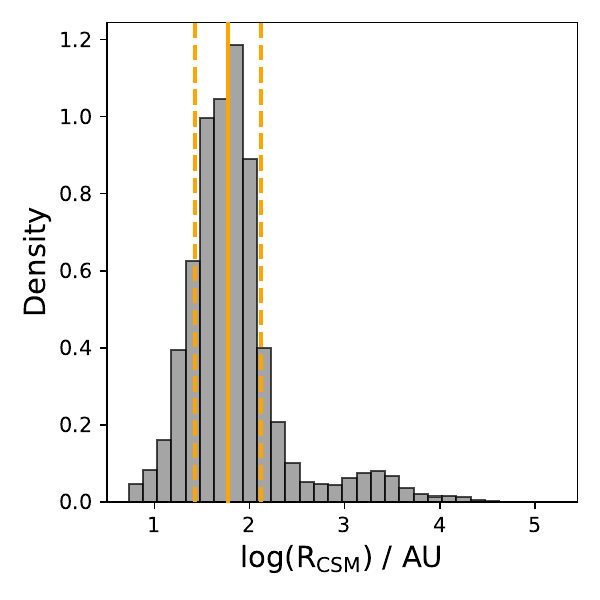}
    \caption{The joint posterior distribution of the CSM radius. The solid vertical line denotes the median value and the dashed vertical lines mark the 16 and 84 percent spread. There peak in the distribution around 100\,AU. We note that there is a an apparent tail to this distribution, but this is due to fits where bimodal solutions were found. Therefore, we do not consider true a true bimodality in the CSM distribution.}
    \label{fig:radcsm}
\end{figure}

Some quantities may be derived from the inferred parameters from \texttt{MOSFiT}. For example, the CSM radius is presented in the bottom right panel of Figure \,\ref{fig:radcsm}. The CSM typically extends to a few 10$^{15}$\,cm, consistent with the CSM radii in the literature \citep{Moriya_2017, Dessart_2022}. When we apply our GMM analysis to this distribution, we do not find multiple components. However there is a peak at around 80\,AU and an apparent tail in the distribution that extends out to 10$^3$\,--\,10$^4$\,AU. As mentioned before, these CSM radii may be a lower limit as the optical light curves of our SNe\,IIn track interaction with relatively close material. More diffuse CSM further from the progenitor may exist from winds or previous episodes of mass loss. 

To test for clustering in two dimensions, i.e. clustering in parameter pairs, we extend this GMM analysis. We fit two-dimensional Gaussian distributions with differing means and covariances to the parameter pair distributions. We do not find any clear examples of multiple components in the parameter pair distributions.

\subsection{Observational Features} \label{sec:obs}

In addition to the parameters derived from \texttt{MOSFiT}, we extract observational features from our SN sample: the rise time, fall time and peak brightness in $r$-band (used to compute the peak $r$-band absolute magnitude). We define rise time as the time taken to rise from $t_{\mathrm{exp}}$ to peak $r$-band brightness as calculated from the model light curves. We define the fall time as the time taken for the transient to fade by 1\,mag from peak brightness \citep[similar analysis was performed out by][to which we compare our findings]{Nyholm_2020}. We calculate the rise time and fall time directly from the \texttt{MOSFiT} model light curves. The subset of SNe\,IIn we use for this analysis is mostly made up of the ZTF\,BTS subsample as it is the largest subsample with consistent use of the same $r$-band filter. In the ZTF\,BTS subsample, there are 89 SNe with a measurable fall time metric. To account for time dilation, we correct these times by a factor of $(1+z)$. The peak magnitudes are additionally corrected for extinction (host and Milky Way) and converted into absolute magnitudes. We do not attempt any k-corrections beyond the cosmological k-correction (which is not a concern given the low redshift distribution of the ZTF\,BTS subsample).

\begin{figure}[!h]
	\includegraphics[width=0.99\columnwidth]{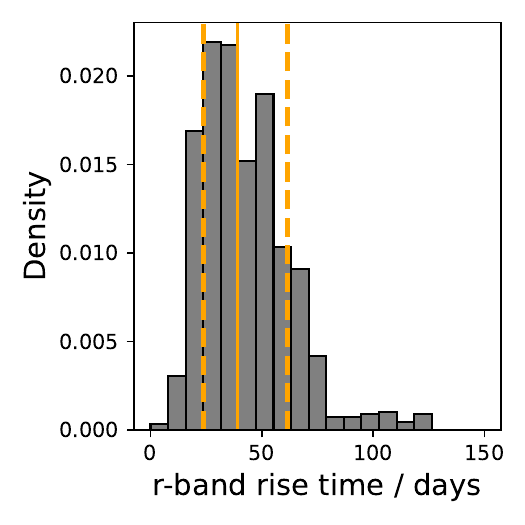}
    \includegraphics[width=0.99\columnwidth]{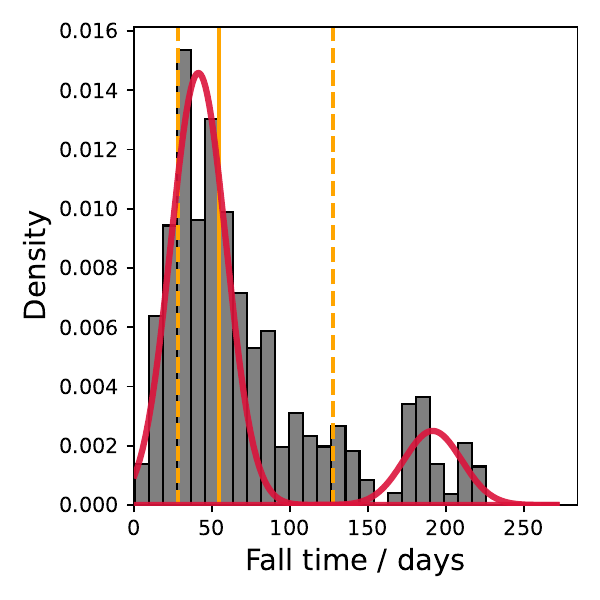}
    \caption{ \textit{top:} $r$-band rise time distribution, defined as the rise from $t_{\mathrm{exp}}$ to peak brightness. The orange vertical lines denote the median (solid line) and the 16 and 84 percent spread (dashed lines). \textit{bottom:} the fall time distribution with the median and spread denoted by the solid and dashed orange lines. Also, in red, are the Gaussian components from the GMM analysis, showing a main component of a few 10s of days and another component of longer lasting SNe.}
    \label{fig:gmm_rise}
\end{figure}

We present the distributions of the observational parameters in Figure \ref{fig:gmm_rise}. We find that the median rise time in the $r$-band and spread is 39$^{+16}_{-23}$\,days. Our rise time distribution does not exhibit clear multimodality, but may have a tail at longer rise times. \citet{Nyholm_2020} used a GMM test to find a possible bimodal population of fast and slow risers in their sample of PTF SNe\,IIn. Those authors find a fast rising population with a mean of $\sim$\,20 days and a slow rising population of SNe\,IIn with a mean rise time of $\sim$\,50 days. Our distribution peaks between these two possible populations and the spread encompasses them. It should, however, be noted that the sample presented by \citet{Nyholm_2020} is significantly smaller than our sample (32 SNe vs 142). Our method of calculating the rise time also differs from those authors. They fit a $\propto$\,t$^2$ power law to the rise of their light curves based on a template rising light curve of one of their SNe, assuming a smooth rise to peak brightness. While our methods are consistent for around half of the common sample, their method may produce somewhat faster rise times compared to our model. The input luminosity is dependent on $n$ and $s$ and our input luminosity is given by, L$_\mathrm{in}$\,$\propto$\,t$^\alpha$, so we typically have $\alpha$\,$<$\,2, slowing the rise to peak. 

%\citep[as][uses a function $\propto\,t^2$]{Nyholm_2020}. 

The bottom panel of Figure \ref{fig:gmm_rise} shows the distribution of the fall times of our SNe\,IIn. Our $r$-band fall time distribution has a median and spread of 56$^{+74}_{-27}$\,days. Our GMM analysis suggests that there is bimodality in the posterior distributions of the fall time. One component has a mean around the median of the distribution at 38 days with a standard deviation of 23 days, and the second component at a longer fall time has a mean of 194 days with a standard deviation of 16 days. We contrast this bimodality with the findings of \citet{Nyholm_2020}.  They find a population of fast and slowly declining SNe\,IIn, where an average decline rate is defined in mag\,d$^{-1}$ and is measured between 0 and 50 days post-peak magnitude. Converting to our fall time metric (i.e. $1/$decline rate), these two populations correspond to a mean fall time of $\sim$\,25\,days and $77$\,days. In other words, our faster peak has a mean in between the two modes of \citet{Nyholm_2020} and our slower component is more than twice as long. It is possible that this difference is related to the definition of fall time (where we do not assume a constant decline rate). Our fall time metric, however, is roughly consistent with the decline rate measured by \citet{Nyholm_2020} for the same objects in their PTF sample. This difference may simply be due to our larger sample; there are three objects with a fall time greater than 200\,days and 13 are over 100\,days. Finally we note that we exclude the very slowly fading, long-lived SNe\,IIn in this analysis. At late times, the contribution of the H$\alpha$ emission line (due to ongoing CSM interaction) starts to dominate over the continuum \citep[e.g., SN\,2005ip, which initially faded linearly followed by a plateau around 150 days post-peak][]{Fox_2020}, we discuss this effect in Appendix\,\ref{sec:synphot}. We also exclude transients that do not fade by 1\,mag within the time-frame of the data\footnote{However, these are compared as a `silver' sample, presented in Appendix\,\ref{sec:silver}. We find that these distributions are consistent with eachother.}.

\begin{table}
	\centering
	\caption{The median values of of properties derived from observational information and also the median value for calculated quantities derived using parameters from \texttt{MOSFiT}.}\label{tab:avparams}
	
	\begin{tabular}{lll} % four columns, alignment for each
		\hline
		Parameter & Median and spread & Method \\
		\hline
        
        M$_{\mathrm{peak,r}}$ & 18.7$\,\pm\,1.0$\,mag & Observed\\
        Rise time &39.2$^{+22.5}_{-15.5}$\,days & Observed \\
        Fall time &55.7$^{+73.5}_{-27.2}$\,days & Observed \\
        R$_{\mathrm{CSM}}$ & 59.6$^{+73.7}_{-32.9}$\,AU & Derived \\
        t$_{\mathrm{CSM}}$ & 2.94$^{+2.37}_{-0.49}$\,years & Derived\\
        log($\dot{M})$ & -0.80$^{+0.83}_{-0.96}$\,M$_\odot$\,yr$^{-1}$& Derived\\
        log($\langle\dot{M}\rangle$) &  --1.32$^{+0.83}_{-1.46}$\,M$_\odot$\,yr$^{-1}$& Derived \\
        v$_{\mathrm{sh}}$ & 2846$^{+2850}_{-1665}$\,km\,s$^{-1}$ & Derived\\
        v$_{\mathrm{CSM}}$ & 370$^{+157}_{-209}$\,km\,s$^{-1}$ & Spectra\\
		\hline

	\end{tabular}
\end{table}

\subsection{Relations Between Pairs of Parameters} \label{sec:pairs}

To assess correlations in our parameters, we use the Pearson correlation coefficient \citep[PCC,][]{Pearson_1907}. Implementing a bootstrapping method similar to \citet{Moriya_2023}, we resample 10$^4$ times. For each resample, we  take one random sample from the posterior distribution of each object and then calculate the PCC. We can then calculate the standard deviation of the PCC (taking this as the uncertainty), and also the p-value. We use the PCC and bootstrapping implementation in \texttt{scipy} \citep{scipy}.

\begin{figure*}[!h]
	\includegraphics[width=0.95\textwidth]{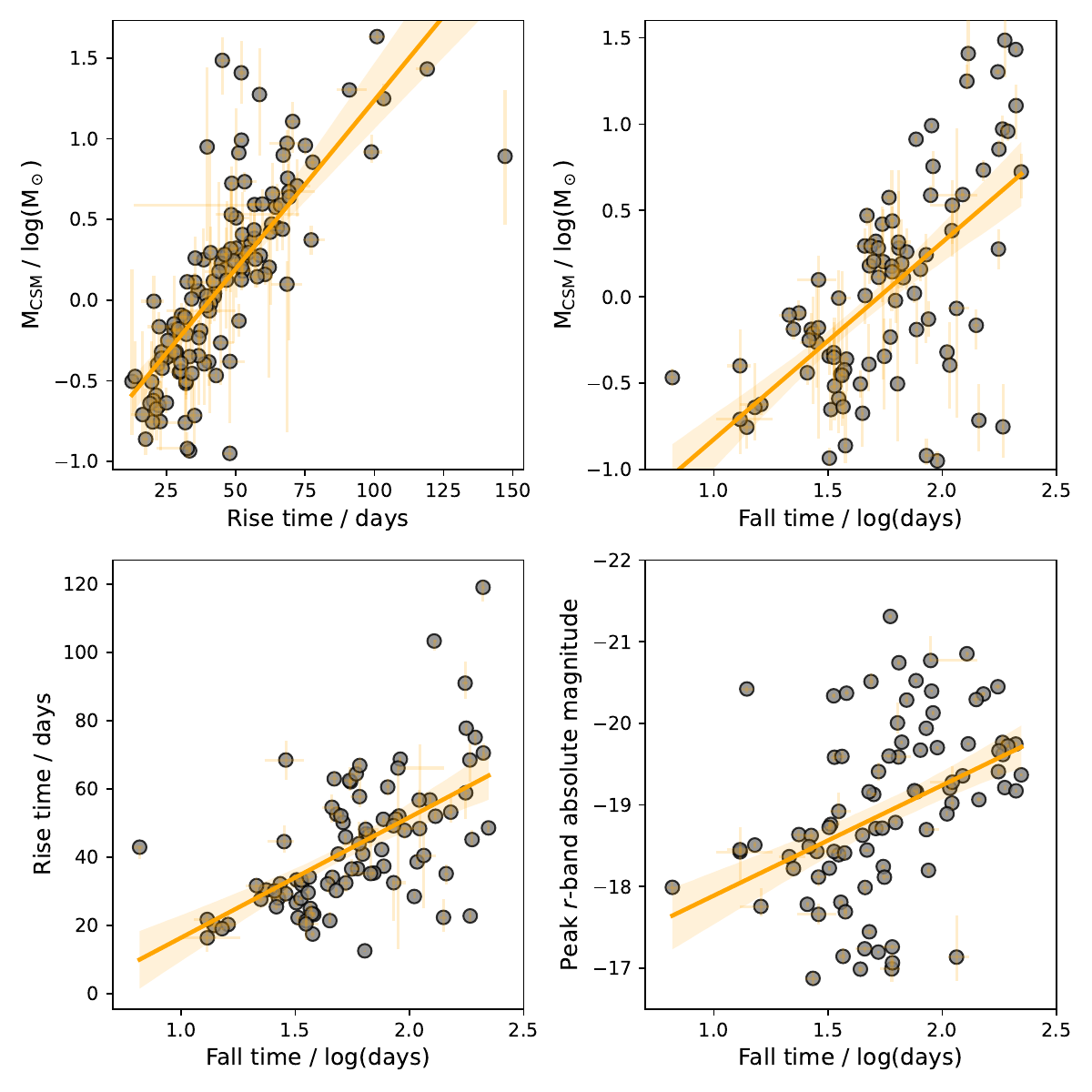}
    \caption{Parameter pair plots between observational and derived parameters from \texttt{MOSFiT}. Also plotted is the best fit line from the PCC test with the 1$\sigma$ uncertainty region shaded for the correlation. \textit{Top left:} the log-log scatter plot of the CSM mass against the $r$-band rise time to peak brightness. \textit{Top right:} the log-log pair plot comparing the CSM mass to the $r$-band fall time. \textit{Bottom left:} the log-log pair plot comparing the $r$-band rise time to peak against the $r$-band fall time, again showing a clear correlation. \textit{Bottom right:} the log-log pair plot comparing the mass-loss rates against the $r$-band fall time.}
    \label{fig:pcc_pairs}
\end{figure*}

\begin{figure*}[!h]
	\includegraphics[width=0.95\textwidth]{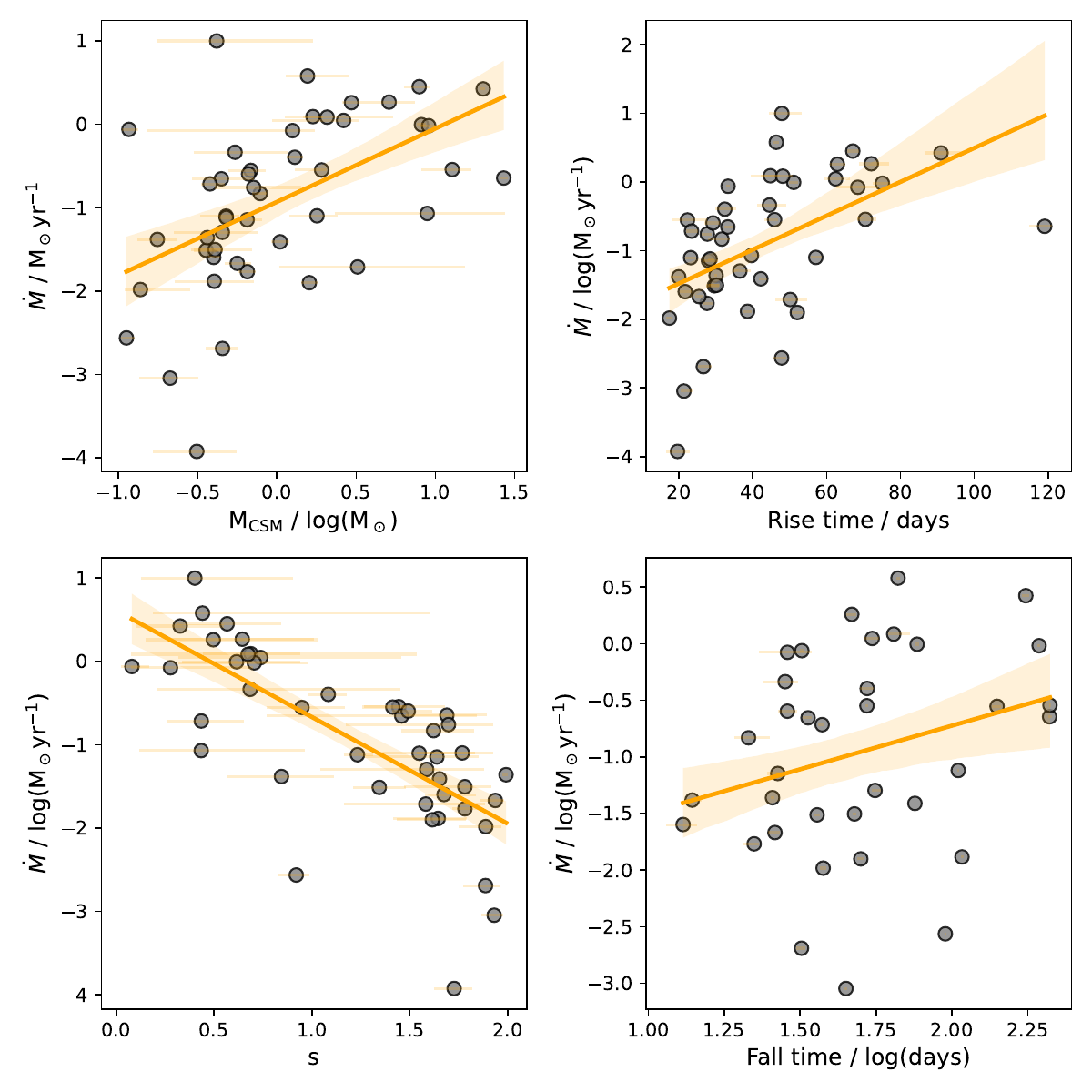}
    \caption{Parameter pair plots showing the correlations between the mass-loss rate and other parameters. Also plotted is the best fit line from the PCC test with the 1$\sigma$ uncertainty region shaded for the correlation. \textit{Top left:} the log-log scatter plot of the CSM mass against the mass-loss rate. \textit{Top right:} the log-log pair plot comparing the $r$-band fall time to the mass-loss rate. \textit{Bottom left:} the log-log pair plot comparing $s$ to the mass-loss rate. \textit{Bottom right:} the log-log pair plot comparing peak $r$-band absolute magnitude to the mass-loss rate.}
    \label{fig:mdot_corr}
\end{figure*}

The most salient correlations (as determined by our bootstrapping method) are shown in Figure\,\ref{fig:pcc_pairs} and Figure\,\ref{fig:mdot_corr}. We find a positive correlation between: the $r$-band rise time and the CSM mass ($p\,<\,0.05$); the fall time and the CSM mass ($p\,<\,0.05$); the rise time and fall time ($p\,<\,0.05$); and the fall time and peak $r$-band absolute magnitude ($p\,<\,0.05$). The mass-loss rate (which we derive from other parameters described in detail in Section\,\ref{sec:mdot}) is positively correlated with the CSM mass ($p\,<\,0.05$) and the fall and rise times ($p\,<\,0.05$). However we note that the correlation with the fall time is not as strong as with the rise time. Finally, we find negative correlations between the mass-loss rate and $s$.

Some of these correlations have been noted in the literature from smaller samples. \citet{Nyholm_2020} note that their fast rising SNe\,IIn tended to decline faster and vice versa. Those authors also show the correlation between the peak $r$-band absolute magnitude and the decline time (with 27 of their objects being used in this analysis). The rise and fall times are positively correlated with the CSM mass, this is likely due to the diffusion timescale in the ejecta-CSM interaction models of \citet{Chatzopoulos_2012} being proportional to the (optically thick) CSM mass as $\tau_{\mathrm{diff}}\,\propto\,M_{\mathrm{CSM-thick}}$. The diffusion timescale governs the rise time, with longer diffusion times producing a longer rise to peak. A longer diffusion timescale also produces a broader peak and slower decline. This may suggest that more massive progenitor systems (assuming massive progenitors have a more massive CSM) produce slower evolving SNe\,IIn with a higher mass-loss rate. There is a weaker, but significant, correlation (with a PCC of $\sim$--0.23) between the peak $r-$band absolute magnitude and the CSM mass. Hence, tentatively, brighter SNe\,IIn may be more likely to be slowly evolving and originating from a massive progenitor that has produced massive CSM. 
 
 A possible bias in this analysis is the lack of sufficient followup for some transients. The population with fall time estimates is smaller due to this lack of follow up, perhaps missing some of the even slower declining transients as they have not decayed by 1\,mag in the time-frame of the observations\footnote{But we do extend our light curve models with these objects, finding a similar distribution. This `silver' sample is presented in Appendix\,\ref{sec:silver}, finding a similar distribution.}. It is unlikely that these observational biases impact the correlations discussed as the well studied, long lived SNe\,IIn have high CSM masses. These long fall time, high CSM mass SNe also positively correlate with the mass-loss rate, a relation which is corroborated by slowly declining SNe\,IIn in the literature \citep[e.g. ASASSN-14il, SN\,2015da and SN\,2017hcc, ][which had mass-loss rates 0.1\,--\,a few\,M$_\odot$\,yr$^{-1}$ and CSM masses of 10s\,M$_\odot$]{Smith_2020_17hcc, Chandra_2022_17hcc, Dukiya_2024, Smith_2024}.

\begin{turnpage}
    
\begin{figure*}[!h]
	\includegraphics[width=1.35\textwidth]{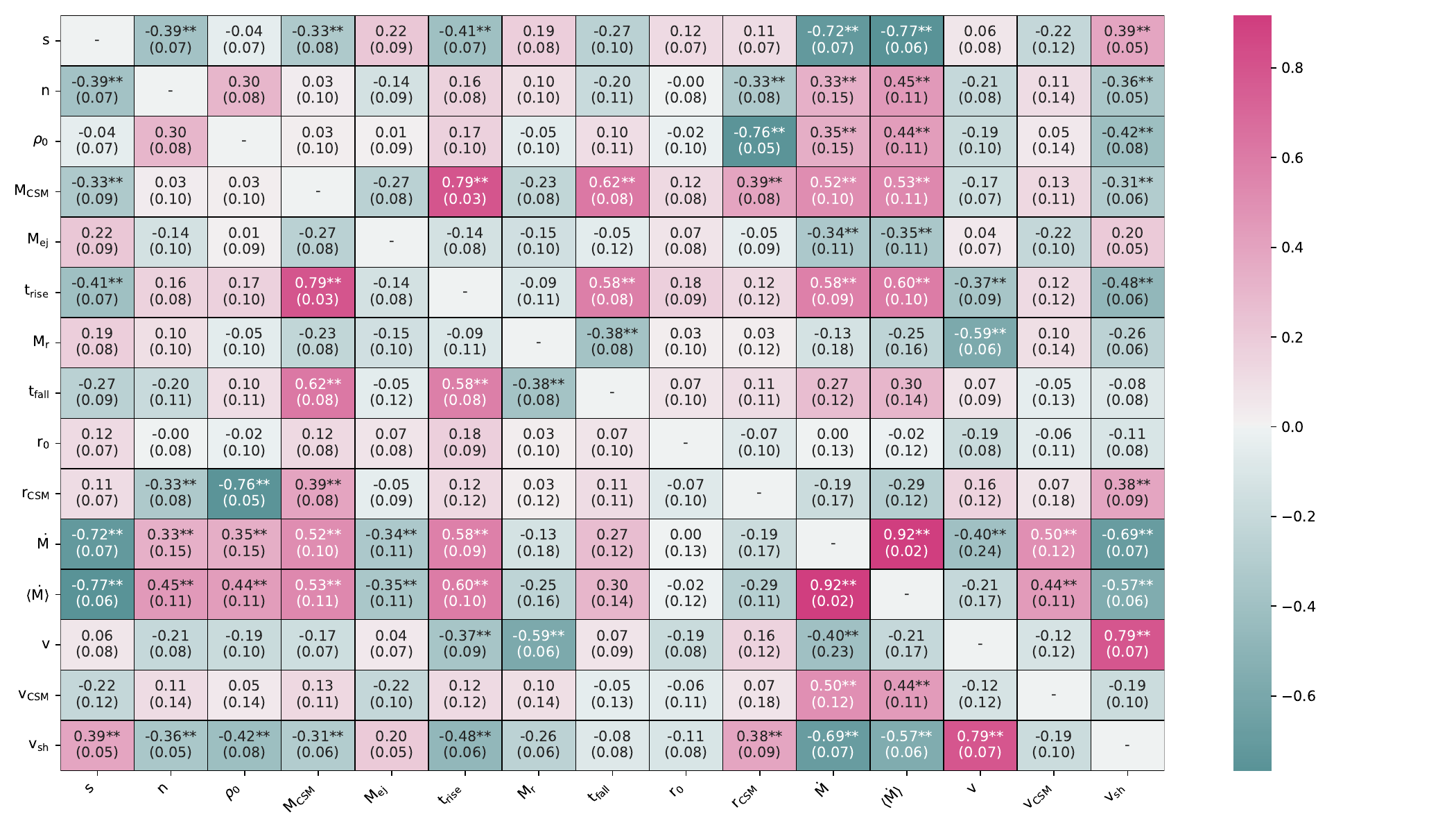}
    \caption{The correlation matrix heatmap for our parameter pairs. Each cell is annotated with the PCC and in parentheses are the 1$\sigma$ uncertainties on the PCC as calculated using our bootstrap method. If the correlation is at least moderately strong (i.e. with a PCC of above 0.3 or bellow $-0.3$) and also significant (with a p-value below 0.05), the PCC is highlighted with `**'.}
    \label{fig:corr}
\end{figure*}

\end{turnpage}

 While the salient correlations are presented in Figures \,\ref{fig:pcc_pairs} and \ref{fig:mdot_corr}, there are a number of significant PCC values between parameter pairs. These are shown in Figure\,\ref{fig:corr}. While there are more parameter pair relations that are `significant' due to their $p$-values, we only highlight the parameter pairs which we determine to have at least a moderate positive or negative correlation (i.e. a PCC over 0.3 or under --0.3). These relations are not further considered. 

 %\newpage

\section{The Photometric Diversity of Type IIn Supernovae} \label{sec:groups}

\begin{figure}[!h]
	\includegraphics[width=0.95\columnwidth]{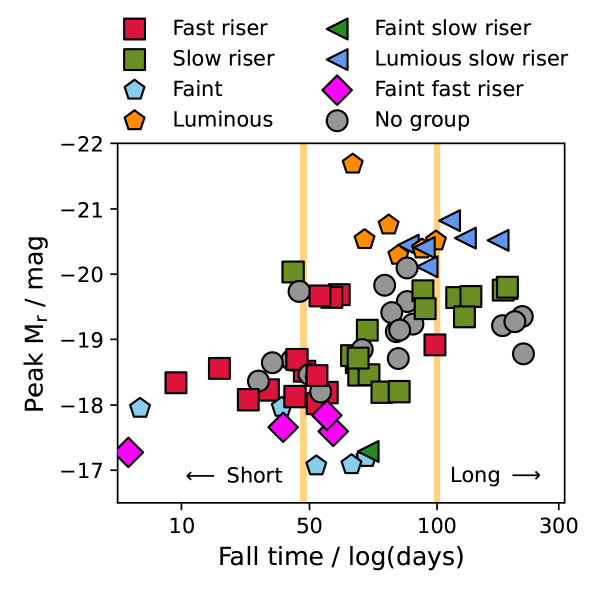}
    \caption{The peak $r$-band absolute magnitude against the $r$-band fall time. The photometric groupings are denoted by different markers. The boundaries between the long fall time and short fall time SNe\,IIn (which we define as over 100 days or under 30 days respectively) is marked by vertical orange lines.}
    \label{fig:peakdur_groups}
\end{figure}

Our \texttt{MOSFiT} modeling has revealed that the SN\,IIn population exhibits great diversity in their physical parameters, reflective of the diverse observable properties. The literature has identified several subgroups of Type IIn SNe: fast/slow-risers, short/long-fall time events, and under/over-luminous events \citep[e.g.][]{Prieto_2008, Stritzinger_2012, hab14, Ofek_2014_lcs, Moriya_2018, Nyholm_2020, Dessart_2022}. In this section, we discuss these photometric groupings which are also summarized in Figure \,\ref{fig:peakdur_groups}.
 
\subsection{Rapidly Declining Supernovae}

In our sample there are 14 SNe\,IIn that display relatively rapid declines from peak, with a fall time of under 30 days \citep[a timescale consistent with the short decliners of][]{Nyholm_2020}. As an example, one of the fastest declining example in our sample is PTF\,10fjh \citep{ptf10fjh_disc, ptf10fjh_class, Ofek_2014}. PTF\,10fjh has a fall time of $\sim$\,15 days. This transient has a low CSM mass of $\sim$\,0.4\,M$_\odot$ which is consistent with our observational correlations presented in Section\,\ref{sec:pairs}. More broadly, the fraction of these short fall time SNe\,IIn is consistent with the fraction of fast decliners in the sample of \citet{Nyholm_2020} when converted into our fall time metric. This is 18 percent of our subsample of 89 SNe\,IIn with a measurable fall time.

As a whole, this subgroup has small CSM masses, with a median of $\sim$\,0.45\,M$_\odot$, again consistent with our correlations between CSM mass and the fall times. These objects also tend to be relatively dim when compared to the full sample, with a median peak $r$-band absolute magnitude of --18.2\, mag. These transients are also typically fast risers, consistent again with our correlations. We discuss the rise time subpopulations in the next section.

\subsection{Slow and Fast Risers}

As found by \citet{Nyholm_2020}, the SNe\,IIn population seems to contain subgroups based on their rise times to peak. These rise times are positively correlated with their fall time as shown by Figure \,\ref{fig:pcc_pairs}, which is likely due to the amount of CSM around the progenitor. One such fast riser in our sample is SN\,2021kwc which was discovered by ZTF \citep{2021kwc_disc, 2021kwc_spec} and had a rise time to peak in the $r$-band of $\sim$\,20 days. We find that SN\,2021kwc has a small CSM mass of $\sim$\,0.2\,M$_\odot$ and also had a short fall time of $\sim$\,25 days. Moreover, SN\,2021kwc is on the lower end of our luminosity distribution with a peak $r$-band absolute magnitude of $\sim$\,--18.0\,mag. Conversely, our slowest rising SN\,IIn is SN\,2019qvr \citep{2019qvr_disc, 2019qvr_class}. This transient takes $\sim$\,120 days to rise to the $r$-band peak. This transient has almost 30\,M$_\odot$ of CSM, with the fall time of $\sim$\,190\,days and bright peak magnitude at M$_{\mathrm{r}}$\,=\,--19.8. The inferred mass-loss rate of SN\,2019qvr is high--$\sim$\,0.3\,M$_\odot$\,yr$^{-1}$. These two objects exemplify our correlations: objects with higher CSM masses evolve slower than SNe\,IIn with low CSM masses.  When considering both of these subgroups, the median CSM mass for the short risers is $\sim$\,0.5\,M$_{\odot}$, while the median CSM mass is $\sim$\,18.6\,M$_{\odot}$ for the slow risers.  

Generally SNe\,IIn rise slowly relative to other SN classes. Indeed, around 34\% percent of our sample have a rise time of 50 days or greater, a somewhat greater proportion than is found by \citet{Nyholm_2020} who find that 25\% of their sample had these longer rise times. Those authors find that around a third of their sample had a rise time of less than 20 days (the mean value of their `short risers'). Our sample has around 25\% of the SNe\,IIn being fast risers, with a rise time of under 30 days. This is contrasted the findings of \citet{Nyholm_2020} who found $\sim$\,65\% of their sample were these faster rising transients. As previously mentioned, this may be a result of their method possibly producing faster rise time estimates.

\subsection{Long-lived supernovae}

The photometric SN\,IIn subclass characterized by a slow light curve decay is sometimes denoted as the `1988Z-like SNe\,IIn', after the archetype \citep[e.g.,][]{Chugai_1994_88z, Stritzinger_2012, hab14}. This subpopulation has also been noted in other, small sample size works \citep[e.g.][]{Kiewe_2012, tadd13, Nyholm_2020}. While famous literature examples were excluded from our analysis due to the H$\alpha$ emission features from CSM-ejecta interaction dominating the $r$-band flux at later times, we can still probe for longer lived transients in our sample. Typically, the photometric data extends a few hundred days post discovery so the late-time H$\alpha$ effect is unimportant in these epochs (or the transients are relatively young). We define a long lived SN as having a long fall time, in excess of 100 days \citep[constistent with the slow decliners identified by][]{Nyholm_2020}. In our sample, 22\% of our events have a long fall time (of the subsample of 89 where the fall time can be measured). This fraction is somewhat higher than what is found by \citet{Nyholm_2020}, who found $\sim$\,15\% of their sample had a fall time of 100 days or greater. This difference is likely due to the differences in measurement, where we measure the time taken to decline by 1\,mag from peak brightness, rather than calculating the average decline rate over 50 days. We also note that our sample of measurable decline times (77) is larger than the sample (27) used in \citet{Nyholm_2020}.

An exemplar of this subclass in our sample is is SN\,2020jhs, which was collected from the YSE sample \citep{2020jhs_disc, 2020jhs_spec}. SN\,2020jhs had a peak $M_r\,=\,-19.8$ and a fall time of of 188\,days. This transient was also a slow riser, with a rise time of $\sim\,71.7$\,days and has a large inferred CSM mass of $\sim$\,9.4\,M$_\odot$. This object exemplifies the slow evolution seen in these luminous, high CSM mass objects. 

As mentioned, the rise and fall times are strongly correlated with the CSM mass and that the fall time is correlated with the peak $r$-band absolute magnitude.  We note that there is a weaker (but statistically significant) correlation between the peak $r$-band absolute magnitude and CSM mass, (PCC\,=\,-0.23). Therefore, while high CSM mass leads to slower photometric evolution, it does not necessitate luminous transients \citep[as seen in the archetypes of this class, SN\,1988Z and SN\,2005ip with absolute magnitudes of --17.6 in the $B$-band and --17.4, unfiltered, respectively][]{Stathakis_1991,Smith_2009_05ip, Smith_2017_05ip}. These long lived SNe\,IIn also tend to have high mass-loss rates (with a median of $\sim$\,0.4\,M$_\odot$yr$^{-1}$). Similarly to the slow risers, the long fall time is due to the longer diffusion timescale allowed by the massive CSM.

Finally, we note that we are unable to define a fall time for our whole sample\footnote{However, we do extend our models in these cases to form a `silver' sample which is presented in the Appendix\,\ref{sec:silver}.}. This is due to the baseline of the photometric data does not always cover 100 days post-peak. This is particularly true for the younger transients in the ZTF\,BTS sample. While this may result in unknown long lived SNe\,IIn being omitted from this work, it is unlikely to affect the strong correlations we have found, with 61\% of the sample with $r$-band data having a measurable fall time.

\subsection{Fainter Transients}

Another photometric subclass that has been identified in the SN\,IIn population are a group of subluminous transients. Well studied examples of this subgroup include the enigmatic SN\,2008S which peaked at M$_{\mathrm{V}}$\,=\,--14.0\,mag \citep[][]{Botticella_2009, Adams_2016}. These fainter transients overlap in brightness space with SN impostors \citep[typically with absolute $V$-band magnitudes --10 to --14\,mag, e.g.,][]{VanDyk_2000, Kochanek_2012, Aghakhanloo_2023_16blu} and may reach the brightness of `normal' SNe\,II, such as the pre-explosion brightening of SN\,1961V \citep[at M$_V$\,=\,--16.5 e.g.,][]{Zwicky_1964, Woosley_2022}. We explicitly exclude confirmed or suspected SN impostors from our analysis, but we do note that some fainter SNe\,IIn are present in our sample. While SNe\,IIn are luminous compared to normal SNe\,II, we define a `faint' SN\,IIn as peaking below M$_r$\,=\,--18, around a magnitude (i.e. at $\sim$\,1$\sigma$) fainter than median of our uncorrected absolute magnitude distribution. 

Our faintest example, SN\,2021aapa was discovered by ZTF \citep{2021aapa_disc, 2021aapa_class} and has a peak $r$-band absolute magnitude of $-16.9$\,mag, typical of the peaks of fainter SNe\,IIP. Around 15 percent of the (uncorrected for Malmquist bias) SNe\,IIn have a peak absolute $r$-band magnitude fainter than $-18.0$. This fraction is consistent with \citet{Nyholm_2020}. These fainter transients tend to have a smaller CSM (e.g., SN\,2021aapa has a CSM mass of $\sim$\,0.6\,M$_\odot$) and are more likely to quickly decline. In the literature, these objects are sometimes associated with having photometric similarities with SNe\,IIP (a tentative plateau may be seen in SN\,2021ras in our sample). These transients, such as SN\,2011ht \citep{Mauerhan13a, Fraser_2013_2011ht} and SN\,2020pvb \citep{Elias-Rosa_2024}, have two explanations accounting for the relatively weak explosions, IIn-like spectra, and associated precursor outbursts. Firstly, a weaker explosion from a low mass progenitor system (8\,--\,10\,M$_\odot$) as a product of an electron-capture SN, with the precursor being a product of unstable burning ejecting some of the stellar envelope \citep[e.g.][]{Dessart_2010, Matsumoto_2022}. However, we note that the only ecSN candidate to date, SN\,2018zd, does not exhibit prolonged interaction signatures \citep{Hiramatsu_2021_18zd}. Secondly, that the progenitor is indeed massive \citep[][put a progenitor detection constraint on SN\,2020pvb of $\lesssim$\,50\,M$_\odot$]{Elias-Rosa_2024}, where the explosion is weakened due to the formation of a black hole, with material falling back \citep[][]{Sollerman_1998, Heger_2003}. It should be noted that our models do not inform on either scenario here. Indeed, in the fallback scenario, it is unknown how much mass is `missing' from the ejecta due to the black hole formation (limits could likely be placed if there was a progenitor detection, however).

\subsection{Bright Transients}

On average, SNe\,IIn are intrinsically luminous compared to other classes of CCSNe due to the efficient conversion of kinetic energy into radiation in the shock interaction \citep{Kiewe_2012}. Furthermore, there are many examples of superluminous (peaking brighter than --20\,mag) SNe\,IIn in the literature such as SN\,2015da, ASASSN-15ua and ASASSN-14il \citep{Smith_2024, Dickinson_2024, Dukiya_2024}. The most luminous SN\,IIn in our sample is SN\,2020vci, which peaked at --21.7\,mag in the $r$-band \citep{2020vci_disc, 2020vci_class}. While this transient does not have a high CSM mass, with $\sim$\,0.6\,M$_\odot$, it has an ejecta mass similar to our median with $\sim$\,17\,M$_\odot$ and also a high ejecta velocity, with $v_{\mathrm{ej}}\,\approx\,1.6\,\times\,10^5$\,km\,s$^{-1}$. Within our sample, around 17 percent have a peak absolute magnitude brighter than --20\,mag in the $r$-band, suggesting that intrinsically very luminous transients are relatively common among SNe\,IIn. This subpopulation has a somewhat heterogeneous set of parameters inferred from \texttt{MOSFiT}, but have a high mass-loss rate (when this can be measured) with a median average mass-loss rate of $\sim$\,3\,M$_\odot$\,yr$^{-1}$.

\subsection{Other Exotic Type IIn Supernovae} \label{sec:rebright}

As large scale transient surveys discover more SNe\,IIn, it has become apparent that not all of these transients smoothly decline after initial peak \citep[including bumpy declines and dramatic rebrightenings, e.g., SN\,2010mc, iPTF13z, SN\,2014C, SN\,2019zrk and SN\,2021qqp][]{Ofek_2014, Nyholm_2017,Margutti_2017,Fransson_2022, Hiramatsu_2024}. Some SNe exhibit dramatic rebrightenings, forming a secondary peak. For example, SN\,2021qqp \citep[for a thorough study of this object, see][]{Hiramatsu_2024}. This luminous SN\,IIn, peaking at an $r$-band magnitude of --19.5\,mag has a slow initial rise to peak. The slow rise lasted for $\sim$\,300 days, followed by a 60 day period of brightening, ending in a rapid rise to peak in the few days prior to maximum brightness. This extreme behavior was linked to dramatic mass loss episodes in the few years preceding explosion, with mass-loss rates of up to 10\,M$_\odot$\,yr$^{-1}$ resulting in the ejecta colliding with dense, massive detached CSM shells. There was also a long lasting precursor event with a peak M$_g$\,=\,--15.6. Using models adopted from \citet{Matsumoto_2022}, it was found that the progenitor suffered two distinct mass loss events, producing 2\,--\,4\,M$_\odot$ of CSM, occurring 0.8 and 2\,yrs prior to the terminal explosion. In our sample, 3 transients were identified as having precursor activity: SN\,2021yaz, SN\,2021ydc and SN\,2022prr \citep[more thorough analysis of the ZTF data, however, may reveal more precursors, e.g.][]{Strotjohann_2020}. Furthermore, some SNe\,IIn in the literature show precursor events that are classified as SN impostors, \citep[e.g. SN\,2009ip][]{Mauerhan13a}. SN\,2021qqp was also identified in our initial sample selection but was omitted from this work due to the double peak and the long precursor occurring directly prior to the SN explosion. 

As our models do not incorporate bumps, secondary peaks or rebrightenings, we exclude these objects from our analysis. Regardless, these striking objects are an important feature in the SN\,IIn landscape, informing on the mass loss history of the progenitor in the years prior to the terminal explosion. While we do not include these objects in our analysis, we present these objects in our full table of SN\,IIn (which includes cut objects) in the online materials. 

\section{Inferred Mass-Loss Rates of Progenitor Systems} \label{sec:mdot}

\begin{figure}[!t]
    \centering
	\includegraphics[width=0.99\columnwidth]{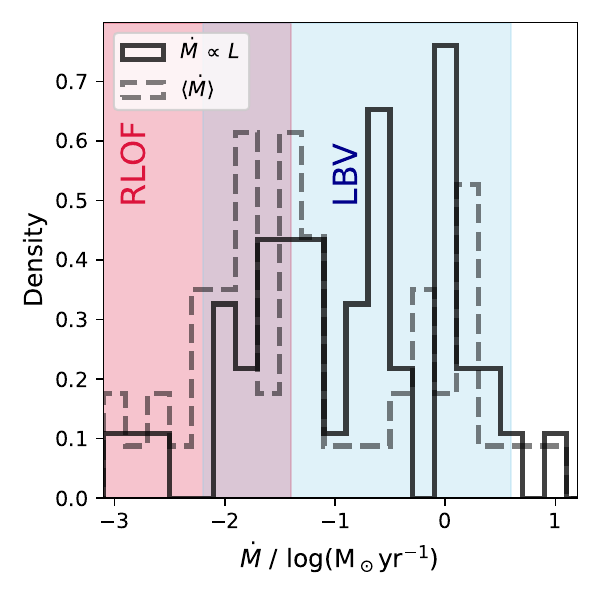}
    \caption{Histogram showing the mass-loss rate distribution of our sample of SNe\,IIn. The solid line shows the mass-loss rate proportional to the peak luminosity. The dashed line shows the average mass-loss rate as calculated using the inferred parameters from \texttt{MOSFiT}. The red shaded region is the mass-loss regime of binary interactions via Roche-lobe overflow (RLOF), and the blue shaded region is the mass-loss rate region of LBV eruptions. In this single dimension, the mass-loss rate of our SNe\,IIn are consistent with both RLOF and the great eruptions of LBVs, with some overlap between these two regions.}
    \label{fig:mdotprogen}
\end{figure}

\begin{figure*}[!ht]
    \centering
	\includegraphics[width=0.99\textwidth]{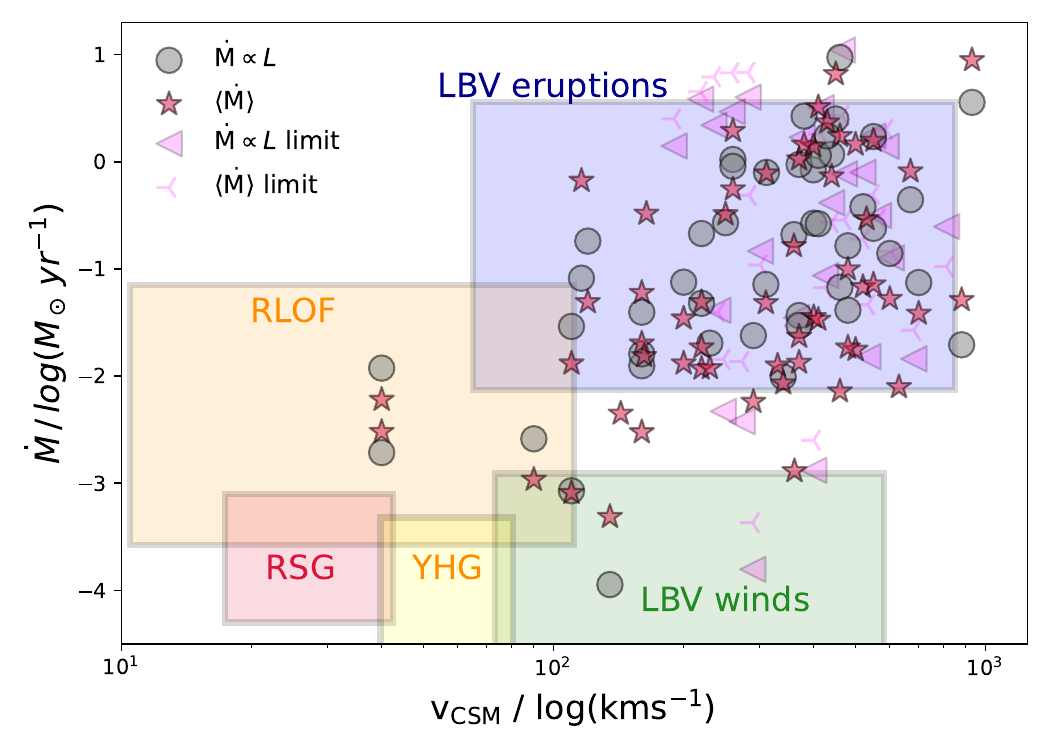}
    \caption{The mass-loss rate distributions of our SNe\,IIn, both the average mass-loss rate and the mass-loss rate proportional to the CSM interaction luminosity (at peak). The shading corresponds to the indicative parameter space corresponding to different mass-loss regimes, against the CSM velocity expected for each mode \citep[reproduced using][]{deJager_1988,vanLoon_2005,Drout_2016, Smith_2017_review}. The blue rectangle is indicative of the high mass rates seen in the great eruptions of LBVs, spanning from 10$^2$\,--\,10$^0$\,M$_\odot$\,yr$^{-1}$. The green region shows the lower mass-loss rates seen from LBV winds. The orange region represents mass loss from Roche-lobe overflow (RLOF, but not including more violent binary events). The red region depicts the mass-loss rates from extreme RSGs and the yellow region shows the mass-loss rates expected from yellow hypergiants (YHGs).}
    \label{fig:mdotscatter}
\end{figure*}

Here, we explore the pre-SN mass loss as inferred from a combination of \texttt{MOSFiT} and observational features. The mass-loss rate of the progenitor before the terminal SN explosion is key to the determination of the progenitor type. In this section, we compare our mass-loss rate estimates with the indicative mass-loss rate/CSM velocities of various progenitor types \citep[as summarized by][]{Smith_2017_review}. 

Progenitor mass loss may occur through massive winds, eruptions, binary interactions or a combination of these effects. The stellar winds of massive evolved stars such as RSGs are radiation pressure driven \citep[e.g. line-driven winds in RSGs;][]{Castor_1975, Vink_2001}. However, the typical mass-loss rates from these objects is insufficient to strip enough mass off a progenitor star to produce the CSM we infer for SNe\,IIn \citep[with mass-loss rates $\sim$\,10$^{-6}$\,M$_\odot$\,yr$^{-1}$;][]{Beasor_2020}. In contrast, LBVs have high line-driven wind velocities reaching a few 10$^2$\,--\,10$^3$\,kms$^{-1}$, producing mass-loss rates 10$^{-5}\,-\,10^{-4}$\,M$_\odot$\,yr$^{-1}$ \citep{Vink_2002, Smith_2004}; this may \textit{still} be too weak to be the sole mass loss mechanism for SNe\,IIn. As previously discussed, LBVs are also known to undergo great eruptions with observed ejecta velocities ranging 10$^2$\,--\,10$^3$\,kms$^{-1}$ which may be due to some eruptive outburst or extreme, optically thick continuum driven winds \citep[e.g.][]{Smith_2004, Kashi_2010}. These great eruptions are distinct from the winds in that they have a mass-loss rate in excess of 1\,M$_\odot$\,yr$^{-1}$\citep{Humphreys_1994, Hillier_2001,Smith_2003} which may be triggered by an interaction with a companion \citep{Kashi_2010}. While the great eruptions may result in short periods of extreme mass loss \citep[it should be noted that less extreme episodic eruptions can lead to more modest amounts of CSM, for example P\,Cygni has $\sim$0.1\,M$_\odot$][]{Smith_Hartigan_2006}, it has been argued that eruptions in single massive stars cannot account for the 10s\,M$_\odot$ of CSM found around some SNe\,IIn. Instead, binary interactions may be critical \citep[e.g., in the cases of $\eta$\,Car and SN\,2015da][]{Kashi_2010, Smith_2024}. These `binary interactions' arise from diverse scenarios, ranging from mergers in a common envelope to collisions and mergers with compact objects \citep[e.g.][]{Soker_2006, Langer_2012, Smith_2018, Schrøder_2020}.

Here, we describe our process for estimating the CSM velocity for our sample of SNe IIn in order to estimate the progenitor mass-loss rates. Although CSM velocities of 100\,kms$^{-1}$ are often assumed \citep[e.g.][]{Moriya_2013,Moriya_2023}, the mass-loss rate is highly sensitive to this velocity. Therefore, we infer the wind velocity using the narrow component of the H$\alpha$ complex. By definition, our sample has at least a classification spectrum publicly available per transient, allowing such a measurement. We collect the available spectra from WISeREP\footnote{\url{https://www.wiserep.org}} and YSE-PZ \citep{YSE-PZ}, with the addition of a small amount of private data from \citet{Ransome_2021} and the PS1 Medium Deep Survey. These data were produced by numerous instruments; the various instruments and respective telescopes are outlined in table\,\ref{tab:spec}.

The CSM velocities are estimated by decomposing the H$\alpha$ profiles into Gaussian components following the procedure from \citet{Ransome_2021}. Because each spectral line is effectively broadened by the instrumental response (i.e. the resolution), we deconvolve the instrumental width from the observed width. In order to estimate the instrumental broadening, we measure the width of lines (sky or host galaxy lines) that we do not expect to be broadened by some other mechanism. The degree of data reduction in each spectrum differs. As such, host or sky lines are not available in every instance, and, typically, the resolution or grating used is not known. We assume a constant instrumental broadening for each spectrograph based on the available measured sky or host lines. We do note, however, that the spectrograph configuration is not known for each spectrum and the resolution of some instruments can vary significantly; for example, the R power of DBSP ranges 10\,--\,10,000. Therefore, our CSM velocity estimates may be considered upper limits. Finally, we note that the CSM may be radiatively accelerated and therefore an overestimate of the CSM velocity \citep[e.g.,][]{Tsuna_2023}. We are able to estimate the CSM velocities of 57 of our SNe\,IIn. Additionally we measure the unresolved upper limits (where the emission line is not resolved given our assumed instrumental responses), we have limits for 27 SNe\,IIn. 

We use the measured velocities to estimate the mass-loss rate for each event. We use two methods: the average mass-loss rate, derived from our \texttt{MOSFiT} parameters (mentioned in Section\,\ref{sec:model}), and the mass-loss rate proportional to the luminosity \citep[e.g.][]{Smith_2017_review, Dickinson_2024}. Using these two methods, we can test if the mass-loss rates using observables and parameters from \texttt{MOSFiT} are consistent (however we would generally expect the mass-loss rates at peak luminosity to be higher than the average mass-loss rate).

We plot the mass-loss rates against the CSM velocity in Figure \,\ref{fig:mdotscatter}, along with the general regions in the phase space representing different mass-loss modes \citep{Smith_2017_review}. Both mass-loss rate distributions are shown in Figure \,\ref{fig:mdotscatter} for comparison. Also shown are the limits using the widths of the narrow components that could not be deconvolved from their instrumental response due to being too narrow to be resolved using the assumed instrumental response. These upper limits on the CSM velocity are roughly consistent with our sample of SNe that have a CSM velocity measurement. This consistency may suggest that we are underestimating the instrumental response in some cases (as typically these spectra are not so high resolution such that the FWHM are almost unaffected after deconvolution) leading to an overestimate in mass-loss rate. Our median (and spread) for the mass-loss rates are log($\langle\dot{M}\rangle$)\,=\,-$1.32^{+0.83}_{-1.46}$\,M$_\odot$\,yr$^{-1}$ and log($\dot{M}$)\,=\,-$0.80^{+0.83}_{-0.96}$\,M$_\odot$\,yr$^{-1}$. As expected, the average mass-loss rate is lower than the mass-loss rate proportional to the $r$-band luminosity at peak. These two measurements are, however consistent with each other within the spread of their respective distributions. 

 We now highlight just the one dimensional mass-loss rate distributions for both of our methods. In Figure\,\ref{fig:mdotprogen} we show our mass-loss rate distribution. Our inferred mass-loss rates span from $\approx\,10^{-3}$\,--\,$10^{0}$\,M$_\odot$\,yr$^{-1}$, with a peak at $\sim$\,10$^{-2}$\,M$_\odot$\,yr$^{-1}$. The mass-loss rate distribution using the peak luminosity is also shown in Figure\,\ref{fig:mdotprogen}. When using equation \ref{eq:smithmdot} (i.e.,  assuming that the $r$-band luminosity is representative of the luminosity due to CSM interaction), we find that the mass-loss rate distribution is largely consistent with the mass-loss rate distribution calculated using equation \ref{eq:avmdot}. The median mass-loss rates are 0.03\,M$_\odot$\,yr$^{-1}$ and 0.08\,M$_\odot$\,yr$^{-1}$ for the average mass-loss rate and mass-loss rate from equation \ref{eq:smithmdot} respectively. While these distributions are similar, the mass-loss rate proportional to the luminosity (equation \ref{eq:smithmdot}) is skewed towards a higher mass-loss rate due to using the peak luminosity (i.e. the highest mass-loss rate assuming the shock and CSM velocities are constant) rather than an average over the CSM extent (equation \ref{eq:avmdot}). The mass loss evolution may be probed over time but having a spectral time-series would be more informative for such a calculation. Where the average mass-loss rate distribution peaks ($\sim$10$^{-2}$\,M$_\odot$\,yr$^{-1}$) is consistent with the lower mass-loss rate of LBV giant eruptions \citep[but is too high for LBV winds, which produce mass-loss rates 10$^{-5}$\,--\,10$^{-3}$\,M$_\odot$\,yr$^{-1}$][]{Cox_1995, Hillier_2001, Najarro_2001, Puls_2008} and binary interactions.

\begin{table}
	\centering
	\caption{The telescopes, instruments, and number of spectra available for our SNe\,IIn for CSM velocity estimation.}\label{tab:spec}	
	\begin{tabular}{llr} 
		\hline
		Telescope & Instrument& $\#$ spec.\\
		\hline
        LT & SPRAT & 16\\
        Palomar 200\,inch& DBSP& 13 \\
        NTT & EFOSC2 & 10 \\
        Keck1 & LRIS & 8\\
        MMT & MMT Blue & 5\\
        Lick 3m & KAST & 4 \\
        UH\,88 & SNIFS&4 \\
        WHT & ISIS & 3\\
        APO\,3.6m & DIS& 2\\
        Magellan & IMACS &2 \\
        MMT & Hectospec & 2\\
        MMT &Binospec & 1\\
        Keck2 & DEIMOS & 1\\
        INT &IDS & 1\\
        NOT &ALFOSC2 &1 \\
        Gemini North & GMOS & 1 \\
        Lijiang\,2.4m & YFOSC& 1\\
        ANU\,2.3m & WiFeS&  1 \\    
		\hline
	\end{tabular}
\end{table}

\section{Host Galaxies of our Type IIn Supernovae} \label{sec:hosts}

Core-collapse SNe generally trace star formation in their hosts. As such, they are often found in spiral-like galaxies. While we do not study the hosts of our sample in detail, we now briefly discuss the Hubble types of our sample. 

Environmental studies have measured the association of SN locations with star formation or stellar population ages using either narrowband photometry or spectroscopy \citep{hab14, tad15,Galbany_2018, Ransome_2022, Moriya_2023}. As previously mentioned, SNe\,IIn seem to inhabit a diverse set of environments, seemingly not being consistent with single massive stars. In Table\,\ref{tab:sample}, we present our sample of SNe\,IIn. In this table, we record the host galaxies of our transients, with this information being gathered from NED \citep{NED} and SIMBAD \citep{simbad}. All but two of the SNe in a classified host are from some class of spiral galaxy, as is the expectation for a CCSN, with one irregular and one elliptical host. In the case of an elliptical host, one may expect a thermonuclear SN. For SNe\,IIn, this possibly suggests SNe\,Ia-CSM e.g. SN\,2002ic \citep{Hamuy_2003}, SN\,2005gj \citep{Prieto_2005}, SN\,2008J \citep{Taddia_2012} and SN2012ca \citep{Fox_2015, Inserra_2016}.

For a massive progenitor, it would be expected that its host galaxy would be an actively star forming spiral galaxy. However after a host galaxy search using SIMBAD, we find that two of the hosts in our sample of SNe\,IIn are elliptical galaxies. Elliptical galaxies, sometimes referred to as `red and dead' galaxies, have lost much of their gas over their evolution and are typically not expected to harbor massive stars due to a lack of active star-formation. However, we have apparent elliptical hosts in our sample. In our sample, one SN\,IIn was found in an apparent elliptical host, SN\,2019bxq in the host 2MASX\,J16575851+7836144. This transient has a total mass of $\sim$\,20\,M$_\odot$ which may preclude a thermonuclear SN.

If the progenitor of this SN\,IIn is indeed massive, then it is possible that the host has underlying star formation. \citet{Irani_2022} present CCSNe from ZTF\,BTS with elliptical hosts. Those authors add three CCSNe from ZTF\,BTS (one SN\,Ic and two SNe\,II) to seven CCSNe with elliptical hosts from the literature. Based on the spectroscopically complete sample of ZTF\,BTS, it is found that elliptical galaxies comprise $\sim$\,0.3\% of CCSN hosts in the local universe. Those authors also find that CCSNe are more offset from the core of the elliptical hosts than SNe\,Ia that have elliptical hosts. Those authors conclude that there may be low surface brightness residual spiral features hosting some star-formation, or there is star-formation encouraged via the introduction of material from flows from the intergalactic medium \citep[see also;][]{Fukugita_2004, Sedgwick_2021}. Therefore, there may be limited star forming activity in some elliptical hosts, possibly producing massive stars. 

A large sample, detailed host study of these SNe\,IIn hosts would be complementary to our results and is the topic of future work. Measuring the properties of the local environments, and linking these data with the physical and observed SN parameters from this work may further establish correlations and progenitor routes in lieu of direct progenitor detections. 

\section{The Progenitors of Type IIn Supernovae} \label{sec:progenitors}

We have modeled 142 SNe\,IIn using \texttt{MOSFiT}. As seen in smaller sample studies \citep[e.g.][]{Kiewe_2012, tadd13, Nyholm_2020}, the overall population is diverse and forms a continuum of observed and derived parameters. Typically, the SNe\,IIn in our sample have a dense inner CSM, $\sim$10$^{-12}$\,g\,cm$^{-3}$, with a CSM density profile that is somewhat flat ($s\,\sim\,1.3$). We infer that the progenitors generally lose a few M$_\odot$ of mass that forms the immediate CSM shell over the course of a few years (assuming the CSM velocity is constant). The progenitors are inferred to be typically massive; the summed CSM and ejecta masses are $\sim24$\,M$_\odot$. Although we again note that this is a lower limit on the progenitor mass (excluding the remnant).

We do not find any clustering in the parameter pairs which could have more clearly pointed to distinct progenitor channels. In posterior distributions of the individual parameters, we only find one example of a population that has multiple components. We find two populations in the fall time distribution, with a long fall time component (centered at around 200 days) as determined by a GMM. These long fall time SNe\,IIn have higher CSM masses on average as inferred by \texttt{MOSFiT}, and likely arise from more massive progenitor systems. We also find that there are is an apparent high CSM tail within the sample (however, not a separate population). This population extends to a few thousand AU and is associated with less dense CSM profiles (a few 10$^{-14}$\,g cm$^{-3}$). However, this apparent population is due to multiple solutions for $\rho_0$ being found for individual objects. Therefore, one should not necessarily consider this a true distinct population.

In addition to the clusters in singular parameters, we also find correlations in the observed light curve features. We find that the rise time and fall time correlate with the inferred CSM mass. These correlations suggest that longer-lived, slower evolving SNe IIn have more CSM and that slower rising SNe\,IIn tend to be slower falling and vice versa. This is to be expected due to the dependency of the diffusion time on the CSM mass in our models. We also find that the slower decliners tend to be intrinsically brighter in the $r$-band, similar to the findings of \citet{Nyholm_2020}. The mass-loss rates also correlate with the CSM mass, the fall time, $s$ and the peak $r$-band absolute magnitude. This suggests that the SNe\,IIn with high mass-loss rates tend to be intrinsically brighter. They also have a more eruptive/outburst like mass loss history producing a massive CSM. These SNe\,IIn then evolve slowly from peak brightness.

 The canonical picture of SN\,IIn progenitors is that they are LBVs, with pre-explosion detections and precursor emission being interpreted as LBV great eruptions. Pre-explosion, progenitor direct imaging is one `smoking gun' clue, although extremely limited. There are (at the time of writing) 6 SNe\,IIn with claimed direct progenitor detections: SN\,1961V \citep[LBV,][]{Goodrich_1989, Filippenko_1995, Smith_2011_imps}; SN\,2005gl \citep[LBV,][]{Gal-Yam_2007, GalYam_2009}; SN\,2009ip \citep[LBV,][]{Smith_2010_09ip, Foley_2011, Mauerhan13a}; SN\,2010jl \citep[possible LBV,][]{Smith_2011_10jl,Fox_2017,Dwek_2017, Niu_2024}; SN\,2015bh \citep[LBV,][]{Thoene_2015, Elias-Rosa_2016} and SN\,2016jbu \citep[LBV or YHG,][respectively]{Kilpatrick_2018, Brennan_2022_progenitor}. 

In this work, we put constraints on the CSM mass and the ejecta mass (therefore the total preexplosion mass). We also estimate the average mass-loss rates of the progenitors. We find that the CSM typically holds a few M$_\odot$ of material (but can be in excess of 10\,M$_\odot$). Furthermore, the progenitors lose mass at a rates of $\sim\,$0.01\,--\,0.1\,M$_\odot$\,yr$^{-1}$ but can be in excess of 1\,M$_\odot$\,yr$^{-1}$. In lieu of our `smoking gun' progenitor detections, we can make inferences on the progenitor channel using these parameters. The massive CSM and generally massive progenitors, along with the high mass-loss rates we find all suggest LBV progenitors. Indeed, this reasoning is employed in the literature. The high CSM mass, precursor events and bumps in the light curve are likely formed by some LBV great eruption event(s) \citep[with some notable examples being SN\,2006tf, SN\,2011fh, SN\,2015da and SN\,2021qqp][]{Smith_2008_06tf, Pessi_2021,Smith_2024,Hiramatsu_2024}. One SN\,IIn with a very high mass-loss rate is SN\,2006tf, which had lost 20\,M$_\odot$ of mass in the few decades leading up to the SN explosion \citep[][]{Smith_2008_06tf}. Furthermore, SN\,2015da was also found to have a high mass-loss rate (up to 0.6\,M$_\odot$\,yr$^{-1}$) and large CSM mass ($\sim$\,20\,M$_\odot$;  \citealt{Tartaglia20, Smith_2024}). Furthermore, the precursor activity and dramatic secondary peak of SN\,2021qqp alludes to extreme mass loss episodes consistent with an LBV progenitor \citep{Hiramatsu_2024}. This reasoning is reflected in SN impostors (which may precede true SNe) where these events are also consistent with LBV great eruptions, with examples being: SN\,1954J \citep{Humphreys_1994, Humphreys_2017}; SN\,2000ch \citep[which periodically explodes, perhaps due to binary interactions,][]{Smith_2011_binlbv, Aghakhanloo_2023}; SN\,2002kg \citep{Weis_2005,Maund_2006, Humphreys_2017}; SN\,2003gm \citep{Kochanek_2012} and AT\,2016blu \citep[periodic and similar to SN\,2000ch][]{Aghakhanloo_2023_16blu}.

 From Figure \,\ref{fig:mdotscatter}, it is apparent that the majority of our SNe\,IIn have CSM velocities and corresponding mass-loss rates that are largely consistent with LBV eruptions. This conclusion is supported by the fact that we find the median CSM profile density parameter $s\,\simeq\,1.3$--i.e., a more eruptive-like CSM profile in general. We do note that while we may overestimate the CSM velocity, even if we assumed a modest value of 100\,km s$^{-1}$, $\sim$\,75\% of our SNe\,IIn would still have mass-loss rates consistent with an LBV great eruption.

  However, we note that in the literature there are some SNe\,IIn with inferred mass-loss rates inconsistent with LBV eruptions. For example, SN\,1988Z, the aforementioned prototypical long-lived SN\,IIn has been suggested to have a clumpy wind-like CSM, with a CSM geometry inconsistent with eruptions \citep{Chugai_1994_88z}. One transient that is presented as representing a possible continuum of mass loss and SN\,IIn progenitors is PTF\,11iqb. This object exhibited weaker CSM interaction a earlier times \citep[flash ionization][]{Khazov_2016, Jacobson-Galan_2022} and stronger interaction at later times--possibly due to interaction with a more distant ring of CSM from more modest mass-loss rates of $\sim\,$10$^{-4}$\,M$_\odot$\,yr$^{-1}$ \citep{Smith_2015}. Those authors draw comparisons to the similar SN\,IIn, SN\,1998S \citep{Fassia_2001, Mauerhan_2012}. Those authors suggest that wind-like mass loss, perhaps coupled with a binary system shaping the CSM, may account for some of these weaker interacting SNe\,IIn \citep[forming a contiuum between SNe\,IIn and SNe\,IIP/L][]{Smith_2009_RSGprog, Stritzinger_2012,Shivvers_2015}. These lower mass-loss rate objects are a seemingly rare exception to the SN\,IIn landscape. Indeed, in our sample, $\sim\,$5\% of the SNe\,IIn have mass-loss rates $\sim$\,10$^{-4}$\,M$_\odot$\,yr$^{-1}$ and have a more wind-like CSM geometry.

  The discussion around progenitor routes is often centered around the initial mass of the progenitors, as this dictates the evolutionary path. Indeed the majority of the literature SN\,IIn and SN impostor examples we have discussed in this section are claimed to be massive LBVs. Without pre-explosion data the best estimate we have in this work of the progenitor mass is the `total mass' (i.e. the ejecta mass added to the CSM mass). This traces the mass of the progenitor in the few decades to hundreds of years prior to the terminal explosion. Most events in our sample have a large total mass (the sum of the ejecta and the CSM masses) with $M_\mathrm{Tot}\simeq\,25$\,M$_\odot$. Hence, our SN\,IIn sample is largely consistent with massive progenitors (e.g., LBVs). The total mass distribution is shown in the lower left panel of  Figure \,\ref{fig:mcsm_mej_totmass_rcsm_hist}. This distribution is skewed to larger total masses, covering a range of total masses from a few to over 50\,M$_\odot$. We do note that the ejecta masses are not as well constrained as the CSM masses and the total mass traces our ejecta mass estimates, so the spread in the distribution of total masses is large. 
  
  In the literature, we see a range of total masses for SNe\,IIn: from  progenitors that may be heavily stripped to the very massive. One example of a possibly highly stripped progenitor in the literature is SN\,1988Z, with subsolar ejecta mass \citep[however, such a degree of mass stripping may require binary mass loss][]{Chugai_1994}. This SN may be an example of a quenched interaction SN\,IIn where M$_{\mathrm{ej}}\,\ll\,\mathrm{M}_{\mathrm{CSM}}$ \citep{Dessart_2024}. In contrast, massive LBVs have been proposed as the progenitors for a number of SNe\,IIn \citep[e.g. SN\,2005gl, SN\,2010jl and SN\,2015da][with masses reaching 50\,M$_\odot$]{Gal-Yam_2007, GalYam_2009,Ofek_2014_10jl,Smith_2024}, and our findings are consistent with these objects. 
  
  Finally, we note again that we cannot rule out an extended mass-loss history due to more distant, diffuse CSM not being dense enough for any interaction features to be observable. Hence, these total masses should be treated as a lower limit on the initial mass of the progenitor.

We have shown that the physical parameters from \texttt{MOSFiT}, paired with observational parameters, for most of our sample are consistent with LBV-like progenitors. There is, however, still tension with other lines of investigation. Chiefly, the environments of SNe\,IIn are inconsistent with the progenitors being single massive stars \citep[with some SNe\,IIn being apparently unassociated with ongoing star formation][]{hab14, Galbany_2018, Ransome_2022}. Identified LBVs in nearby galaxies (the Small and Large Magellanic Clouds and M\,33) seem well associated with ongoing star formation \citep{Kangas_2017}. Some of this discrepancy may be explained by considering some massive stars may be in low surface brightness H\,II regions. It has also been suggested that LBVs may also form as the product of mergers and mass accretion in binary systems with lower-mass stars \citep{Podsiadlowski_1992, Justham_2014,Tombleson_2015}, which may be corroborated when considering the extremely isolated SN\,2009ip \citep[][]{Smith_2011_binlbv, Kashi_2013,Mauerhan13a, Soker_2013}. We do not perform a detailed environmental analysis in this work, but we do note that almost all of the SNe in our sample with a classified host \citep[many SN\,IIn hosts are low surface brightness/dwarf galaxies e.g.][]{Ransome_2022} are in actively star forming galaxies. A detailed analysis of the local environments of our SNe\,IIn is a topic of future work.

Recently, there has been growing evidence that at least some LBVs, and the great eruptions associated with them, may be the product of binary interaction \citep{Kashi_2010,Smith_2011_binlbv, Soker_2013, Kashi_2013, Aghakhanloo_2023_16blu, Aghakhanloo_2023}. If this is the case for at least a subset of LBV-like progenitors, this may alleviate some of the tension with the aforementioned environmental inconsistencies \citep{hab14, Galbany_2018, Ransome_2022}. As our sample is consistent with LBV progenitors, we now briefly discuss the evidence for LBVs in binaries. Observationally, the CSM geometry around SNe\,IIn is inferred to be often asymmetric. Such asymmetry may be revealed via the analysis of the H$\alpha$ profiles or measurements from polarimetry. \citet{Bilinski_2024} present multi-epoch spectropolarimetric observations for a sample of 14 SNe\,IIn. Those authors find that at peak optical light, there is an intrinsic polarization for most of their SNe\,IIn sample. This polarization level reached as high as 6\% in the case of SN\,2017hcc--the highest level of polarization seen in any SN. Those authors also note that their results suggest that SNe\,IIn are the most strongly polarized of any SN class. \citet{Bilinski_2024} conclude that for such persistent asymmetric CSM profiles to be present, eruptive mass loss into spherical shells is unlikely. Rather, mass loss due to binary interaction or eruptive mass loss that is in some way enhanced in an equatorial torus is favored. \citet{Brennan_2022_progenitor} also argue that their models of the progenitor of SN\,2016jbu are consistent with a massive binary system.  %We will now discuss arguments that the high mass-loss rates and CSM masses found around some of our SNe\,IIn may require some interaction in a massive binary system as opposed to the eruptions of a single massive star.

As previously discussed, the mass-loss rates from steady line-driven winds are insufficient to be the sole mass loss mechanism for SNe\,IIn. The large CSM masses around some SNe\,IIn require extreme mass loss rates, which we see in our sample but also seen in the literature \citep[e.g. ASASSN-14il, SN\,2015da, ASASSN-15ua][]{Dukiya_2024, Smith_2024, Dickinson_2024}. When considering the asymmetry seen in the CSM, it has been suggested that the high mass-loss rates may require violent interactions within a binary. Mass loss (and perhaps the SN itself) in binary systems may manifest from mechanisms such as: violent mass transfer events aside from normal Roche-lobe overflow; mergers of massive stars; mergers of mass stars with a compact object; common envelope interactions or collisions \citep{Soker_2006, Chevalier_2012,SmithArnett_2014, Schrøder_2020,Matsuoka_2024, Ercolino_2024, Schneider_2024}. The semi-periodic eruptive episodes seen in SN impostors (with likely LBV progenitors) or other massive outbursts have also been linked to binary interactions \citep[][]{Kashi_2010, Smith_2011_binlbv,Soker_2013, Aghakhanloo_2023, Aghakhanloo_2023_16blu}. Indeed a Galactic analog is the aforementioned $\eta$\,Car, which has at least one companion and is entombed in the highly asymmetric Homunculus nebula \citep[with the 19th century Great Eruption possibly being some merger event,][]{Smith_2003, Smith_2010, Kashi_2010, Smith_2011_binlbv}. These observational clues lead to the argument that as well as the continuum driven (or other mechanism) eruptions in LBVs, violent binary interactions may be key into understanding the mass loss around SN\,IIn progenitors. 

We note that even with binary interactions, an LBV progenitor may still be required (given the observed progenitors and observed LBV outbursts). We do note that our models can not inform on the mass-loss mechanisms outside of the mass-loss rate estimates. Nonetheless, these are important considerations in the discussion of SN\,IIn progenitors \citep[indeed, most massive stars, and LBVs in particular are in binaries, e.g.][]{Sana_2012, Mahy_2022}. We do, however, caution that the regions shown in Figure\,\ref{fig:mdotscatter} are indicative based on empirical measurements. With more theoretical work, an additional illustrative region may be added for LBVs suffering violent binary interactions (or the great eruption region may be split into single star/binaries as there are likely degeneracies or overlap). Specifically, models that can reproduce a CSM forming rapidly and/or episodically would help solidify our understanding of these systems. Observationally, the binary scenario for individual objects could be confirmed with late time imaging that reveals a surviving companion.

 To summarize this discussion, the typical set of parameters from \texttt{MOSFiT} and other measured observational quantities from the photometry indicate that most of our progenitors are consistent with LBV progenitors. We do, however, note that our measured CSM velocities may be upper limits and overestimate the mass-loss rates. Typically, our mass-loss rates and CSM velocities are too high to be attributed to line-driven winds and Roche-lobe overflow mass loss in binaries. However, there is observational evidence \citep[e.g.][]{Bilinski_2024} that many SNe\,IIn are in binary systems and mass loss mechanisms in these systems may be more violent than the Roche-lobe overflow scenario. The mass-loss rates of binary interactions are poorly understood and may extend a large region of the mass-loss rate parameter space that we show in Figure \ref{fig:mdotscatter} \citep[][]{Langer_2012, smith14}. Some of our SNe\,IIn have mass-loss rates higher than what are typically understood to be from the great eruptions of LBVs (i.e. a few M$_\odot$\,yr$^{-1}$); such extreme mass-loss rates are attributed to dramatic mass loss events due to binary interactions, beyond the mass-loss rates expected from Roche-lobe overflow \citep{Kashi_2010,Soker_2013,SmithArnett_2014, Smith_2024, Dickinson_2024}. In this study we have demonstrated that SNe\,IIn exhibit a diverse range of properties. When we consider the masses, CSM geometries, ejecta geometries and mass-loss rates, we find that our sample is most consistent with massive progenitors, likely LBVs. These findings are consistent with findings in the literature, where direct progenitor detections, high CSM masses and high mass loss rates converge on LBVs being a main progenitor path for SNe\,IIn.

\section{Future Prospects and Discovery Rates in LSST } \label{sec:future}

Large surveys have made great strides in the discovery and classification of SNe. At the time of writing, there are $\sim\,$10$^4$ classified SNe. The Legacy Survey of Space and Time (LSST) conducted at the Vera C. Rubin Observatory is scheduled to start observations in 2025 \citet{LSST, LSST2}, and is expected to increase the discovery rate of SNe to 10$^6$ a year.

LSST has great potential to construct large samples of rare transients, such as SNe\,IIn. In order to estimate the SN\,IIn rate in LSST, we use our SN\,IIn sample to simulate a realistic population of transients observable with LSST, following \citet{Villar_2018}. First, we generate model light curves using the \texttt{MOSFiT} CSM models in the LSST \textit{ugrizy} filter set using the priors from this work. We use rejection sampling of the simulated light curves to match the observed luminosity function with our Malmquist corrected luminosity function. For these generated light curves, we use a log-uniform distribution of host galaxy extinction from 10$^{-5}$\,--\,1\,mag, consistent with the observed extinction observed in this sample. We inject simulated light curves into a LSST-like simulation, using \texttt{OpSim}\footnote{v2.1.0 \url{https://github.com/lsst/rubin_sim}}. In \texttt{OpSim}, we inject 100 generated SNe\,IIn into 20 equally-spaced redshift bins spanning $z\,=\,$0.01\,--\,1.5. These SNe are injected uniformly across the sky. For each SN at each redshift bin, we determine if $>=\,10$ points were detected at SNR$\,>=\,10$ in any of the LSST \textit{ugrizy} filters; if they are then this is considered a detection. The number of detected SNe\,IIn is then multiplied by a volumetric rate at each redshift \citep[see Equation. 2 in ][]{Villar_2018}. This volumetric rate follows the models of \citet{Strolger_2015}, assuming that SNe\,IIn comprise around 4 percent of CCSNe \citep[e.g.][]{Cold_2023}. The resultant annual SN\,IIn rate in LSST is $\sim$\,1.6\,$\times$\,10$^{5}$\,yr$^{-1}$. This is a much higher rate than the  estimate from \texttt{PLAsTiCC} of $\sim$\,6\,$\times$\,10$^4$\,yr$^{-1}$ \citep{Kessler_2019}, likely due to an artificially dim luminosity function used in the \texttt{PLAsTiCC} training set.

Clearly, not all of these objects will be able to be spectroscopically followed. Therefore to build new, large samples of SNe\,IIn, photometric classifiers are required for rapid classification \citep[e.g. SuperRAENN, Superphot and Superphot+][]{Villar_2020,Hosseinzadeh_2020, deSoto_2024}. Anomaly detection on real time data may be able to find the peculiar light curve bumps or pre-cursor activity we see in many of our SNe\,IIn \citep[e.g. LAISS][]{Aleo_2024}. Furthermore, LSST may even aid in the direct progenitor detection of SNe\,IIn, elucidating the progenitor paths of these enigmatic transients \citep[e.g.][]{Strotjohann_2024}.

\section{Conclusions and Summary} \label{conc}

In this paper we have performed the first systematic light curve modeling of a large sample of 142 SNe\,IIn, mostly collected from large surveys such as ZTF, PTF, YSE and PS1-MDS. Our light curve models from \citet{Chatzopoulos_2012} and implemented via \texttt{MOSFiT} allow us to infer the physical parameters of these enigmatic and highly diverse transients. These parameters, along with other observational features, were used to explore correlations between parameter pairs and also to calculate mass-loss rates when accompanied by spectral information. Our conclusions can be summarized as follows:

\begin{enumerate}
    \item Our large sample of SNe\,IIn show a broad continuum of inferred parameters. However, these events typically have high mass-loss rates, large CSM and preexplosion masses, and a CSM geometry suggestive of eruptive mass loss. These parameters are all consistent with LBV progenitors.
    \item These SNe have typical mass-loss rates of $\sim$\,10$^{-2}$\,M$_\odot$\,yr$^{-1}$. These mass-loss rates are consistent with the values in the literature and are typical of LBV progenitors. 
    \item The density profiles of the CSM skew toward $s\,=\,2$ but the inferred population distribution spans a wide range of geometries. The median value of $s\,\approx\,1.3$ suggests that the mass loss of SNe IIn progenitors are typically not undergoing steady-state, wind-like mass loss.
    \item The inferred CSM mass surrounding SNe IIn is typically 1\,M$_\odot$. This distribution does, however, extend to larger CSM masses (in excess of 10\,M$_\odot$).
     \item Ejecta mass is not particularly well-constrained by out models. However, the median ejecta mass of $\sim\,20\,$M$_\odot$ suggests massive progenitors.
    \item The density profile of the inner ejecta, $n$ has a broad posterior distribution, with a median of $\sim\,$9.3 at the lower end of our physically-informed prior. This is indicative of a more stripped or compact progenitors such as LBVs.
    \item Our SNe have very dense inner CSM profiles, with an average $\rho_0\,\simeq\,1.5\,\times\,10^{-12}$\,g\,cm$^{-3}$. This value is consistent with the range found in the literature.
    \item The $r$-band rise time is strongly, positively correlated with the CSM mass-- i.e. the more CSM the progenitor has around it, the slower its rise to peak. The fall time of the $r$-band light curve is also strongly positively correlated with the CSM mass, with higher CSM masses resulting in a more slowly fading transient. This is consistent with the findings of \citet{Nyholm_2020}. 
    \item Similarly, the $r$-band fall times and rise times are positively correlated, indicating that slowly rising SNe\,IIn also slowly fade. There is a negative correlation between the $r$-band fall time and the peak absolute $r$-band magnitude, suggesting that the more luminous a SN\,IIn is, the slower it will fade. These correlations are due driven by the CSM in each SN. 
    \item We find multimodality in the fall time, with a long fall time subgroup which is centered around 200 days and a main grouping centered around 50 days. These long fall time SNe\,IIn tend to have massive CSM.
    \item We can group our SNe into several broad photometric categories which form a continuum. There is a small population of fast decliners, with $\sim\,20$ percent of our sample having fall times of less than a month. While our temporal coverage is typically not long enough to confirm, there are several examples of slowly declining SNe\,IIn in our sample. These may be consistent with the observed population of long-lived SNe\,IIn. Within our sample, around 15 percent of the transients are very luminous, with a peak $r$-band absolute magnitude of at least --20\,mag. Conversely there is a population of fast risers, which makes up around 25\% of the sample, a smaller proportion than found by \citet{Nyholm_2020}. 
    \item The host galaxies of our SNe\,IIn are primarily star forming spiral galaxies (as expected for massive progenitors) with one exception of an elliptical host.
    \item Our sample of SNe is consistent with the distribution of the peak $r$-band absolute magnitude found in the literature \citep[e.g.][]{Kiewe_2012}, with a Malmquist corrected average (and standard deviation) of --18.7\,$\pm\,1.0$\,mag.  
    \item Finally, using our Malmquist corrected luminosity function, we simulate the discovery rates of SNe\,IIn in the upcoming Legacy Survey of Space and Time. We estimate that the SN\,IIn discovery rate will be $\sim$\,1.6\,$\times$\,10$^5$\,yr$^{-1}$.
    
\end{enumerate}

For the first time, utilizing a large sample and systematic modeling, we have found that the majority of the progenitors of SNe\,IIn are consistent with LBVs. The high mass-loss rates, large CSM masses and total masses, along with the typical CSM geometries are all suggestive of massive progenitors that suffer eruptive mass-loss events (whether as a single star or due to binary interaction). In the upcoming years, we expect the SNe\,IIn sample to increase by over three orders of magnitude--opening the possibility to better understand the explore the energetics, environments and progenitor channels of these events.

\begin{acknowledgments}

C.L.R.\ and V.A.V.\ acknowledge previous support from the Charles E.\ Kaufman Foundation through the New Investigator Grant KA2022-129525. The Villar Time-domain Astronomy Data Lab acknowledges additional support through the National Science Foundation under AST-2433718, AST-2407922 and AST-2406110, as well as an Adamant Fellowship for Emerging Science Research. This work is additionally supported by NSF under Cooperative Agreement PHY-2019786 (The NSF AI Institute for Artificial Intelligence and Fundamental Interactions, \url{http://iaifi.org/}).

The Pan-STARRS1 Surveys (PS1) and the PS1 public science archive have been made possible through contributions by the Institute for Astronomy, the University of Hawaii, the Pan-STARRS Project Office, the Max-Planck Society and its participating institutes, the Max Planck Institute for Astronomy, Heidelberg and the Max Planck Institute for Extraterrestrial Physics, Garching, The Johns Hopkins University, Durham University, the University of Edinburgh, the Queen's University Belfast, the Harvard-Smithsonian Center for Astrophysics, the Las Cumbres Observatory Global Telescope Network Incorporated, the National Central University of Taiwan, the Space Telescope Science Institute, the National Aeronautics and Space Administration under Grant No. NNX08AR22G issued through the Planetary Science Division of the NASA Science Mission Directorate, the National Science Foundation Grant No. AST–1238877, the University of Maryland, Eotvos Lorand University (ELTE), the Los Alamos National Laboratory, and the Gordon and Betty Moore Foundation.

This research has made use of the NASA/IPAC Extragalactic Database (NED),
which is operated by the Jet Propulsion Laboratory, California Institute of Technology,
under contract with the National Aeronautics and Space Administration.

YSE-PZ was developed by the UC Santa Cruz Transients Team with support from The UCSC team is supported in part by NASA grants NNG17PX03C, 80NSSC19K1386, and 80NSSC20K0953; NSF grants AST-1518052, AST-1815935, and AST-1911206; the Gordon \& Betty Moore Foundation; the Heising-Simons Foundation; a fellowship from the David and Lucile Packard Foundation to R. J. Foley; Gordon and Betty Moore Foundation postdoctoral fellowships and a NASA Einstein fellowship, as administered through the NASA Hubble Fellowship program and grant HST-HF2-51462.001, to D. O. Jones; and a National Science Foundation Graduate Research Fellowship, administered through grant No. DGE-1339067, to D. A. Coulter.

\end{acknowledgments}


\vspace{5mm}
\facilities{Pan-STARRS, ZTF, PTF}


\software{astropy \citep{astropy},  
          \texttt{MOSFiT} \citep{Guillochon_2017}, 
          numpy \citep{numpy},
          pandas \citep{pandas},
          scikit \citep{scikit-learn}
          }



\appendix

\section{Testing \texttt{MOSFiT} CSM Models with the Superluminous Models of Dessart et al. 2015} \label{sec:dessart}

It has been noticed that the CSM-ejecta interaction models implemented in \texttt{MOSFiT} may overestimate CSM masses in some cases. \citet{Nicholl_2020} used \texttt{MOSFiT} to model the light curve of the SLSN-II, SN\,2016aps. Those authors initially argued that the transient may be an example of a PISN from a very massive star. Along with a large radiated energy budget of $\sim$\,10$^{52}$\,erg and a slow evolution timescale, modeling suggested that SN\,2016aps could be explained by $\sim$\,100\,M$_\odot$ of ejecta colliding with $\sim$\,100\,M$_\odot$ of CSM. However, \citet{Suzuki_2021} note that the models for SLSNe-IIn from \citet{Dessart_2015} can reproduce a similar luminosity and photometric timescale evolution to SN\,2016aps with much more modest parameters (with M$_\mathrm{CSM}$\,$\sim$\,17\,M$_\odot$ and M$_{\mathrm{ej}}$\,$\sim$\,10\,M$_\odot$). Those authors go on to model SN\,2016aps with radiation-hydrodynamic simulations, finding that SN\,2016aps was still had a very massive progenitor, but consisted of 30\,M$_\odot$ of ejecta colliding around 8\,M$_\odot$ of CSM.

\begin{figure}[!htb]
\centering 
	\includegraphics[width=0.45\columnwidth]{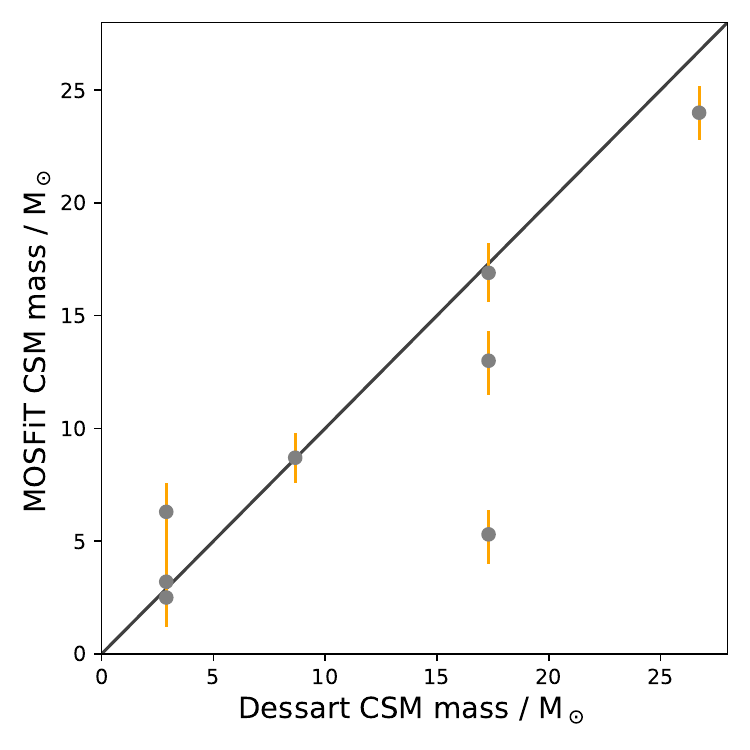}
    \caption{Scatter between the CSM masses inferred by \texttt{MOSFiT} and the models of \citet{Dessart_2015}. Error bars are at the 1$\sigma$ level and the 1:1 line is plotted in black. }
    \label{fig:dessart}
\end{figure}

To test the validity of the \texttt{MOSFiT} inferences for SNe\,IIn which verge on the superluminous (i.e. $\sim$\,--20\,mag), we take the models of SLSNe-IIn from \citet{Dessart_2015} and fit the light curves with \texttt{MOSFiT}. In these models, $\sim$\,10\,M$_\odot$ of ejecta interacts with varying CSM masses at various maximum ejecta velocities. The inner CSM being at $R_0\,\approx\,70\,\mathrm{AU}$. In our \texttt{MOSFiT} fits, we leave the CSM mass and ejecta velocity as free parameters, but fix the others to match that of \citet{Dessart_2015}. The comparisons of the CSM mass values from our fits with the values from \citet{Dessart_2015} are shown in Figure\,\ref{fig:dessart}. Within the 1$\sigma$ uncertainties, 4 out of the 8 of the trials are consistent with the \citet{Dessart_2015} models. The other trials however are typically within a few 10\% of the \citet{Dessart_2015} value. The one set of parameter that does not agree well is the Xe10m6 trial which had high ejecta and SN kinetic energies, $\sim$\,17\,M$_\odot$ of CSM and a high ejecta velocity of $\sim\,30,000\,$kms$^{-1}$. This configuration results in a very high inner CSM density of $\sim$\,10$^{-8}$\,g\,cm$^{-3}$, which is well outside of our allowed range.

\section{Luminosity Function} \label{sec:malm}

\begin{figure*}[!htb]
	\includegraphics[width=0.95\textwidth]{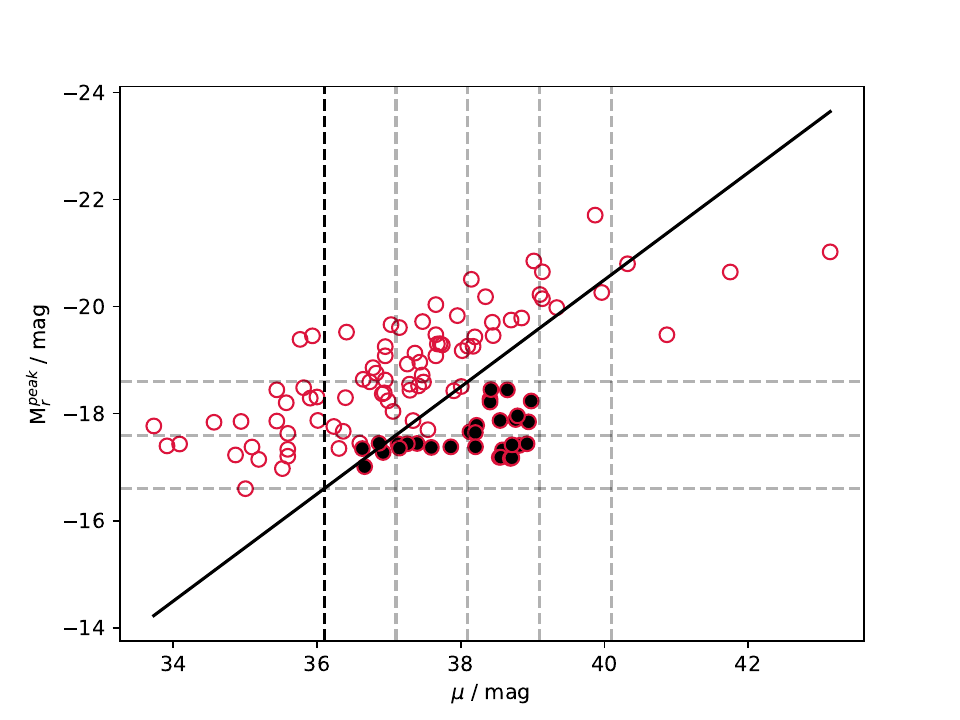}
    \caption{A Miller diagram, showing the peak $r$-band absolute magnitudes of our SNe\,IIn (the subsample that share a comparable $r$-band filter) against their distance moduli. The solid diagonal line is the absolute $r$-band magnitude cut-off for the ZTF\,BTS. The vertical dashed black line is the distance modulus at which the absolute magnitude of the faintest member of the sample, intercepts with the ZTF\,BTS magnitude cut limit. The grey dashed lines define the magnitude and distance modulus bins that are used in the simulation of the missing SNe\,IIn. The empty circles are the SNe\,IIn used in this analysis and the circles filled in black are the simulated missing SNe\,IIn from our Malmquist bias correction.}
    \label{fig:malmquist}
\end{figure*}

Our sample suffers from Malmquist bias, meaning that our sample is biased towards SNe that are observationally brighter \citep{Malmquist_1922}. Some of of our sample is constructed from magnitude limited surveys which inherently contain Malmquist bias, with around 60\% of the sample coming from ZTF\,BTS which has a cut for peak magnitude but is spectroscopically complete. In order to correct this bias in our luminosity function (absolute $r$-band magnitude, the most comparable measure in our sample), we implement the method outlined in \citet{Richardson_2014, Taddia_2019, Nyholm_2020} which was used on magnitude limited surveys such as PTF/iPTF \citep{Taddia_2019, Nyholm_2020} and samples from numerous sources \citep[][]{Richardson_2014}. 

To set up our Malmquist bias correction, we find the absolute magnitude limit based on a selected survey limiting magnitude. In this case, as the majority of our sample is from the ZTF\,BTS, we use the apparent magnitude cut-off of 19.5\,mag in the $r$-band (this is shown as the diagonal line in Figure \ref{fig:malmquist}. Any SNe found to the diagonal left (i.e. brighter than the limit) is considered part of a `complete' sample. Then, the intrinsically faintest member of the sample is identified. In this case, the faintest SN in our BTS sample has an $r$-band absolute magnitude of --17.0 mag. The intersection between this faintest SN and the ZTF\,BTS magnitude cutoff tells us to which distance modulus our sample is complete to (i.e., 35.69). The sample is then split between --17\,mag and --18\,mag as our intrinsic distribution of `faint' SNe, and another between --18\,mag and --19\,mag. These distributions, while not complete in the sense that all SNe are present, are taken to be representative of the luminosity distribution. The fraction of luminous SNe to the fainter ones is then used to simulate the missing SNe at larger distance moduli, iteratively for each of the subsequent 2 distance bins. Before the correction for the Malmquist bias, the distribution is roughly Gaussian with a mean peak $r$-band absolute magnitude and spread of $-19.2\,\pm\,1.0$ mag; with correction, it is $-18.7\,\pm\,1.0\,$mag, both of which are consistent with the average found by \citet{Kiewe_2012}. Our simulated missing SNe are represented by the filled circles in Figure \,\ref{fig:malmquist}. 

\section{Individual Object Posterior Distributions} \label{sec:indposts}

Due to there being apparent `pile-up' in the joint posterior distributions shown in Section\,\ref{sec:params}, we plot the individual transient posteriors and compare against the full sample. Many of our parameters probe the full prior distribution, showing a slight preference for the higher/lower end of the prior range (e.g. $s$ and $n$). The velocity posterior distributions are centered around the average value seen in in the joint posterior. The apparent pile up to the extremes of the prior distributions is less pronounced here, suggesting an additive effect. We also calculate the fraction of the individual posterior distributions where the modal bins are at the extreme bounds of the prior distribution (i.e. the first and last bins).

\begin{figure*}[!htb]
	\includegraphics[width=0.3\textwidth]{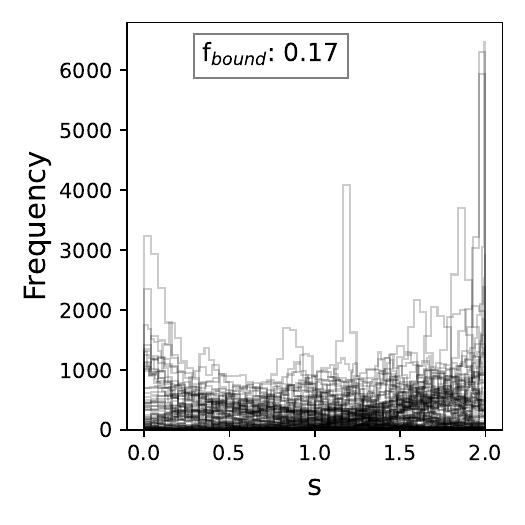}
    \includegraphics[width=0.3\textwidth]{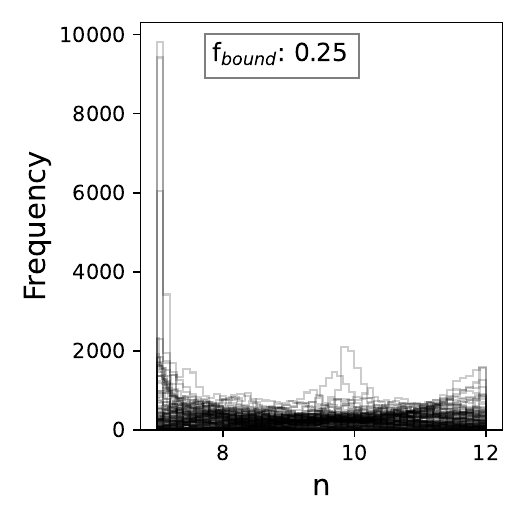}
    \includegraphics[width=0.3\textwidth]{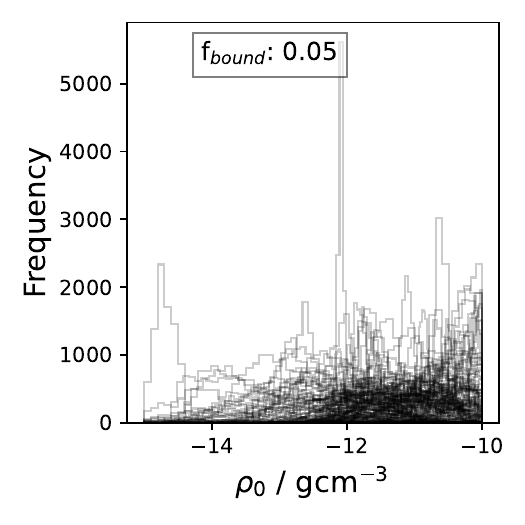}
    \includegraphics[width=0.3\textwidth]{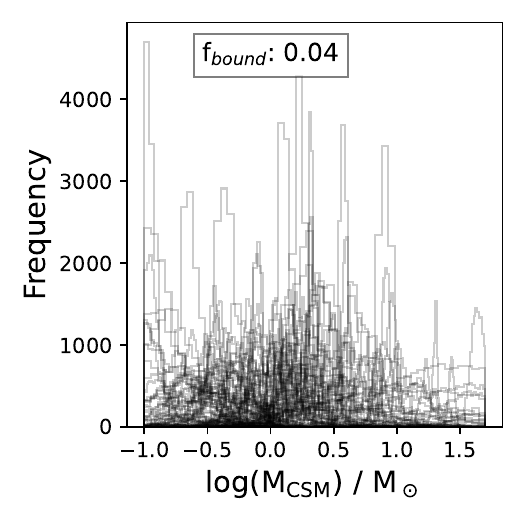}
    \includegraphics[width=0.3\textwidth]{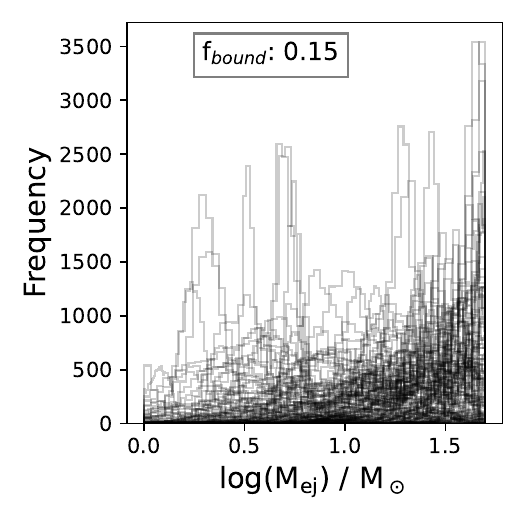}
    \includegraphics[width=0.3\textwidth]{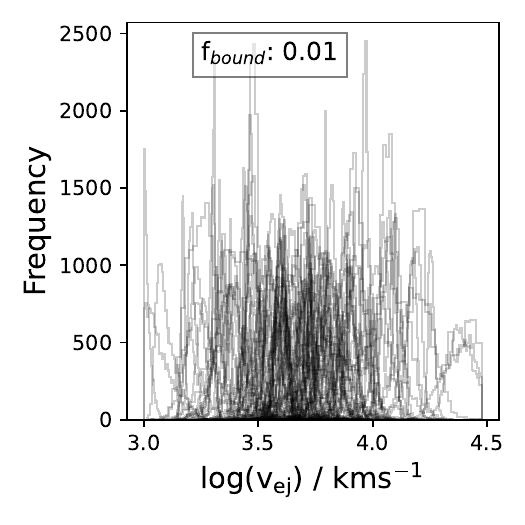}
    \includegraphics[width=0.3\textwidth]{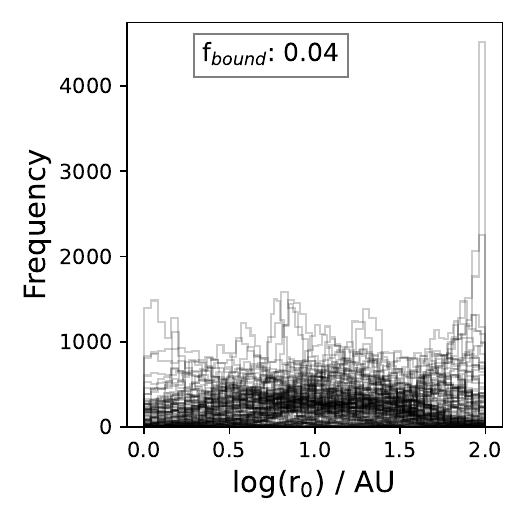}
    \caption{(\textit{Top left:}) The individual posteriors of $s$. While there is a slight apparent preference towards $s\,=\,2$, there is a spread over all values of $s$, suggesting we are probing a diverse set of geometry parameters. (\textit{Top middle:}) the individual posterior distributions for $n$. These distributions show a preference for smaller $n$ values but again, generally span the full prior distribution. (\textit{Top right:}) the individual posterior distributions for $\rho_0$. These distributions show a preference for a denser inner CSM. (\textit{Bottom left:}) The individual posterior distributions for the CSM mass. Here the distributions prefer a CSM mass of around 1\,M$_\odot$. (\textit{Bottom middle:}) The individual posteriors for the ejecta mass. Our distributions skew to the higher end of the prior distribution, but also probe smaller masses. (\textit{Bottom right:}) The individual posterior distributions of the ejecta velocity. These distributions are centered around $\sim\,4000$\,km\,s$^{-1}$. This is representative of the Gaussian-like joint posterior distribution. Alos shown in each of these plots is the fraction of the realizations with modal values at either bound of the prior distribution.}
    \label{fig:ind_post}
\end{figure*}

\section{Assessing the Contribution of the H$\alpha$ Line to the $r$-band Flux} \label{sec:synphot}

The CSM interaction model we implement does not explicitly consider the flux contribution from to the strong H$\alpha$ emission lines. In the local universe, the H$\alpha$ line is in the $r$-band. We assess this contribution in order to determine if there is a significant systematic difference between the light curve fits and observations due to unaccounted Balmer emission.

In order to estimate this contribution, we perform synthetic photometry on a representative sample of SN\,IIn spectra using \texttt{synphot} \citep[version 1.1.2][]{synphot}. We use the spectra for the SNe\,IIn in \citet{Ransome_2021} which were collated from the OSC \citep{OSC}. In total, 139 spectra are used. These SNe\,IIn have multiple spectral epochs spanning a range  $\sim$\,10\,--\,10$^3$\,days. An $r$-band filter in \texttt{synphot} was used for the synthetic photometry on the spectra. The flux from the full $r$-band bandpass (5400\,--\,7000\AA) was measured  for each spectra. Then the flux from these spectra with the H$\alpha$ region masked out and the spectrum interpolated to the continuum, effectively removing the H$\alpha$ emission. The flux differences between the unaltered and altered spectra were assessed using the fractional difference of flux in the $r$-band bandpass, where the difference in counts is normalized by the full $r$-band flux. 

\begin{figure}[!tb]
\centering
	\includegraphics[width=0.5\columnwidth]{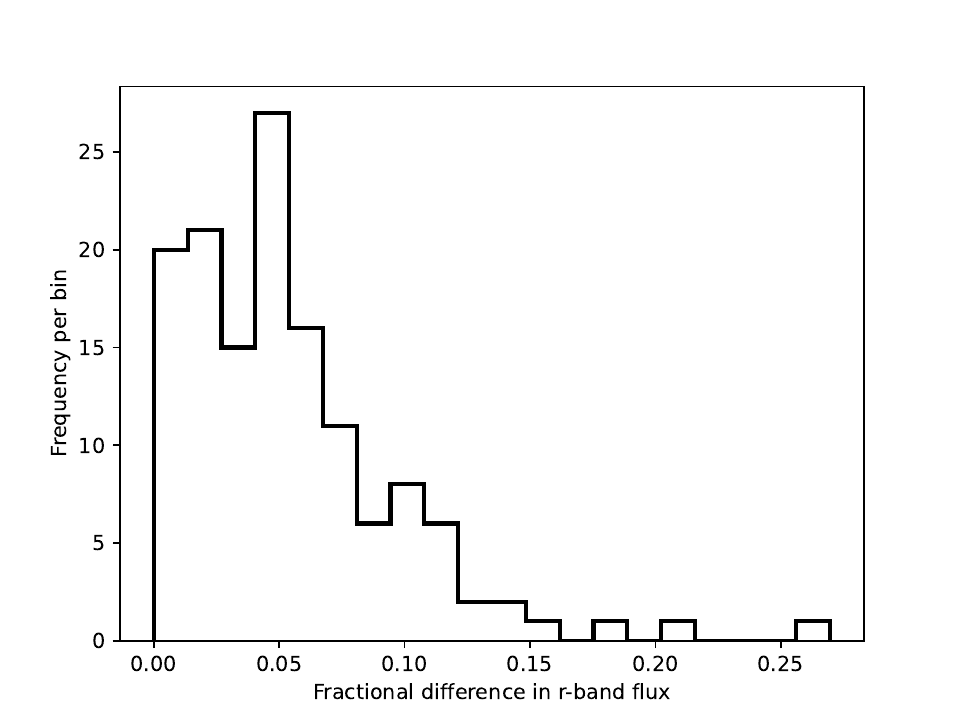}
    \caption{Histogram showing the fractional difference in $r$-band flux--i.e., the fractional difference between synthetic photometry measurements of SN\,IIn spectra and spectra with the H$\alpha$ lines removed. }
    \label{fig:diffs}
\end{figure}

The fractional flux differences between the spectra within 100 days of the explosion are shown in Figure \,\ref{fig:diffs}. We find that 60\% of the fractional differences are below the 5\% level and 84\% of the sample are below the 10\% level. We find that when the differences are assessed over longer timescales, especially in the case of the long-lasting SNe\,IIn such as SN\,2005ip and SN\,2006jd, these differences are more appreciable. When these data are available, we see that the fractional differences may increase with time. At late times for the long-lasting SNe\,IIn (thousands of days post explosion), the fractional differences increase and may begin to dominate over the continuum. 

In this paper, our \texttt{MOSFiT} models are fitting for a duration of a few hundred days post-peak brightness, so the typical fractional flux differences are largely negligible.

\section{Fall Time Silver Sample} \label{sec:silver}

\begin{figure*}[!htb]
	\includegraphics[width=0.95\textwidth]{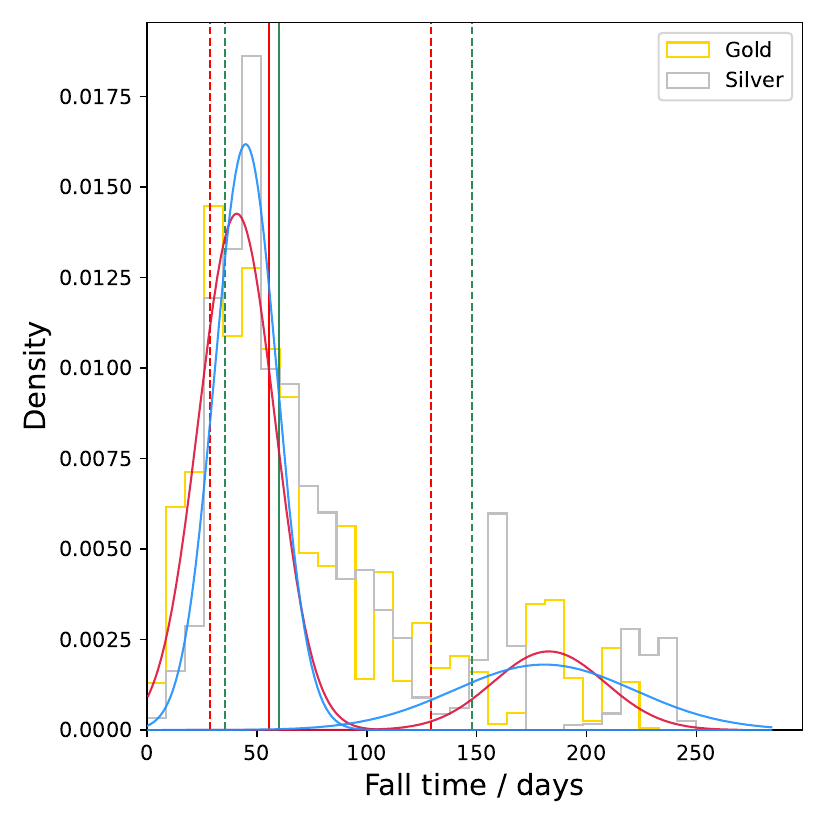}
    \caption{The fall time distribution of the `silver' SNe\,IIn where the data does not show a decline of 1\,mag from $r$-band peak. This distribution is shown as a silver histogram The median is shown by the solid green vertical line and the bounds of the spread are highlighted by dashed vertical green lines. The blue Gaussians are the components found by a GMM analysis, and show a main shorter fall time component and a longer fall time component. Also plotted for comparison is the main sample, or `gold' sample where a fall time measurement was possible. This is shown as a gold histogram, with red GMM compoenents and red lines demarking the median and bounds of the spread.}
    \label{fig:silver}
\end{figure*}

In our analysis measuring the fall time from peak, we exclude SNe\,IIn that do not have data covering a decay of 1\,mag from the $r$-band peak. Here we extend our models beyond the time range of the data to estimate the fall times for a `silver' sample (albeit, less constrained, with a median spread of $\sim$\,5 days for the silver sample, compared to $\sim\,$1 day for the gold sample). There are 46 `silver' SNe\,IIn for this analysis. A histogram of the fall time distribution is shown in Figure\,\ref{fig:silver}. This distribution is consistent with the main set of fall time measurements presented in Section\,\ref{sec:obs}. The median fall time and spread is  $\sim$\,60$^{+88}_{-24}$ days. While the individual objects may be less constrained in terms of the fall times when the model light curves are extended, this set shows a similar distribution to the full set, with two components found by a GMM.

\section{Excluded Transients}

Some sub-classes of SNe\,IIn are omitted in our analysis as they likely deviate from the CCSN regime of SNe\,IIn. Notably, in the gold sample of \citet{Ransome_2022}, SN\,2008S is a low luminosity, nearby transient in the Fireworks Galaxy, NGC\,6946 \citep{Arbour_2008}. Due to the low energy of this SN, it was initially proposed to be a SN impostor. However, \citet{Adams_2016}, using \textit{Hubble Space Telescope} and \textit{Spitzer} data, conclude that the progenitor had disappeared. We exclude this transient as there is a possibility that the progenitor is heavily dust enshrouded. Other events classed as impostors were not included in our light curve modeling, with these objects being cataloged by \citet{Ransome_2021}, e.g.,SN\,1997bs, SN\,2000ch and SN\,2013fs.

The thermonuclear explosion of a white dwarf embedded in a dense CSM gives rise to the thermonuclear SN\,IIn subclass, or SNe\,Ia-CSM \citep[less commonly known as SNe\,IIa or SNe\,Ian,][]{Dessart_2024}. As these are not CCSNe, they are not an appropriate application of our models, we exclude some transients in the gold sample of \citet{Ransome_2021}, including SN\,2005gj \citep{Prieto_2005}, SN\,2008J \citep{Taddia_2012}. We still include the superluminous SN\,2006gy but we do note that recent work suggests that SN\,2006gy may have thermonuclear origins \citep[based on late time spectroscopy,][]{Jerkstrand_2020}. 

\section{Target lists} \label{sec:targets}

\begin{longrotatetable}
    \movetabledown=3mm

    %\begin{center}
        
    \begin{longtable*}{llllllr}
    
    \caption{The full sample of spectroscopically confirmed SNe\,IIn that we use in this work. These transients conform to the schema outlined in Section\,\ref{sec:sample}. This table shows the transient name, the host, discovery date, redshift, J2000 coordinates and the source of the discovery/spectroscopic classification.\textit{* as presented by \citet{Ransome_2021}.} }\label{tab:sample}
%\hline
    %\multicolumn{7}{l}{}
    \\
    \hline
    SN Name & Host & Host type & $z$ & R.A (J2000) & Dec (J2000)  & Reference \\
    \hline
    \endfirsthead
    %\hline
    \multicolumn{7}{l}{\tablename\ \thetable\ --- Continued from previous page\ldots}\\
    \hline
    SN Name & Host & Host type & $z$ & R.A (J2000) & Dec (J2000) & Ref. \\
    \hline
    \endhead
    \hline
    \endlastfoot
    \hline
    \multicolumn{7}{r}{Continued on next page\ldots}\\
    \endfoot
            SN\,1989C & MCG+01-25-25& SB & 0.0063 & 09:47:45.49 & 02:37:36.10  & P.~Challis*\\
            SN\,1994W & NGC\,4041& SAB &  0.0040 & 12:02:10.92 & 62:08:32.70  & \citet{Ransome_2021}  \\
            SN\,1994Y & NGC\,5371& SABbc & 0.0085  & 13:55:36.90 & 40:27:53.40  &\citet{1994y} \\
            SN\,1995N & MCG-02-38-17& IB & 0.0062  & 14:49:28.29 & -10:10:14.40  & \citet{1995n} \\
            SN\,1998S &  NGC\,3877& SA & 0.0030 & 11:46:06.13 & 47:28:55.40  & \citet{1998s} \\
            SN\,2005db & NGC\,214& Sbc & 0.0151 & 00:41:26.79 & 25:29:51.60  & \citet{2005db} \\
            SN\,2006gy & NGC\,1260& SB0a & 0.0192 & 03:17:27.06 & 41:24:19.51  & \citet{2006gy}  \\
            SN\,2008B & NGC\,5829& - & 0.0188 & 15:02:43.65 & 23:20:07.80  & \citet{2008b} \\
            \hline
            SN\,2018fdt &Z\,197-30& SBc & 0.0550 & 17:04:44.33 & +38:14:08.00 &  \citet{2018fdt_disc}, \citet{2018fdt_spec} \\
            SN\,2018khn & WISEA\,J085619.17+523252.7& -& 0.0910 & 08:56:18.01 & +52:32:58.00   & \citet{2018khn_disc}, \citet{2018khn_spec} \\
            SN\,2018kyv & WISEA\,J130104.80+262103.2& - &  0.0940 & 13:01:04.95& +26:21:03.80 & \citet{2018kyv_disc}, \citet{2018kyv_disc} \\
            SN\,2018leh & UGC\,2949& SB:ab & 0.02396 &04:05:03.30 & +25:15:42.90   & \citet{2018leh_disc}, \citet{2018leh_spec}\\
            SN\,2018lmy & -  & - & 0.0520 &18:24:27.77  & +46:37:09.50 &  \citet{2018lmy_disc}, \citet{2018lmy_spec}\\
            SN\,2019bxq & 2MASX\,J16575851+7836144& E & 0.0140 &16:57:58.51 & +78:36:13.60 &  \citet{2019bxq_disc}, \citet{2019bxq_spec}\\
            SN\,2019dvv & LEDA\,89698& S & 0.0305 & 12:23:17.77 & +19:43:23.30	 & \citet{2019dvv_disc}, \citet{2019dvv_spec}\\
            SN\,2019hgy & WISEA\,J174820.57+481206.9& -  & 0.0360 &17:48:20.50  & +48:12:07.00 &\citet{2019hgy_disc}, \citet{2019hgy_spec}\\
            SN\,2019krt & SDSS\,J165256.08+204305.6& -& 0.0300 &16:52:54.76 &+20:43:03.50&\citet{2019krt_disc}, \citet{2019krt_spec} \\
            SN\,2019kud & SDSS\,J144429.05+335919.1& SBc& 0.0324 &14:44:29.73 &  +33:59:10.08&\citet{2019kud_disc}, \citet{2019kud_spec} \\
            SN\,2019lkr & LEDA\,2756618&-& 0.0300 & 16:05:07.98& +73:35:04.00 & \citet{2019lkr_disc}, \citet{2019lkr_spec} \\
            SN\,2019pgu & -  &- & 0.1055 &	16:18:42.83 & +67:54:00.30 &\citet{2019pgu_disc}, \citet{2019pgu_spec} \\
            SN\,2019qny & LEDA\,1083065& - & 0.0480 & 03:32:59.94 & -02:46:41.50 & \citet{2019qny_disc}, \citet{2019bxq_spec} \\
            SN\,2019qt & - &- & 0.0350 & 	14:59:10.65 & +43:49:11.60 & \citet{2019qt_disc}, \citet{2019qt_spec} \\
            SN\,2019qvr & - & - & 0.0750 & 	03:41:21.69 & +22:49:20.80 & \citet{2019qvr_disc}, \citet{2019qvr_spec} \\
            SN\,2019rz & UGC\,3445& SB:b & 0.01885 & 06:50:25.81 & +43:03:11.60 & \citet{2019rz_disc}, \citet{2019rz_spec} \\
            SN\,2019sxv & 2MASX\,J22195786+2537447& - & 0.0400 & 22:19:58.25 & +25:37:46.70  &\citet{2019sxv_disc}, \citet{2019sxv_spec} \\
            SN\,2019vkl & WISEA\,J015630.94+182623.5& - & 0.0640 &01:56:30.79  & +18:26:23.80 &\citet{2019vkl_disc}, \citet{2019vkl_spec} \\
            SN\,2019wmf & WISEA\,J102954.55+704710.0& - & 0.0600 & 10:29:53.88& +70:47:09.60 &\citet{2019wmf_disc}, \citet{2019wmf_spec} \\
            SN\,2019wnc & LEDA\,1183665& - & 0.0216 & 10:10:56.36 & +01:04:17.10 &\citet{2019wnc_disc}, \citet{2019wnc_spec} \\
            SN\,2020cn & - & - & 0.0770 & 21:28:10.73 & +79:06:36.30 & \citet{2020cn_disc}, \citet{2020cn_spec} \\
            SN\,2020fhw & WISEA\,J150029.88+171410.8& - & 0.1116 & 15:00:29.89 & +17:14:10.50 & \citet{2020fhw_disc}, \citet{2020fhw_spec} \\
            SN\,2020hem & 2MASX\,J15024011+0918137& - & 0.0935 & 15:02:40.15 & +09:18:13.80 & \citet{2020hem_disc}, \citet{2020hem_spec} \\
            SN\,2020rc & - & - & 0.0810 & 11:50:34.07 & -04:20:22.70 & \citet{2020rc_disc}, \citet{2020rc_spec} \\
            SN\,2020rno & - & - & 0.0640 & 01:18:51.79 & +18:40:09.50 &\citet{2020rno_disc}, \citet{2020rno_spec} \\
            SN\,2020sj & - & - & 0.0770 & 09:26:19.87 & -07:28:30.00 &\citet{2020sj_disc}, \citet{2020sj_spec} \\
            SN\,2020tis & WISEA\,J235725.57+250330.7& - & 0.0577 & 23:57:24.78 & +25:03:48.60 &\citet{2020tis_disc}, \citet{2020tis_spec} \\
            SN\,2020tyk & - & - & 0.0870 & 01:37:58.38 & +20:00:03.00 &\citet{2020tyk_disc}, \citet{2020tyk_spec} \\
            SN\,2020vci & - & -  & 0.1930 & 15:04:22.63& +51:04:56.80 &\citet{2020vci_disc}, \citet{2020vci_spec} \\
            SN\,2020vou & WISEA\,J233743.84+215110.8& - & 0.1190 & 23:37:43.80 & +21:51:14.60 &\citet{2020vou_disc}, \citet{2020vou_spec} \\
            SN\,2020xpo & GALEXASC\,J023225.46-125741.7& - & 0.0750 & 02:32:25.65 & -12:57:41.00&\citet{2020xpo_disc}, \citet{2020xpo_spec} \\
            SN\,2020yy & SDSS\,J141853.90+630945.& - & 0.0675 & 14:18:53.91 & +63:09:45.00 &\citet{2020yy_disc}, \citet{2020yy_spec}\\
            SN\,2020zos & WISEA\,J050632.07+074535.3& - & 0.1400 & 05:06:32.08 & +07:45:36.40 & \citet{2020zos_disc}, \citet{2020zos_spec} \\
            SN\,2020aafb & WISEA\,J015128.98+224947.2& - & 0.0780 & 01:51:28.95 & +22:49:44.10 &\citet{2020aafb_disc}, \citet{2020aafb_spec}\\
            SN\,2020aaut & LEDA\,1221074& - & 0.0757 & 12:09:12.54 & +02:15:24.30 & \citet{2020aaut_disc}, \citet{2020aaut_spec}\\
            
            SN\,2021bwf & SDSS\,J163457.25+562855.8& - & 0.1940 & 16:35:02.90 & +56:28:38.40 &\citet{2021bwf_disc}, \citet{2021bwf_spec}\\
            SN\,2021bxo & Mrk\,1209Ê& Sp & 0.0339 & 08:04:00.58 & +10:00:33.50 &\citet{2021bxo_disc}, \citet{2021bxo_spec} \\
            SN\,2021ezt & 2MASX\,J16583923+2629311& - & 0.0502 & 16:58:38.98 & +26:29:32.40  & \citet{2021ezt_disc}, \citet{2021ezt_spec}\\
            SN\,2021fel & WISEA\,J165158.55+623404.8& - & 0.0680 & 16:51:58.52 & +62:34:04.30  &\citet{2021fel_disc}, \citet{2021fel_spec}\\
            SN\,2021fpn & ZwCl\,1012-0047&- & 0.0424 & 10:13:47.65&	-00:54:55.00  &\citet{2021fpn_disc}, \citet{2021fpn_spec} \\
            SN\,2021gpw & WISEA\,J132154.84+164439.7&- & 0.0750 & 13:21:54.85&	+16:44:39.70  &\citet{2021gpw_disc}, \citet{2021gpw_spec}\\
            SN\,2021hsn & - &- & 0.0540 & 19:22:16.83&	+56:21:00.20 & \citet{2021hsn_disc}, \citet{2021hsn_spec}\\
            SN\,2021hur & LEDA 2773190& - & 0.0300 & 13:30:59.36 &	+77:12:52.60  &\citet{2021hur_disc}, \citet{2021hur_spec}\\
            SN\,2021iui & SDSS\,J141659.24+321406.7& - & 0.1060 & 14:16:59.19 & +32:14:06.50 &\citet{2021iui_disc}, \citet{2021iui_spec} \\
            SN\,2021kat & - & - & 0.1013 & 19:50:31.65 & +57:59:28.00 &\citet{2021kat_disc}, \citet{2021kat_spec} \\
            SN\,2021kqv & UGC\,10132& SB:b & 0.0510 & 15:56:52.03 & +78:27:48.90 &\citet{2021kqv_disc}, \citet{2021kqv_spec} \\
            SN\,2021kwc & NGC\,5231& SA & 0.0218 &13:35:47.91 & +02:59:59.30 &\citet{2021kwc_disc}, \citet{2021kwc_spec}\\
            SN\,2021kwj & WISEA\,J181949.35+561011.3& - & 0.0250 &18:19:49.50 & +56:10:01.20 &\citet{2021kwj_disc}, \citet{2021kwj_spec} \\
            SN\,2021lhy & WiggleZ\,R22J213215370-00310110&- & 0.1420 & 21:32:14.94	&-00:30:05.10  &\citet{2021lhy_disc}, \citet{2021lhy_spec} \\
            SN\,2021osr & 2MASX\,J21345199-0726529& -& 0.0855 & 21:34:51.91&	-07:26:54.50  & \citet{2021osr_disc}, \citet{2021osr_spec}\\
            SN\,2021qeq & - &- & 0.0550 &  16:54:26.77&	+53:43:26.30 &\citet{2021qeq_disc}, \citet{2021qeq_spec}\\
            SN\,2021qug & J220232.70-164537.8&- & 0.0570 & 22:02:32.45	&-16:45:36.40  &\citet{2021qug_disc}, \citet{2021qug_spec}\\
            SN\,2021ras & 2MASX\,J17385176+2622447&- & 0.0290 & 17:38:51.92& 	+26:22:44.20  &\citet{2021ras_disc}, \citet{2021ras_disc} \\
            SN\,2021srg & WISEA\,J231817.23+145004.& -& 0.0690 &23:18:17.26	&+14:50:04.80  &\citet{2021srg_disc}, \citet{2021srg_spec}\\
            SN\,2021ukt & UGC\,505& S & 0.01288 & 00:49:24.86&	-01:45:58.70  & \citet{2021ukt_disc}, \citet{2021ukt_spec}\\
            SN\,2021uru & LEDA\,864812& - & 0.0540 &02:52:22.67&	-18:41:47.50  &\citet{2021uru_disc}, \citet{2021uru_spec}\\
            SN\,2021vzp & 2MASX\,J03473546+0252586& - & 0.0310 &03:47:35.55&	+02:52:58.40 &\citet{2021vzp_disc}, \citet{2021vzp_spec}\\
            SN\,2021wrr & WISEA\,J171107.70+722919.8& - & 0.0480 & 17:11:07.79	&+72:29:19.80 &\citet{2021wrr_disc}, \citet{2021wrr_spec} \\
            SN\,2021yaz & - & -& 0.0390 &02:30:37.00 &+25:26:07.50  & \citet{2021yaz_disc}, \citet{2021yaz_spec} \\
            SN\,2021ydc & WISEA\,J012517.64+222324.8&- &0.0550&01:25:17.62 &	+22:23:25.30 &\citet{2021ydc_disc}, \citet{2021ydc_spec}\\
            SN\,2021yys & 2MASX\,J07254250+4449209&- & 0.0433 & 07:25:42.21 & +44:49:25.10 &\citet{2021yys_disc}, \citet{2021yys_spec} \\
            SN\,2021yyy & WISEA\,J224007.46-050012.4&- &0.0930 &22:40:07.55	&-05:00:12.50   &\citet{2021yyy_class}, \citet{2021yyy_spec} \\
            SN\,2022fnl & WISEA\,J153342.47+434445.5&- & 0.1035 & 15:33:42.47&	+43:44:45.40  &\citet{2022fnl_disc}, \citet{2022fnl_spec} \\
            SN\,2022jie & LEDA\,2669503& - & 0.0748 & 12:42:48.12	&+64:34:29.60  & \citet{2022jie_disc}, \citet{2022jie_spec} \\
            SN\,2022mds & WISEA\,J173233.61+431623.3& -& 0.0745 &17:32:33.90&	+43:16:27.10  &\citet{2022mds_disc}, \citet{2022mds_spec} \\
            SN\,2022mma & Z\,104-58& Sc & 0.0380 & 14:39:01.49	&+15:59:11.70 & \citet{2022mma_disc}, \citet{2022mma_spec}\\
            SN\,2022myl & Z\,169-7& - & 0.0371 & 16:51:02.37 &	+30:39:52.30 & \citet{2022myl_disc}, \citet{2022myl_spec} \\
            SN\,2022oeh & LEDA\,2273049&- & 0.0300 & 19:16:50.01&	+45:57:18.60 & \citet{2022oeh_disc}, \citet{2022oeh_spec}\\
            SN\,2022paz & - & -& 0.0660 &18:05:09.01&	+32:09:00.90  &\citet{2022paz_disc}, \citet{2022paz_spec}\\
            SN\,2022prr & NGC\,6745& -& 0.0152& 19:01:41.90& +40:45:03.70  & \citet{2022prr_disc}, \citet{2022prr_spec}\\
            SN\,2022pss & - & -& 0.0600 & 01:28:19.22	&+20:23:45.30  & \citet{2022pss_disc}, \citet{2022pss_spec} \\
            SN\,2022rhl & - &- & 0.1180 & 19:20:44.21 & +46:52:54.70 & \citet{2022rhl_dsic}, \citet{2022rhl_spec} \\
            SN\,2022tbh & LEDA 2706824 & - & 0.0520 & 17:25:09.72 & +67:49:45.77 & \citet{2022tbh_disc}, \citet{2022tbh_spec} \\
            SN\,2022ymc & - &- & 0.0280 & 03:13:40.62 & -01:16:32.21 & \citet{2022ymc_disc}, \citet{2022ymc_spec} \\
            SN\,2022zyd & - &- & 0.0630& 02:08:33.20 & +52:49:36.37 & \citet{2022zyd_disc}, \citet{2022zyd_spec} \\
            SN\,2023adz & - &- & 0.0500 & 11:32:37.72 & +68:23:48.74 & \citet{2023adz_disc}, \citet{2023adz_spec} \\
            SN\,2023awp & NGC\,5936 &Sc & 0.0136 & 15:30:01.54  & +12:59:15.56 & \citet{2023awp_disc}, \citet{2023awp_spec} \\
            SN\,2023erg & - &- & 0.0670 & 15:30:05.64 & +35:40:19.92 & \citet{2023erg_disc}, \citet{2023erg_spec} \\
            SN\,2023iex & 2MASX\,J21395531+2439332 &- & 0.0290 & 21:39:56.07  &  +24:39:34.67 & \citet{2023iex_disc}, \citet{2023iex_spec} \\
            SN\,2023kqw & SDSS\,J143643.35+120523.8 &- & 0.0547 & 14:36:43.33 &  +12:05:24.26 & \citet{2023kqw_disc}, \citet{2023kqw_spec} \\
            SN\,2023meo & SDSS\,J143643.35+120523.8 &- & 0.0540 & 21:40:59.12 & +24:00:26.28 & \citet{2023meo_disc}, \citet{2023meo_spec} \\
            SN\,2023nof & 2dFGRS\,TGS855Z331 &- & 0.0692 & 22:39:44.18 &  -15:50:04.38 & \citet{2023nof_disc}, \citet{2023nof_spec} \\
            SN\,2023pfj & - &- & 0.1110 & 02:17:12.15 & +01:35:56.35 & \citet{2023pfj_disc}, \citet{2023pfj_spec} \\
            SN\,2023usc & - &- & 0.0600 & 05:21:31.65  &  +00:28:20.72 & \citet{2023usc_disc}, \citet{2023usc_spec} \\
            \hline
            SN\,2019uit & NGP9\,F378-0626697&- & 0.0710 & 12:50:16.09& +21:20:15.40 & \citet{2019uit_disc}, \citet{2019uit_spec} \\
            SN\,2020jhs & - & -& 0.0650 & 09:28:14.09& +25:40:13.30 & \citet{2020jhs_disc},\citet{2020jhs_spec} \\
            SN\,2020qmj & LEDA\,1280605&- & 0.0220 & 00:44:06.006 &+05:15:35.93  & \citet{2020qmj_disc}, \citet{2020qmj_spec} \\
            SN\,2020tan & SDSS\,J010944.57+002024.3& -& 0.0790 &01:09:44.599& +00:20:24.93 & \citet{2020tan_disc}, \citet{2020tan_spec} \\
            SN\,2020uaq & SDSS\,J153452.65+105839.1& -& 0.1150 & 15:34:49.030& +10:58:37.34 & \citet{2020uaq_disc}, \citet{2020uaq_spec}\\
            SN\,2020utm & LEDA\,1078124&	- & 0.0440 & 03:15:27.45 &-03:00:38.19 &\citet{2020utm_disc}, \citet{2020utm_spec}\\
            SN\,2020ybn & - & -& 0.0960 & 06:11:02.260& -22:17:50.28  &\citet{2020ybn_disc}, \citet{2020ybn_spec} \\
            SN\,2021bmv &  - & - & 0.0900 & 04:26:38.67  & -11:53:45.72 &\citet{2021bmv_disc}, \citet{SN2021bmv_spec}\\
            SN\,2021aapa & MCG-02-11-023  & - & 0.0320 & 22:04:24.170 &-19:32:28.63 & \citet{2021aapa_disc}, \citet{2021aapa_spec}\\
            \hline
            PTF\,10abui & - & -& 0.0516&06:12:18.46	&-22:46:15.60 & \citet{Nyholm_2020}\\
            PTF\,10acsq & SDSS\,J080133.14+464553.0& - &0.1730 & 08:01:33.17	&+46:45:52.50 & \citet{Nyholm_2020} \\
            PTF\,10cwl & - & -& 0.0845& 12:36:22.06&	+07:47:38.00 &\citet{ptf10cwl} \\
            PTF\,10cwx & 2dFGRS\,TGN321Z210 &- &0.0731 &12:33:16.53&	-00:03:10.60 & \citet{Nyholm_2020} \\
            PTF\,10ewc & - & -& 0.0542 &14:01:59.08&	+33:50:11.60  &\citet{Nyholm_2020} \\
            PTF\,10fjh & UGC\,10547 & SAb & 0.0321 &16:46:55.36&+34:09:34.70 & \citet{ptf10fjh_disc}, \citet{ptf10fjh_class} \\
            PTF\,10flx & WISEA\,J164658.92+642650.0&- & 0.0674 &16:46:58.28&	+64:26:48.50 & \citet{Nyholm_2020} \\
            PTF\,10gvd & - & -& 0.0693 &16:53:02.12&	+67:00:08.90  &\citet{Nyholm_2020}\\
            PTF\,10gvf & SDSS\,J111344.87+533749.6&- & 0.0810 &11:13:45.24&	+53:37:44.90& \citet{Nyholm_2020} \\
            PTF\,10oug & - & -& 0.1501 &17:20:44.79&	+29:04:25.60 &\citet{Nyholm_2020} \\
            PTF\,10qwu & - &- & 0.2259 &16:51:10.36&	+28:18:06.20 & \citet{Nyholm_2020}\\
            PTF\,10tel & A\,J172130+4807& -& 0.0349 & 17:21:30.68	&+48:0:47.40& \citet{2010mc_disc}, \citet{2010mc_spec}\\
            PTF\,10tyd & LEDA\,1818789&	- & 0.0633 &17:09:19.41	&+27:49:08.60 & \citet{Nyholm_2020} \\
            PTF\,10weh & - & -& 0.1379 & 17:26:50.46	&+58:51:07.40 & \citet{ptf10weh}\\
            PTF\,11fzz & - & -& 0.0813 &11:10:46.68&	+54:06:18.80 & \citet{Nyholm_2020}\\
            PTF\,11oxu & WISEA\,J033834.32+223242.7& - & 0.0878 &03:38:34.38&	+22:32:59.40 &\citet{ptf11oxu}\\
            iPTF\,13agz & SDSS\,J143432.09+250941.5& - &0.0572 &14:34:32.12 &+25:09:43.60 & \citet{Nyholm_2020}\\
            iPTF\,13asr & - & -& 0.1543 &12:47:28.61&	+27:04:03.60 &  \citet{Nyholm_2020}\\
            iPTF\,13cuf & - &- & 0.2199 &02:04:52.97&	+14:37:59.70  &  \citet{iptf13cuf}\\
            iPTF\,14bpa & - &- & 0.1220 & 15:26:59.96&	+24:41:17.50&   \citet{Nyholm_2020}\\
            iPTF\,15aym & MCG+09-22-059& -&0.0334 &13:26:26.67&	+55:23:43.40  & \citet{Nyholm_2020}\\
            iPTF\,15bky & NGC\,5837& S &0.0288 &15:04:40.80	&+12:37:43.40 & \citet{iptf15bky_spec}\\
            iPTF\,15blp & - & -& 0.1949 &16:27:15.21&	+41:08:58.10  & \citet{Nyholm_2020} \\
            iPTF\,15eqr & WISEA\,J040115.10+331700.1& -& 0.0467&04:01:15.67&	+33:16:58.30 & \citet{iptf15eqr_disc}\\
            iPTF\,16fb & WISEA\,J102209.12+152822.4& -& 0.0811&10:22:09.25	&+15:28:19.20  & \citet{iptf16fb_disc}, \citet{iptf16fb_spec}\\
            \hline
            PSc\,130812 & - &- &0.0141 &12:22:32.93    &	+47:19:48.36& \citet{villar2019supernova} \\
            PSc\,130942 & - &- &0.0087&16:01:26.26     &	+55:14:25.80 & \citet{villar2019supernova} \\
            PSc\,150582  & - & -& 0.0099&10:47:16.30    &	+59:01:26.04 & \citet{villar2019supernova} \\
            PSc\,300140  & - &- &0.0406& 23:30:48.91    &	-01:12:26.64 & \citet{villar2019supernova} \\
            PSc\,350633 & - & -&0.0059 & 10:45:19.15    &	+59:04:54.84  & \citet{villar2019supernova} \\
            PSc\,360030 & - &- &0.0142 &12:20:37.27	&46:31:22.10  & \citet{villar2019supernova} \\
            PSc\,430005 & - & -& 0.0564&22:17:39.86    	&+00:37:35.40  & \citet{villar2019supernova} \\
            PSc\,450025 & - & -&0.0237&02:23:33.53    	&-05:32:01.68  & \citet{villar2019supernova} \\
            PSc\,450266  & - & -& 0.0259& 08:40:55.46      &	+44:56:41.64   & \citet{villar2019supernova} \\
            PSc\,460001  & - & -& 0.0077& 03:33:04.15    &-28:12:49.32  & \citet{villar2019supernova} \\
            PSc\,480845 & - & - &0.0135 &12:29:08.59       &	+46:52:47.28   & \citet{villar2019supernova} \\
            PSc\,500063 & - & - & 0.0118&12:25:06.67  &	+47:55:33.96  & \citet{villar2019supernova} \\
            PSc\,530085 & - & - &0.0661 & 22:12:43.46  &-00:05:07.44 & \citet{villar2019supernova} \\
            PSc\,550221  & - & - &0.0243 & 02:21:52.39    &	-03:26:00.60  & \citet{villar2019supernova} \\
            PSc\,580280 & - & - &0.0074 &03:32:54.12& -28:14:24.00 & \citet{villar2019supernova} \\
            PSc\,580289 & - & - &0.0373 &02:28:34.32    	&-04:54:46.08   & \citet{villar2019supernova} \\

    \end{longtable*}
   % \end{center}
\end{longrotatetable}

\begin{longrotatetable}
    \movetabledown=3mm
\begin{scriptsize}
\begin{longtable*}{lllllll}
\caption{Table of SNe\,IIn excluded from our sample. This table shows the transient name (survey name or IAU name), the reason why the transient was excluded from our sample, the J2000 coordinates and the source of the discovery/spectroscopic classification. \textit{* as presented in \citet{Ransome_2021}}. Note that in this table, we report some objects as having `No data'. This means that there was no easily accessible public data, not necessarily that there is no photometric data existent in the literature.}
\label{tab:samplecut}

\\
\hline
SN Name & Host & Host type & $z$ & R.A (J2000) & Dec (J2000)  & Ref. \\
\hline
\endfirsthead
\multicolumn{7}{l}{\tablename\ \thetable\ --- Continued from previous page\ldots}\\
\hline
SN Name & Host & Exc. reason & $z$ & R.A (J2000) & Dec (J2000) & Ref. \\
\hline
\endhead
\hline
\endlastfoot
\hline
\multicolumn{7}{r}{Continued on next page\ldots}\\
\endfoot
        SN\,1987B & NGC\,5850& No data &0.0085  & 15:07:02.92 & 01:30:13.20 & \citet{Schlegel_1996} \\
        SN\,1989R & UGC\,2912& No data & 0.0180 & 03:59:32.56 & 42:37:09.20  & P.~Challis* \\
        SN\,1993N & UGC\,5695& No data &   0.0098 & 10:29:46.33 & 13:01:14.00 & \citet{Ransome_2021}  \\%A.~Filippenko\\
        SN\,1994ak & NGC\,2782& No data & 0.0085 & 09:14:01.47 & 40:06:21.50  & \citet{1994ak} \\
        SN\,1995G & NGC\,1643& No data & 0.0160 & 04:43:44.26 & -05:18:53.70  & \citet{1995g} \\
        SN\,1997eg & NGC\,5012& No data & 0.0087 & 13:11:36.73 & 22:55:29.40  & \citet{1997eg} \\
        SN\,1999eb & NGC\,664& No data & 0.0180 & 01:43:45.45 & 04:13:25.90  & \citet{1999eb}  \\
        SN\,1999el &NGC\,6951& No data & 0.0047 & 20:37:18.03 & 66:06:11.90 & \citet{1999el} \\
        SN\,2000P & NGC\,4965& No data & 0.0075 & 13:07:10.53 & -28:14:02.50  &\citet{2000p}  \\
        SN\,2000eo & MCG-02-09-03& No data  & 0.0100 & 03:09:08.17 & -10:17:55.30  & \citet{2000eo}  \\
         SN\,2003G & IC\,208& No data & 0.0120 & 02:08:28.13 & 06:23:51.90  & \citet{2003G}  \\
        SN\,2005gl & NGC\,266& No data & 0.0155 & 00:49:50.02 & 32:16:56.80  & \citet{2005gl}  \\
        SN\,2005ip & NGC\,2906& Long-lived & 0.0072 & 09:32:06.42 & 08:26:44.40 & \citet{2005ip}  \\
        SN\,2005kj & A084009-0536& No data & 0.0160 & 04:40:09.18 & -05:36:02.20  & \citet{2005kj}  \\
        SN\,2006jd & UGC\,4179& Long-lived & 0.0186 & 08:02:07.43 & 00:48:31.50  & \citet{2006jd}  \\
        SN\,2008J & MCG-02-07-33& Ia-CSM & 0.0159 & 02:34:24.20 & -10:50:38.50  & \citet{2008J}  \\
        SN\,2008S & NGC\,6946& Imp? & 0.0002 & 20:34:45.35 & 60:05:57.80 & \citet{2008s}  \\
        SN\,2009ip & NGC\,7259& No data & 0.0059 & 22:23:08.30 & -28:56:52.40 & \citet{2009ip} \\
        SN\,2009kn & MCG-03-21-06& No data & 0.0143 & 08:09:43.04 & -17:44:51.30  & \citet{2009kn}  \\
        SN\,2010bt & NGC\,7130& No data & 0.0162 & 21:48:20.22 & -34:57:16.50 & N.~Elias-Rosa*\\
        SN\,2010jl & UGC\,5189A& Long-lived & 0.0107 & 09:42:53.33 & 09:29:41.80  &\citet{2010jl}\\
        SN\,2010jp & A\,061630-2124& No data & 0.0092 & 06:16:30.63 & -21:24:36.30  & \citet{2010jp}  \\ 
        SN\,2011A & NGC\,4902& No data &  0.0089  & 13:01:01.19 & -14:31:34.80 & \citet{2011a} \\
        SN\,2011ht & UGC\,5460& No data & 0.0036 & 10:08:10.56 & 51:50:57.12  & \citet{2011ht} \\
        SN\,2012ca & ESO\,336-G9& Ia-CSM & 0.0190 & 18:41:07.25 & -41:47:38.40  & \citet{2012ca} \\
       SN\,2013fc & ESO\,154-G10& No data  & 0.0186 & 02:45:08.95 & -55:44:27.30  & \citet{2013fc}  \\    
        SN\,2013gc & ESO\,430-G20& No data & 0.0034 & 08:07:11.88 & -28:03:26.30 & A.~Reguitti* \\
         SN\,2015da &  NGC\,5337& Long-lived & 0.0072 & 13:52:24.11 & 39:41:28.60  & J.~Zhang*\\
        SN\,2016bdu & -  & No data & 0.0170&  13:10:13.95 & 32:31:14.07  &N.~Elias-Rosa*\\
        \hline
        SN\,2018jdo & MCG+06-06-007& Short baseline & 0.0390 & 02:17:48.70 & +35:48:29.10 &  \citet{2018jdo_disc}, \citet{2018jdo_spec} \\
        SN\,2019cac & 2MASX\,J13504376-0230249& Spec & 0.0467& 13:50:43.89 & -02:30:24.90  &\citet{2019cac_disc}, \citet{2019cac_spec} \\
        SN\,2019cqw & 2MASX\,J09060852+6731212& Short baseline & 0.0490& 09:06:16.33 & +67:30:50.1 &\citet{2019cqw_disc}, \citet{2019cqw_spec} \\
        SN\,2019ctt & - & Spec& 0.0460 & 10:00:42.29	&+12:02:23.40  & \citet{2019ctt_disc}, \citet{2019ctt_spec} \\
        SN\,2019dde & GAMA\,50846& Spec & 0.0570&14:28:12.03 & -01:36:15.00 & \citet{2019dde_disc}, \citet{2019dde_spec} \\
        SN\,2019ejb & -  &Spec- & 0.1180 & 14:08:28.58	 & +29:16:11.10 & \citet{2019ejb_disc} \\
        SN\,2019gjs & UGC\,9634& Short baseline & 0.0430 &  14:58:57.16& +20:03:10.50  &\citet{2019gjs_disc}, \citet{2019gjs_spec}\\
        SN\,2019mom & - & Short baseline & 0.0488 & 01:55:36.52	 & +53:35:30.80 & \citet{2019mom_disc}, \citet{2019mom_spec} \\
        SN\,2019njv & 2MASX\,J20195707+1522402& Short baseline & 0.0146 &20:19:57.19 & +15:22:38.70& \citet{2019njv_disc}, \citet{2019njv_spec} \\
        SN\,2019smj & WISEA\,J074940.69+050411.1& Double peaked & 0.0600 & 07:49:40.71 & +05:04:27.10  & \citet{2019smj_disc}, \citet{2019smj_spec}\\
        SN\,2019tlv & -  & Short baseline & 0.0440 & 00:19:22.34 & +21:46:32.9 & \citet{2019tlv_disc}, \citet{2019tlv_spec} \\
        SN\,2019tpl & 2MASX\,J01000084-0306377& Short baseline & 0.0760 & 01:00:00.68 & -03:06:30.20  &\citet{2019tpl_disc}\\
        SN\,2019wrt & WISEA\,J051648.34-072834.4&Short baseline  & 0.0573 & 05:16:47.98 & -07:28:43.20 & \citet{2019wrt_disc}, \citet{2019wrt_spec}\\
        SN\,2019zrk & UGC\,6625& Double peaked & 0.0362 & 11:39:47.40 & +19:55:46.70 & \citet{2019zrk_disc}, \citet{2019zrk_spec}\\
        SN\,2020acct & NGC\,2981& Double peaked & 0.0350 & 09:44:56.05 & +31:05:45.70 & \citet{2020acct_disc}, \citet{Angus_2024} \\
        SN\,2020cke & WISEA\,J084717.88-203139.0& Short baseline & 0.0350 & 08:47:17.91 & -20:31:39.30  &\citet{2020cke_disc}, \citet{2020cke_spec}\\
        SN\,2020hcr & 2MASX\,J13542244-0521311& Spec & 0.0510 & 13:54:22.59 & -05:21:25.20 & \citet{2020hcr_disc, 2020hcr_spec}\\
        SN\,2020hfn & MCG-02-50-007& Short baseline & 0.0249 & 19:48:47.98 & -10:34:22.60 & \citet{2020hfn_disc}, \citet{2020hfn_spec}  \\
        SN\,2020km & WISEA\,J043711.04+722725.6& Spec & 0.1029 & 04:37:08.94 & +72:27:37.00 & \citet{2020km_disc}, \citet{2020km_spec}\\
        SN\,2020qmj & LEDA\,1280605& Spec & 0.0220 & 00:44:06.00 & +05:15:35.80 & \citet{2020qmj_disc}, \citet{2020qmj_spec} \\
        SN\,2020qpo & - & Spec & 0.0530 &17:16:41.36&	+57:54:12.30 &\citet{2020qpo_disc}, \citet{2020qpo_spec}\\
        SN\,2020qyy & SDSS\,J155554.94+362451.7& Short baseline & 0.0680 & 15:55:53.01 & +36:24:35.40 &\citet{2020qyy_disc}, \citet{2020qyy_spec}\\
        SN\,2020xkx& - & Double peaked & 0.0415 & 23:20:28.17 & +22:59:12.80 &\citet{2020xkx_disc}, \citet{2020xkx_spec} \\
        SN\,2021cvd & UGC\,8195& Short baseline & 0.0230 & 13:06:23.22 & +29:39:27.90  &\citet{2021cvd_disc}, \citet{2021cvd_spec}\\
        SN\,2021qqp & LEDA\,1741591& Double peaked& 0.0410 &22:32:40.41	&+25:34:34.80   &\citet{2021qqp_disc}, \citet{2021qqp_spec}\\
        SN\,2022mgr & - & Spec & 0.0680 &18:32:37.01	&+20:36:09.40 &\citet{2022mgr_disc}, \citet{2022mgr_spec}\\
        SN\,2022gzi & WISEA\,J174604.84+421634.5& AGN? & 0.0890 &17:46:04.84	&+42:16:34.50  &\citet{2022gzi_disc}, \citet{2022gzi_spec}\\
        SN\,2022hev & 2MASX\,J13143311-1025416& Short baseline& 0.0540 &13:14:36.72&	-10:25:38.80 & \citet{2022hev_disc} \\
        SN\,2022hsu & UGC\,11946& Spec & 0.0180 &22:11:37.72	&+46:18:40.00 & \citet{2022hsu_disc}, \citet{2022hsu_spec} \\
        SN\,2022iaz & LEDA\,860052& Spec & 0.0670 &12:30:31.31	&-19:04:42.10   &\citet{2022iaz_disc}  \\
        SN\,2022iep & LEDA\,2164253& Spec & 0.0250 & 16:29:41.96	& +40:20:05.70  &\citet{2022iep_disc}, \citet{2022iep_spec} \\
        \hline
        SN\,2020bwr & SDSS\,J163459.26+361227.7& Short baseline& 0.0606 &16:34:59.33 &+36:12:27.43 & \citet{2020bwr_disc}, \citet{2020bwr_spec} \\
        SN\,2020noz & Z\,42-125& Short baseline & 0.0250 & 12:29:00.25 &+07:50:58.42  & \citet{2020noz_disc}, \citet{2020noz_spec} \\
        SN\,2020rdu & - & Spec& - &15:57:02.51 &+24:34:50.72  & \citet{2020rdu_disc} \\
        SN\,2021xre & - & Spec& 0.0600 &00:15:07.19    &+17:29:55.38  & \citet{2021xre_disc}, \citet{2021xre_spec}  \\
        \hline
        PTF\,09tm & 2MASX\,J13465543+6133179& Short baseline & 0.0349 & 13:46:55.94&	+61:33:15.60 & \citet{Nyholm_2020}\\
        PTF\,09uy & - & Short baseline& 0.3135& 12:43:55.80&	+74:41:08.10  & \citet{Nyholm_2020}\\
        PTF\,09bcl & - & Short baseline& 0.0621& 18:06:26.78&	+17:51:43.00  & \citet{Nyholm_2020}\\
        PTF\,10achk & 2MASX\,J03055776-1031246&Short baseline &0.0325 & 03:05:57.54	&-10:31:21.00 & \citet{Nyholm_2020} \\
        PTF\,10vag & - & Short baseline& 0.0517 &21:47:18.48&	+18:07:51.5  &\citet{Nyholm_2020}\\
        PTF\,10xgo & - & Short baseline& 0.0336 &21:55:57.38&	+01:19:14.10 & \citet{Nyholm_2020}\\
        PTF\,11mpg &  SDSS\,J221736.67+003647.6& Spec& 0.0933 & 22:17:36.66	&+00:36:48.40 & \citet{Nyholm_2020}\\
        PTF\,11qnf & UGC\,3344& Bumpy & 0.0148 & 05:44:54.14&	+69:09:06.90 &\citet{Nyholm_2020}\\
        PTF\,11qqj & WISEA\,J095801.73+004315.2& Spec& 0.0931 & 09:58:01.64&	+00:43:14.70 & \citet{Nyholm_2020}\\
        PTF\,11rfr & - &Short baseline & 0.0675 & 01:42:16.98&	+29:16:25.70 &\citet{Nyholm_2020}\\
        PTF\,11rlv & WISEA\,J124934.15-092042.7& Short baseline& 0.1323 & 12:49:34.04&	-09:20:40.50 & \citet{Nyholm_2020}\\
        PTF\,12cxj & LEDA\,2282052& Short baseline	 & 0.0356 & 13:12:38.68&	+46:29:06.30 & \citet{Nyholm_2020}\\
        PTF\,12frn & WISEA\,J162200.01+320938.9&Short baseline & 0.1365 & 16:22:00.16	&+32:09:38.90 & \citet{Nyholm_2020}\\
        PTF\,12glz & - & Short baseline& 0.0793& 15:54:53.04&	+03:32:08.50 & \citet{Nyholm_2020}\\
        PTF\,12ksy & - & Short baseline& 0.0314& 04:11:46.09&-12:28:00.80  & \citet{Nyholm_2020}\\
        iPTF\,13aki & - & Short baseline & 0.1610 & 14:35:34.35&	+38:38:31.00 &  \citet{Nyholm_2020}\\
        iPTF\,14bcw & - & Short baseline& 0.1206& 13:48:41.18&	+35:52:17.10  & \citet{Nyholm_2020}\\
        \hline
        PSc070763 & - & -&0.0165 & 23:29:48.70&	-00:17:29.80  & \citet{villar2019supernova} \\
        PSc110468 & - & Short baseline& 0.0179 &10:00:45.53&	02:01:24.20 & \citet{villar2019supernova} \\
        PSc120067 & - & Short baseline&0.0071 & 10:56:21.02&	57:39:55.40 & \citet{villar2019supernova} \\
        PSc360359 & - & Spec& - & 14:07:39.94&	52:14:37.70 & \citet{villar2019supernova} \\
        PSc380012 & - & Short baseline& 0.0241&10:02:21.34	&01:08:57.50   & \citet{villar2019supernova} \\
        PSc390712 &-  & Short baseline&0.0081 &16:05:02.62&	56:13:54.50  & \citet{villar2019supernova} \\
        PSc420248 & - & Short baseline& 0.0244& 02:23:54.62&	-03:36:47.90 & \citet{villar2019supernova}\\
        PSc580234 & - & Short baseline & 0.0253&10:03:44.64&	02:45:47.90   & \citet{villar2019supernova} \\

\end{longtable*}
\end{scriptsize}
\end{longrotatetable}

\bibliography{bibliography}{}
\bibliographystyle{aasjournal}



\end{document}